\documentclass[12pt,preprint]{aastex}

\newcommand{\smpy}{\mathrm{\,M_\odot\,yr^{-1}}}
\newcommand{\percc}{\mathrm{\,cm^{-3}}}
\newcommand{\msun}{\mathrm{\,M_\odot}}
\newcommand{\sgra}{Sgr~A$\textrm{*}$}

\renewcommand {\deg}   {\mbox{$^\circ$}}

\newcommand   {\arcs}  {\mbox{$^{\prime\prime}$}}

\newcommand   {\kms}   {\mbox{km\,s$^{-1}$}}
\renewcommand {\ga}    {\mbox{\rlap{\hbox{\lower5pt\hbox{$\sim$}}}\hbox{$>$}}}
\renewcommand {\la}    {\mbox{\rlap{\hbox{\lower5pt\hbox{$\sim$}}}\hbox{$<$}}}

\begin{document}
\pagenumbering{arabic} 
\def\kms {\hbox{km{\hskip0.1em}s$^{-1}$}} % km/s
\voffset=-0.8in

\def\msol{\hbox{$\hbox{M}_\odot$}}
\def\lsol{\hbox{$\hbox{L}_\odot$}}
\def\kms{km s$^{-1}$}
\def\Blos{B$_{\rm los}$}
\def\etal   {{\it et al.}}                     % et al
\def\psec           {$.\negthinspace^{s}$}
\def\pasec          {$.\negthinspace^{\prime\prime}$}
\def\pdeg           {$.\kern-.25em ^{^\circ}$}
\def\degree{\ifmmode{^\circ} \else{$^\circ$}\fi}
\def\ut #1 #2 { \, \textrm{#1}^{#2}} % \ut unit p  unit^p 
\def\u #1 { \, \textrm{#1}}          % \u unit     unit
\def\nH {n_\mathrm{H}}
\def\ddeg   {\hbox{$.\!\!^\circ$}}              % Degrees over dot
\def\deg    {$^{\circ}$}                        % Degrees symbol
\def\le     {$\leq$}                            % <=
\def\sec    {$^{\rm s}$}                        % Second of time
\def\msol   {\hbox{$M_\odot$}}                  % Solar mass
\def\i      {\hbox{\it I}}                      % italic I
\def\v      {\hbox{\it V}}                      % italic V
\def\dasec  {\hbox{$.\!\!^{\prime\prime}$}}     % Arcseconds over dot
\def\asec   {$^{\prime\prime}$}                 % Arcseconds symbol
\def\dasec  {\hbox{$.\!\!^{\prime\prime}$}}     % Arcseconds over dot
\def\dsec   {\hbox{$.\!\!^{\rm s}$}}            % Second over dot
\def\min    {$^{\rm m}$}                        % Minutes of time
\def\hour   {$^{\rm h}$}                        % Hours of time
\def\amin   {$^{\prime}$}                       % Arcminutes symbol
\def\lsol{\, \hbox{$\hbox{L}_\odot$}}
\def\sec    {$^{\rm s}$}                        % Second of time     
\def\etal   {{\it et al.}}                     % et al.
\def\la{\lower.4ex\hbox{$\;\buildrel <\over{\scriptstyle\sim}\;$}}
\def\ga{\lower.4ex\hbox{$\;\buildrel >\over{\scriptstyle\sim}\;$}}
\def\refitem{\par\noindent\hangindent\parindent}
\oddsidemargin = 0pt \topmargin = 0pt \hoffset = 0mm \voffset = -17mm
\textwidth = 160mm  \textheight = 244mm
\parindent 0pt
\parskip 5pt
%\pagestyle{empty}

%\slugcomment{subbmitted to  ApJ June 7, 2009}
%\slugcomment{to be submitted to ApJ}
%\slugcomment{Submitted ApJL}
\shorttitle{Sgr A*}
\shortauthors{}
%\input psfig.sty
% and Star Formation within the S Cluster} 

\title{Sgr A* and its Environment: Low Mass Star Formation,\\
the Origin of X-ray Gas and Collimated Outflow}
\author{F. Yusef-Zadeh$^1$, M. Wardle$^2$, R. Sch\"odel$^3$, 
D. A. Roberts$^1$, W. Cotton$^4$,\\ 
H. Bushouse$^5$,  R. Arendt$^6$, 
 \& M. Royster$^1$}
\affil{$^1$CIERA, Department of Physics and Astronomy Northwestern University, Evanston, IL 60208}
\affil{$^2$Department of Physics and Astronomy,  Macquarie University, Sydney NSW 2109, Australia}
\affil{$^3$Instituto de Astrofi$'$sica de Andaluci$'$a (CSIC), Glorieta de la Astronomica, 18008 Granada,  Spain} 
\affil{$^4$National Radio Astronomy Observatory,  Charlottesville, VA 22903}
\affil{$^5$Space Telescope Science Institute, 3700 San Martin Drive, Baltimore, MD  21218}
\affil{$^6$NASA GSFC,  Code 665, Greenbelt, MD 20771} 

%CRESST, University of Maryland-Baltimore County, Baltimore, MD 21250 and 
%\affil{$^5$Dept of Physics, University of Alberta, 4-183 CCIS, Edmonton AB T6G 2E1, Canada}
%and Research Centre for Astronomy, 
%Astrophysics \& Astrophotonics,
%and Center for Interdisciplinary Research in Astronomy, 
%Radio Continuum Observations of the Galactic Center:\\

\begin{abstract} 
We present high-resolution multiwavelength radio continuum images of the region within 150\arcs\ of \sgra, revealing a 
number of new extended features and stellar sources in this region. First, we detect a continuous 2\arcs\ east-west ridge of 
radio emission, linking \sgra\ and a cluster of stars associated with IRS 13N and IRS 13E. The ridge suggests that an 
outflow of east-west blob-like structures is emerging from \sgra.
In particular, we find 
arc-like radio structures within the ridge with morphologies suggestive of photoevaporative protoplanetary disks.  We 
use infrared K$_s$ and L$'$ fluxes to show that the emission has similar characteristics to those 
 of a protoplanetary disk 
irradiated by the intense radiation field at the Galactic center. This suggests that star formation has taken 
place within the S cluster 2\arcs\ from \sgra. We suggest  that the diffuse X-ray emission associated with Sgr A* is 
due to an expanding hot wind produced by the mass loss from  
B-type main sequence  stars, and/or the disks of 
photoevaporation of low mass young stellar objects (YSOs) 
at a rate $\sim 10^{-6}\msol\,\mathrm{yr}^{-1}$. The proposed model naturally reduces the inferred 
accretion rate and is an 
alternative to the inflow-outflow style models to explain the underluminous nature of Sgr A*.
Second, on a scale of 5$''$ from Sgr A*, we detect new cometary radio and infrared 
sources  at a position angle PA$\sim50^\circ$ which is similar to that of two other cometary sources X3 and X7, all of 
which face Sgr A*. In addition, we detect a striking tower of 
radio emission at a PA$\sim50^\circ-60^\circ$ along the major axis of the Sgr A East SNR shell
on a scale of 150$''$ from Sgr A*.  
We suggest  that  the cometary sources and the tower feature 
are tracing  interaction sites of a mildly relativistic jet
from Sgr A*
with the atmosphere of stars and the nonthermal Sgr A East shell at a PA$\sim50-60^\circ$
 with   $\dot{M}\sim1\times10^{-7}\smpy$, and opening angle 10 degrees.
Lastly, we suggest that the east-west ridge of radio emission  traces  an outflow 
that is  potentially associated with past flaring  activity from Sgr A*. 
The  position angle of the outflow driven by  flaring activity is close to $-90^\circ$ which is different than 
the PA$\sim60^\circ$ of the radio ridge. 
\end{abstract} 

%This  diffuse X-ray emission which  coincident with the S cluster can also be explained by 
% from disks of young stars by tidal truncation 

\section{Introduction}

A 4$\times 10^6$ \msol\, black hole is coincident with the compact nonthermal radio source \sgra\ at the center of the 
Galaxy (Ghez \etal\, 2008; Gillessen \etal\, 2009; Reid and Brunthaler 2004). The estimated mass accretion rate onto 
Sgr A* is several orders of magnitude smaller than the rate at which young, windy stars in the innermost 0.5 pc supply 
mass to the Bondi radius of Sgr A* (Coker \& Melia 1997; Cuadra \etal\, 2006, 2008). $Chandra$ observations have 
characterized the X-ray emission surrounding \sgra\ to be spatially extended with a radius of $\sim$1.5\arcs\ 
(Baganoff \etal\, 2003; Wang \etal\, 2013). The X-ray luminosity is interpreted as arising from a radiatively 
inefficient accretion flow (RIAF, e.g. Yuan \etal\, 2004; Moscibrokzka \etal\, 2009). In this model, a fraction of the 
gaseous material accretes onto \sgra\ and the rest is driven off as an outflow from \sgra\ (e.g., Quataert 2004; 
Shcherbakov and Baganoff 2010; Wang \etal\, 2013). Another  mechanism that may reduce the accretion rate is interaction 
with a 
jet or an outflow limiting the amount of gas falling onto Sgr~A* (Yusef-Zadeh \etal\, 2014a), thus modifying the 
 accretion flow. In this picture, the interaction of the outflow with the surrounding gas or the 
atmosphere  of mass-losing stars can provide an estimate of the power of the outflow. 

Two different types of activity  are associated with Sgr A*. One
is flaring 
on hourly time scale at multiple wavelengths (e.g., Baganoff \etal\, 2001; 
Genzel \etal\, 2003). 
Observations of Sgr A* have detected a time delay at submm, mm, and  radio wavelengths consistent with a scenario  
in which plasma blobs expand away from the disk, becoming visible at successive 
longer  wavelengths
as the optical depths become of order unity  effects (Yusef-Zadeh \etal\,  2006, 2008, 2009; Marrone \etal\,  2008; 
Eckart \etal\, 2008; Brinkerink \etal\, 2015).  
The other is  a jet-driven outflow (e.g., Falcke \& Markoff 2000). 
Unlike the flare activity, 
the  existence of a jet from Sgr A* has not been firmly established because of the complex
  thermal and nonthermal structures in this confused  region of the Galaxy.
At least five independent investigations based on X-ray, near-IR, and radio observations have suggested that a jet 
is emanating from \sgra. These studies have found  discrepant values for the jet position angle (PA) and 
inclination  (Markoff \etal\,  2007; Broderick \etal\,  2011; Zamaninasab \etal\,  2011; Muno \etal\, 2008; 
Li \etal\, 2013; Yusef-Zadeh \etal\,  2012; Shahzamanian \etal\, 2015). 
It is possible that   some  of the gas approaching Sgr A* is
pushed away as part of an expanding hot plasma driven by flaring  and jet activity,    
resulting Sgr A*'s low radiative efficiency.
Thus, the presence of  collimated 
structures  from \sgra\,  is critical in distinguishing between the competing accretion and outflow models.
 
The Galactic center is 
a challenging region in which to image  a radio jet or a flare 
close to \sgra, because of the limited spatial resolution and dynamic 
range  caused by confusing sources, scatter broadening, and intrinsic temporal 
variability of \sgra\, on hourly time scales (e.g., Bower \etal\, 2014).
Here we present  sensitive observations of the Galactic center at multiple radio frequencies
obtained using the improved broad-band capability of the VLA, 
 finding  new radio structures interpreted to be associated with Sgr A* activity.

On a scale of a few arcseconds  from   Sgr A*, 
we identify  a ridge of east-west radio emission which bends toward the SW in the direction away from Sgr A*. 
This ridge,  which is detected to the west of Sgr A*,   shows a number of blobs and arc-like 
features surrounded by a diffuse plume-like structure.  
We interpret that the  plume-like feature as arising  from flaring activity, thus produces  
an  outflow from 
the direction of Sgr A* on a scale of $\sim$0.1 pc, with an opening angle of $\sim35^\circ$.
The position angle of the outflow driven by flaring activity of  Sgr A* is consistent with the   east-west elongation of 
Sgr A* observed on milliarcsecond (mas) scale  (Bower \etal\, 2015).   
A large number of  stars being members of 
the so-called ``S-cluster''\footnote{A loosely defined term 
consisting of $\sim$30 stars within a projected distance of 1$"$ from  Sgr A*, 2/3 of which 
are spectroscopically classified as  O/B stars with orbital periods of a dozen  to few hundred years 
and the rest older stars (e.g., Genzel \etal\, 2003). 
We point out that the stars in this region 
have different spectral types and ages and probably heterogeneous origins} 
  also lie 
along the ridge.   
We  compare  infrared  and radio images of the ridge and argue that bow-shock  structures 
detected within the inner 2$''$ of Sgr A* are   proplyd candidates. We show that 
radio  sources with comparable scale size to those associated with  proplyds 
can not be simply blobs of dusty ionized gas but are associated with a  reservoir  of 
hot dust surrounded by ionized gas.

%the West  side of the new radio features is  more prominent as 

On a larger  scale,  within two arcminutes NE of Sgr A*, 
we  find  new cometary sources (F1, F2 and F3)  pointing toward Sgr A*  
and a large-scale tower-like structure associated with the Sgr A East supernova remnant (SNR) 150$''$ from Sgr A*.
The position angle of these new structures
is similar to two other IR-identified cometary sources, X3 and X7,  found to the SW of Sgr A* (Muzic \etal\, 2007, 2010). 
One interpretation that we put forth is 
that these features could be the  result of a
jet from Sgr A*  interacting with the atmosphere  of dusty  stars near 
Sgr A* and with the  Sgr A East shell, respectively.

In addition, we suggest that 
the diffuse X-ray emission  centered on  Sgr A* arises  through hot gas
created by the collision of stellar winds from B stars in the S-star cluster or 
young low-mass stars (c.f. Loeb 2004). 
This nuclear wind   created by  mass-losing stars near Sgr A* produces hot expanding X-ray gas (c.f. Quataert 2004)
that excludes the shocked winds from O and WR stars in the central parsec of the Galaxy and prevents accretion 
onto Sgr A*.  
Meanwhile Sgr A* accretes material from the cluster winds
at a much lower rate potentially explaining
the low luminosity of Sgr A* without the ejection of a large fraction of the accreted material.

%The position angle of the plume-like structure of Sgr A* is different than the position angle of 
%cometary and tower-like sources.  

%We interpret these features as tracers of  an outflow from Sgr A* that 
%interacts with the surrounding  medium. 

%These new observations suggest that the collimated hot plasma  is   interacting 
%with the orbiting gas associated with the Eastern arm of the mini-spiral HII region as well as stars 
%associated with the young  S cluster and the evolved cluster. 

% We also 
%detect two compact radio sources that coincide with two near-IR-identified S stars. 

%We argue that the interaction of the outflow from Sgr A* with the S-cluster increases the 
%mass-loss of S stars by ablating the atmosphere of these stars. The increased mass-loss can collectively provide a 
%more spherical outflow that in turn reduces accretion onto Sgr A*.

%with the S-stars within the inner few arcseconds of Sgr A*, as well as 
%Second, we present structural details of the ionized gas associated with the 
%mini-spiral suggesting that an outflow must have shaped the morphology of ionized gas orbiting Sgr A*.

\section{Observations and Data Reduction}
\subsection{Radio Data}

Multi-wavelength radio continuum observations were carried out with the Karl G. Jansky Very Large Array 
(VLA)\footnote{Karl G. Jansky Very Large Array (VLA) of the National Radio Astronomy Observatory is a facility 
of the National Science Foundation, operated under a cooperative agreement by Associated Universities, Inc.} in 
its A-configuration at 44, 34.5, 8.5, 5.5 and 1.4 GHz during March and April 2014. Table 1 gives columns of the 
date, center frequency, bandwidth, the number of subbands (IF), the number of channels, and the spatial 
resolution of each observation. 
In all 
observations, we used 3C286 to calibrate the flux density scale,  3C286 and J1733-1304 (aka NRAO530) to 
calibrate the bandpass, and J1744-3116 to calibrate the complex gains. The broad 8 GHz bandwidths at 34 and 44 
GHz, 2 GHz bandwidths at 5 and 14 GHz, and 1 GHz bandwidth
 at 1.4 GHz provide   a significant improvement over earlier  observations that had only 100 MHz of bandwidth.
We observed  Sgr A* using the 3-bit sampler system at 34 and 44 GHz, which provided full polarization correlations. 
 
%details of which will be discussed elsewhere.  

\subsection{Infrared  Data}

Details of the  near-IR observations and data reduction of the Galactic center at $K_s$  and $L'$ bands, 
central wavelengths 2.18, and 3.8 $\mu$m, respectively, were recently given in Yusef-Zadeh \etal\, (2015). 
These observations used adaptive optics and were acquired with 
VLT/NACO\footnote{Based on observations made with ESO Telescopes at the La
Silla or Paranal Observatories under programs ID 089.B-0503}, with a pixel scale of 0.027$"$ per pixel.
$L'$-band observations were obtained 
in speckle mode using 
five fields with different pointing and
depths. Field\,1 was centered on \sgra, and Fields 2--5 were offset by approximately 20\arcs\ to the northeast,
southeast, southwest, and northwest, respectively.
Standard near-IR  image data reduction was applied 
followed by combining  individual pointings into large mosaics. In the case of
the $L'$ image, the speckle holography technique, as described in Sch\"odel \etal\,  (2013), was applied to the
thousands of obtained speckle frames to create final high-Strehl images for each pointing. 
Finally, we calibrated the images astrometrically by using the positions and proper motions of SiO maser stars in the
Galactic center (Reid \etal\ 2007). 

Imaging data in M$'$-band were taken with NACO/VLT in 2003, 2004, and 2006. We retrieved the data from the ESO archive. 
Chopping was used for background subtraction. The images were flat-fielded and corrected for bad pixels. Since the 
chop throw was small in all observations (due to technical limitations at the VLT) the images from the individual 
epochs show strong negative residuals from stellar and diffuse sources in the off-target chop positions. However, some 
dithering was applied during the observing runs and the initial pointing as well as the chopping angle were different 
for the different observing epochs. Therefore, residuals could be effectively removed by averaging the images from 
all observing epochs; the images were sub-pixel shifted onto a common position of the centroid of the star IRS 16C 
and median combined. Any remaining artifacts from chopping were effectively removed by rejecting the lowest 20\% at 
each pixel before calculating the median value. Approximately 12000 frames, each with 0.056 s integration time, were 
combined, corresponding to an accumulated exposure time of about 800 s. The photometry was calibrated by assuming 
constant extinction between the L and M bands and the same magnitudes at both bands for the sources IRS 16C and IRS 
16NW. Their L'-band fluxes were taken from Sch\"odel \etal\,  (2010). The uncertainty of the zero point was estimated to 
be 0.15 mag.

%The results of these multi-wavelength observations  will be given elsewhere.  

%Radio continuum observations were carried out with the Karl G. Jansky Very Large Array (VLA) in its A-configuration at           
%34.5 GHz on April 15, 2014 (14A-229). We observed a field of view with a radius of XX## centered on Sgr A* in Q-band (XX
%mm) using the 3-bit sampler system, which provided full polarization correlations in 4 basebands, each 2 GHz wide,
%centered at 41.20, 42.80, 44.48, 46.30 GHz. Each baseband was composed of 16 subbands 128 MHz wide. Each subband was
%comprised of 64 channels, each 2 MHz wide.  We used 3C286 to calibrate the flux density scale and used both 3C286 and
%J1733-1304 (aka NRAO530) to calibrate the bandpass. We used J1744-3116 to calibrate the complex gains.

%The 1$\sigma$ rms sensitivity  is  7.8 $\mu$Jy.
%The main reason for the improved sensitivity of this VLA  data is the large 8 GHz bandwidth  available
%with new correlator  compared to the 100 MHz bandwidth used in earlier observations.
%Our 44 GHz variability was strong, thus the image at this wavelength has a limited dynamic range. 

\section{Results}

Because of better sensitivity to detect weak radio emission, we first present newly recognized  
features within a few arcseconds of Sgr A* and then  compare  the positions of 
near-IR identified stellar sources with radio data. Lastly, we present large
structures on a scale between a few arcseconds to  two arcminutes surrounding 
Sgr A* and identify new features associated with the Sgr A East SNR shell. 

\subsection{Small Scale  Radio  Features}

\subsubsection{The Sgr A* Radio Ridge}

Figure 1a,b show images of the inner 10\arcs\ of \sgra\ at mean frequencies (wavelengths) 34.5 GHz (9mm) and 14.1 GHz 
(2cm). Prominent stellar and ionized features are labeled. These images reveal a ridge of emission to the 
west of north of  \sgra\ extending for about 1.5--2\arcs\ at position angles ranging between $\sim$90 and $-125^\circ$. This ridge of 
emission,  which was previously detected in low resolution images at 8 GHz (Wardle and Yusef-Zadeh 1992) links Sgr A* to the 
western edge of the minicavity, an ionized feature with a diameter of 2\arcs. The western edge of the minicavity 
coincides with the two stellar clusters IRS 13N and IRS 13E. The  \sgra\ radio ridge consists of a number of blob-like and 
arc-like structures with angular scales of 0.2--0.5\arcs. The two arc-like structures coincide with source $\epsilon$ 
which has been detected in earlier narrow bandwidth images with low spatial resolutions at 15 GHz (Yusef-Zadeh \etal\, 
1990; Zhao  \etal\, 1991). To illustrate the asymmetric nature of the ridge structure, 
Figure 1c shows a saturated image of the region 
shown in Figure 1a,b. We interpret that this structure has
 an opening angle of $\sim35^\circ$ 
pointed in the direction toward  W and SW from Sgr A*. Figure 1d 
shows another 34.5 GHz rendition of the ridge of emission in reverse color, but at higher resolution than that of 
Figure 1a.  The images at multiple frequencies confirm the reality of blob and arc-like structures along the ridge. 
Note the diffuse plume-like structure, as drawn schematically on Figure 1a, 
surrounding the blob and arc-like structures. 
The widening of the plume-like 
structure away from Sgr A* suggests  that a plume of  gaseous material is moving outward  and
expanding away, 
suggesting  that Sgr A* is responsible for the ridge.
Proper motion measurements of the brightest source in the ridge indicate 
high velocity ionized gas moving away from Sgr A* to the SW (Zhao \etal\, 2009).  
This radio ridge appears distinct  from the ionized gas associated with the mini-spiral HII region 
orbiting Sgr A* and is not confused 
with numerous dusty stellar sources and diffuse emission   found in mid-IR images of this region. 
Future high resolution proper motion, polarization 
and spectral index measurements of the ridge will provide additional constraints in 
the claim that this feature is associated physically to Sgr A*. 
 
\subsubsection{Radio Emission from the S Cluster}

The region within $\sim$2\arcs\ of Sgr A* where the ridge of radio emission and the plume-like structure are detected, 
is adjacent to the S-cluster. 
This cluster consists mostly of young, early type stars with orbital periods of 10 to a few 100 years.
The kinematics of the stars in the cluster, which following other workers we refer to as  $``$S stars$''$, 
have been used to 
measure the mass of Sgr A* (Ghez \etal\, 2008; Gillessen \etal\,  2009).  Figures 2a,b show 
contours of 34.5 GHz emission and a grayscale image of 
a region  of $3.5''\times2.5''$ centered on Sgr A*. 
Figure 2c shows contours of 5.5 GHz  
emission  superimposed on a  grayscale image. 
The crosses on Figure 2a,c correspond to eight radio sources RS1--8 found in the plume-like region to the 
west of Sgr A*.  

The trajectories of young stars are known from proper motion 
measurements in the near-IR. To see if these  stars have radio continuum counterparts, their positions at the epoch of 
our 34 GHz observation on March 9, 2014 (2014.19) have been calculated based on proper motions and orbital 
accelerations derived from 
near-IR observations (Gillessen \etal\, 2009; Lu \etal\, 2006; Yelda \etal\ 2014). Tables 2, 3  give the 
positions of S cluster  and their corresponding positional 
uncertainties at the epoch of 2014.19 and from two different catalogs (Gillessen \etal\, 2006; Lu \etal\, 2009). 
Table  4
gives the predicted X/Y offsets (in 
arcsecond) of  stars relative to Sgr A* for the epoch 2014.19. These tables list the positions of the 28, 31 and  
117 
stars identified by Gillessen \etal\, (2009), Lu \etal\, (2006) and Yelda \etal\, 2014,  respectively.   
Table 5 lists Gaussian-fitted 
positions of 8 radio sources (RS1--8) embedded within the diffuse extended emission associated with 
 the Sgr A* ridge and the plume-like 
structure at 34 GHz (see Figs. 1 and 2).
Entries in columns 1 to 9 give the name of the source at 34 GHz, alternative  names in the literature, the RA and Dec,
 the angular distance from Sgr A* in 
increasing order, positional accuracy, the peak intensity and  the flux density 
that are associated with these sources.

%An edge-enhanced filtered image of this figure clearly shows the bow shock structures with both sources RS5 and RS6 
%facing roughly in the direction of Sgr~A*.

Figure 3a,b superimpose contours of radio emission on a 34 GHz grayscale image, with crosses indicating 
the positions of the  stars in the S cluster for which orbits have been determined by Gillessen \etal\, (2009) and Lu \etal\,  
(2009), respectively.  
Many of the S-stars, as noted in Figure 3a,  are projected against the bright, scatter-broadened radio source Sgr A* and 
the diffuse emission from the Sgr A* ridge, thus they can not be discerned  at radio wavelengths. 
The comparison of 
radio sources (RSs) listed in Table 5 with stellar sources  given in Table 2 
indicates that RS1, RS2 and RS3 lie within the 
1$\sigma$ position of S33, which is an early type star (Gillessen \etal\,  2009).  These radio sources 
appear compact but they are embedded within the extended emission from the ridge of emission (see Fig. 2b) 
so it is not 
clear if these sources are radio  counterparts to S33 unless we measure the proper motion of radio sources.  
If these radio sources are not randomly coincident with S33 and are associated with stars, we  
can determine the 
mass loss rate from the ionized winds. 
The radio emission is assumed to arise  from S33 with a  flux density of $\sim0.2$ mJy at 34.5 GHz  
and is due to a spherically-symmetric, homogeneous  wind of fully--ionized 
gas expanding with a constant terminal velocity $\sim$700 \kms\, (Panagia \& Felli 1975). 
The   mass loss rate of S33 is estimated to be $\dot{M} = 2\times10^{-6}$ \msol yr$^{-1}$. 
Clumpiness of the  ionized wind would  reduce this estimate. 

%Because of the strong emission from Sgr A*, extended emission associated with the Sgr A* ridge, as well as two 
%extended arc-like sources, we can only place upper limits 100 $\mu$Jy beam$^{-1}$ at 35 GHz for many members of the S 
%cluster located to the west of Sgr A*.  The S-stars S33 and S87 in Figure 3a (Gillessen \etal\, 2009) have possible 
%radio counterparts with flux densities at a level of $\sim0.2-0.5$ mJy at 34 GHz. To confirm the association of radio 
%sources with the S stars, the comparison of radio proper motion measurements with those at near-IR are needed and 
%proper motion of radio sources are not available.

Figure 3c shows a grayscale 34 GHz image in negative  and labeled with the positions of those  stars for 
which orbits have been determined by Yelda \etal\, (2014). We note several prominent young stars, 
IRS~16SW, IRS~16NW, IRS~16C, that are not members of the S-cluster but have radio counterparts (Yusef-Zadeh \etal\, 
 2013). Figure 3d shows the distribution of stars over the area shown in Figure 3c at 3.8$\mu$m. The 
crosses correspond to the position  of stars (Yelda \etal\, 2014) at the epoch that radio data were 
taken. Apart from  stars labeled on this figure, 
the spectral classification of remaining faint sources is unknown. 
These faint sources  are 
possibly  late-type stars associated with  the evolved nuclear cluster or young low-mass stars (see $\S4$).
Five near-IR stellar sources A--E are  labeled on Figure 3d and  will be discussed below.

\subsubsection{Bow Shock-like Radio Sources RS5 and RS6}

Figure 4a,b,c show the relative position of 34.5 GHz sources RS5 -- RS8 and 3.8 $\mu$m and 2.18 $\mu$m sources A -- D with 
respect to each other.  Two of the newly detected radio sources in the ridge, RS5 and RS6 located $\sim1.6''$ from Sgr A* (see 
Fig. 2), are partially resolved, with size scales ranging between 850 to 1200 AU. A close-up views of these arc-like 
structures are shown in Figure 4a whereas Figures 4b,c show contours of 3.8 and 2.18$\mu$m emission superimposed on a 34.5 GHz 
image, respectively. A comparison of radio with near-IR images reveals that the arc-like structures RS5 and RS6 have 3.8$\mu$m 
counterparts with a bow shock morphology. There are two near-IR stellar sources, S1-22 (Lu \etal\, 2009) and a stellar source 
A, as labeled on Figure 4c.  S1-22 lies at the apex of a bow shock-like structure and is an early type star which is projected 
against the extended radio emission associated with RS6. Star A has no radio counterpart, and is not identified 
in any catalogs of early type stars near Sgr A*. 
It is unlikely that star A 
is  associated with RS6 for two  reasons. First,  proper motion measurements indicate 
that star A is moving to 
the NW with a velocity of 117$\pm16$ \kms\, (Sch\"odel \etal 2009). The proper motion of ionized gas associated with the blob 
$\epsilon$ which coincides with RS5, RS6 and RS7 is $\sim338\pm21$ \kms\, to the SW (Zhao \etal\, 2009). Second, 
 it 
is unlikely that a late-type star could have a stellar wind strong enough to produce the observed 
 stand-off distance of the bow-shock's 
apex,  $\sim$6 milliparsec (mpc).  This  requires a Wolf-Rayet star  (Tanner 
\etal\, 2005; Sanchez-Bermudez \etal\, 2014). Thus, it is unlikely that star A is associated with the bow-shock structure RS5. 

As for S1-22, the proper motion data gives a tangential velocity of $\sim$326 \kms\, to the SE (Yelda \etal\, 2014). 
Radio 
proper motion measurements have a coarse arcsecond spatial resolution , 
giving the proper motion of RS5, R6 and RS7, when compared to 
those of near-IR sources. In spite of  the difference in resolution between radio and near-IR proper motion measurements, 
the magnitude of radio proper motion of RS5, RS6 and RS7 is similar to that of S1-22. Thus, it is possible that S1-22 is 
physically associated with RS6 with the standoff distance of $\sim$4 mpc. 

%Alternatively, RS5 and RS6 could be both part of a bowshock structure centered on star A. The apparent reverse 
%curvature of RS6 is caused by confusion with fainter emission from S1-22. The single bowshock is variable, and 
%currently happens to be brighter in the wings than at the apex. The proper motion measurements could be affected by 
%the variability and reflect a pattern velocity rather than the true space velocity.

In addition to star A, three 3.8$\mu$m sources B, C and D, labeled on Figures 4b,c, are not 
found in the catalogs of early type stars. 
RS7 and RS8 are likely  to be radio counterparts 
to source B, and C, respectively. A  radio source with a flux density of 58 $\mu$Jy at 34 GHz is detected at the 
position of source D which lies in an extended region associated with the ridge.  
The bright radio sources B and D are 
clearly stellar sources since they are detected at L$'$ and K$_s$ bands. 
The offsets seen in the position of radio and 
near-IR sources could have a contribution from  proper motion of individual sources. 
The near-IR images at 2.18 and 3.8$\mu$m were 
taken during June and September 2012 whereas the 34.5 GHz data was taken on March 9, 2014.
The L$'$ sources associated with RS5, RS7 (or source A), RS8 (or source B), and D, 
lie  to the  SSE of their radio counterparts, suggesting either 
correlated motion, coincidence, or a systematic error in the image registration arising because 
the radio and infrared  images are taken at different epochs.
We compared Gaussian fitted positions of B, C and D at 34.5 GHz and L$'$ band and found  
3.4 and 3 sigma offsets to the north  in the positions of stars B and D and their radio counterparts, respectively. 
We also compared the positional offset for IRS 16C and its radio counterpart and determined  
 that this implies a proper motion of IRS 16 that is roughly twice 
higher the actual value determined at radio and infrared wavelengths (Lu \etal\,  2009; Yusef-Zadeh \etal\, 2015a). 
Thus, we can not establish that radio sources are counterparts to stellar sources. 
Given the offsets in positions and the non-detection of source C in K$_s$,  it is just possible that 
the radio sources
are gas blobs and are not directly associated with  stars. However, it is 
not clear how gas blobs near  Sgr A* could survive the tidal shear of  Sgr A* unless they have densities 
that withstand  the tidal shear or that  they are transient (Yusef-Zadeh \etal\, 2015c). 

%The positional offsets for star B are 2.6$\pm21.9$ and 71.7$\pm21.3$ mas whereas for source D the offsets are 
%21.4$\pm15.9$ and 49.20.3$\pm16.7$ mas in RA and Dec, respectively. We also compared the positional offset of IRS 16NE 
%and its radio counterpart and found 24.6$\pm0.6$ mas and 40.5$\pm1.2$ mas in RA and Dec, respectively. The positional 
%offsets for stars B and D are within a factor of two of IRS 16NE, thought with large errors. Thus, If radio sources 
%are counterpart to stellar sources, stars B and D are moving to the north with velocities that are comparable to a 
%$\sim1000\pm300$ \kms. There was no significant offset detected in the position of RS8 and source C within 
%33.6$\pm19.8$ and 39.3$\pm28.8$ mas in RA and Dec, respectively

%To compare a near-IR and radio continuum images, Figure 2b shows a composite image at 34 GHz in red, 14 GHz in green 
%and 3.8$\mu$m in blue colors. A number of near-IR identified stellar sources as well as diffuse dusty features at 
%3.8$\mu$m have radio continuum counterparts (Yusef-Zadeh \etal\, 2013) whereas the ridge of emission is best detected 
%at radio wavelengths.
%The bottom two panels show the ridge of emission at 14 and 44 
%GHz with resolutions of XX and 0.1". The dynamic range of the 44 GHz image is limited by phase errors due to variability 
%of Sgr A*.  

\subsection{Large Scale Features}
  
\subsubsection{Cometary  Sources F1 and F2}

There are two known cometary sources, X3 and X7,  lying within 3$''$ to the SW of Sgr A* 
at near-IR wavelengths. 
X3 is located $\sim3''$ 
from Sgr A* (Muzic \etal\, 2007, 2010) 
showing tail-head structure pointing toward Sgr A*. 
It  has a radio counterpart at 44 GHz (source 16 in Yusef-Zadeh \etal\, 2014b) and 
 a  peak flux density  0.22 mJy beam$^{-1}$ at 34.5 GHz. 
The second cometary source  X7 (Muzic \etal\, 2007)   
is  identified in Figure 3d as source E at near-IR.   
This   cometary source also 
points  toward  the direction of Sgr A*. 
A compact 34.5 GHz source with a peak flux density of 100 $\mu$Jy is detected at the position 
of source E  0.7$''$ from Sgr A*. 
This source lies too close to Sgr A* where the noise increases near the bright source  Sgr 
A*, thus, structural details of this source are not  clear at radio wavelengths. 
We detect a third radio source with a cometary morphology similar to X3 and X7. 
Unlike X3 and X7, this source,  which we denote  F1,   
lies  4.8$''$ to the NE of Sgr A*. 
Figure 5a,b show grayscale 
contours of this cometary  feature at 34.5 GHz and 3.8$\mu$m, respectively. 
F1  has an extent of 0.65$''\times0.25''$ (length $\times$ width) 
with an integrated flux density of $\sim2$ mJy  and background subtracted peak intensity 
of 416 mJy per  $(0.12'')^2$ at 34.5 GHz and 3.8$\mu$m, respectively.

Figure 5c shows a large  view  of the 
region  which includes X3, X7,  the  radio cometary feature (F1) and the  MIR cometary  feature (F2) at 
4.68 $\mu$m in reverse color.
We note a gap in the region to the NE of Sgr A* where F1 is detected. 
This gap  appears to be devoid of dust emission at MIR. 
The MIR  gap can also be identified at radio in  Figure 1b. However, 
the lack of short   {\it uv} spacings may contribute in suppressing  the emission from 
 the bright source Sgr A*.  
A close up view of infrared  emission at M$'$ band 
from the inner $3''\times3''$ 
of Sgr A* is shown in the inset to the right of Figure 5c.  
An additional cometary feature is detected at MIR 
to the NE of IRS 16C in the inset. This source,  which we call F2,  is  well within the MIR gap and lies along 
the direction where X3, X7 and  F1  are located. 

Lastly,  an additional  radio source $\sim13''$ NE of Sgr A* shows  a head-tail structure 
pointing toward Sgr A*. This radio feature shows a tail feature with an extent of $\sim0.6''$ at the  
PS$\sim57^\circ$. The peak flux density of this source which we call  F3 is 
0.32 mJy beam$^{-1}$ at 34.5 GHz with a spatial resolution 
of 89$''\times46''$ mas.  
This source has been detected   in the L$'$ band as IRS 5 SE (Perger \etal\, 2008).
This source  is  interpreted as a stellar bow-shock resulting from the interaction of a masss-losing stars 
and the minispiral. 

\subsubsection{Sgr A East Tower}

The nonthermal radio source Sgr A East 
is  a young shell-type  SNR  with an angular size of 
2.7$'\times3.6'$ and a spectral index $\alpha$=0.76, where 
the flux density S$_\nu\propto\nu^{-\alpha}$
(Ekers \etal\, 1983; Yusef-Zadeh \& Morris 1987; 
Pedlar \etal\, 1989).  Thermal X-ray emission is concentrated in  the interior of the remnant suggesting 
that  Sgr A East  is a mixed morphology SNR interacting with the 50 \kms\,  molecular cloud 
 (Maeda \etal\, 2002; Park \etal\, 2005).  The thermal X-ray emitting plasma  has 
two  components,  characterized by temperatures of 1 and 6 keV and  corresponding  electron density of 4.7 
 and 0.6  cm$^{-3}$, respectively. 
(Park \etal\, 2005; Koyama \etal\, 2007).  A candidate neutron  star CXOGC J174545.5--285829 (the cannonball) 
detected in  X-ray and radio has also been associated with the remnant (Park \etal\, 2005; Zhao \etal\, 2013). 

Our broad band 1.5 GHz image of Sgr A East provides a wealth of details associated with Sgr A East and the surrounding 
environment.  Figures 6a,b show large scale views of Sgr A East displayed with two different grayscale 
levels at 1.5 GHz. The major axis of the Sgr A East shell is along the Galactic plane. A number of new features are 
detected in this complex region. Here we describe three new features. One is a distorted region to the NE of the Sgr A 
East shell, 100$''$ E and 80$''$ N of Sgr A*.  A striking tower-like structure with an extent of 100$''$ appears to 
emerge from a gap 
in the brightness distribution of Sgr A East. The base of this tower is about 
20$''-25''$ across with mean flux density of $\sim2.8$ mJy per 1.39$''\times0.6''$ beam at 1.5 
GHz. The base  narrows as it extends to the NE with a position angle of $\sim50^\circ$. A schematic diagram of 
these features is shown in Figure 6c. The tower is terminated by  two bow-shock-like structures. Grayscale contours 
of these  are displayed in Figure 7a. 

The second feature is the  polarized source P1 which was first identified at 8 GHz (Yusef-Zadeh 
\etal\, 2012). At  1.5 GHz, P1  is resolved into two linear structures
 (see the total intensity image in Figure 6b), that appears to cross each other at right angles. One of the linear 
features has a PA$\sim50^\circ$, as shown in Figure 7b. Lastly, the region surrounding the polarized source P4 
(Yusef-Zadeh \etal\, 2012) is shown at 1.5 GHz in Figure 7c.  This elongated feature extents for 20$''$ at 1.5 GHz with 
a position angle of $\sim50^\circ$. A number of blob-like structures were previously reported to the NE of this 
feature (Yusef-Zadeh \etal\, 2012).
 
%Figure 9c shows another linear structure with a PA$\sim50^\circ$. This is also the region 
%where a linearly polarized feature was detected at 8 GHz (Yusef-Zadeh \etal\, 2013).

\subsection{Other Features}

\subsubsection{A Semi-linear Feature}
We identify a striking semi-linear radio 
continuum feature  projected 
perpendicular to the radio Sgr A* 
ridge\footnote{The Sgr A* radio  ridge should be distinguished from the mid-IR Sgr A* ridge discussed  
by Sch\"odel \etal\, (2011)}.  
Figure 8a shows a 5.5 GHz image that reveals  roughly 
uniform brightness 
$\sim0.5$ mJy per 0.5$''\times0.27''$ beam. The semi-linear feature appears to arise from the ionized bar 
as it curves concave toward Sgr A*, crosses the ridge at an angular distance of $\sim$1\arcs\ west of Sgr A*, and 
continues to the north of Sgr A*. To highlight  this structure, white dashed lines are drawn along this radio feature. 
The semi-linear continuous structure has  an extent of 5\arcs\ and  width of 0.5\arcs,  becoming 
wider and more diffuse 
as it extends to the north of Sgr A*.  The Northern arm shows a discontinuity, 
as it  approaches Sgr A*, 
 which is best seen at $\alpha, \delta\, (J2000)=17^h\, 45^m\, 40.18^s, -29^\circ\, 
0'\, 29''$ in Figures 1a-c. 
The elongated features makes a 90$^\circ$ 
change to the south in its direction as it approaches Sgr A* (see the schematic diagram in Fig. 9). 
The kinematics of ionized gas show that the radial 
velocity of this elongated feature is close to  zero \kms\, but 
changes by $\sim$200 \kms\, close to the  location of the discontinuity (Zhao \etal\, 2009). 

%Figure 1c points to this gentle discontinuity. 

%At the morphology of an elongated structure runs along the inner edge of the North arm.

%In spite of highly confused region where multiple structures coexist near Sgr A*, 

\subsubsection{The Minicavity and the  Bar}

The bar of ionized gas lies a few arcseconds to the south of Sgr A* where the Eastern and Northern arms of 
Sgr A West cross  each other. High spatial resolution images of the ionized bar 
show a minicavity of ionized gas  with high velocity dispersion 
(Lacy \etal\,  1991; Roberts \etal\,  1996; Zhao \etal\,  2009).
Figure 8b shows a larger area  of the Eastern arm and the minicavity at 34 GHz. 
The mini-cavity and the Eastern  arm are distinct from the North arm. 
High resolution observations  have shown that the minicavity extends further to the SE (Zhao \etal 1991). 
The new images show that a gap separates the eastern and western halves of the minicavity and that 
the   North and East arms are 
two distinct features associated with the eastern and western halves of the mini-cavity, respectively.   
The structure of the Eastern arm has  a wavy appearance 
with a wavelength of  2$''-4''$ as its extends along the western half of the minicavity. 
Figure 9 shows a schematic diagram of  the   small and large-scale features 
revealed in the inner 50\arcs\ of 
Sgr A*. 

\subsubsection{A Bent Filament}

Figures 10a,b   show the large-scale area  at 1.5 and 5.5 GHz, respectively. 
 A striking filamentary structure is seen within a cavity of ionized gas 
to the north of Sgr A*.
This bent and narrow  filament  resembles  the nonthermal radio filaments seen throughout the Galactic center (e.g., 
Yusef-Zadeh \etal\, 2004; Nord \etal\,  2004). 
The southern end of this filament  is pointed toward Sgr A*. 
Figure 10c,d show details  of  the bent filament 
at 1.5 GHz at  two different resolutions indicating that the 
filament may experience  another bend at $\sim2''$ to the N of Sgr A*. 
Because of the large angular size of Sgr A* 
induced by  scatter broadening at 1.5 GHz, it is not clear if there is additional change in the position angle 
of the bent filament in the inner 1$''$ of Sgr A*.  
Figure 10e shows grayscale contours of 8.9 GHz emission from the filament. 
The surface brightness of the filament at 1.5 GHz is $\sim3$  mJy beam$^{-1}$ as it bends to the NW (see Fig. 10). 
The  length  of  this filament at 1.5 GHz is  $\sim25''$.
The width of the filament is unresolved at  1.5 GHz but is partially resolved at roughly  0.3\arcs\, at 8.9 GHz. 
The cavity  in which  this bent filament lies, as best revealed in Figure 10c, 
is dark  and devoid of diffuse ionized gas at 1.5 GHz. This cavity has  a cone-shaped 
structure  and the cometary supergiant star IRS 7 (Rieke and Rieke 1989; Yusef-Zadeh \etal\, 1989; 
Serabyn \etal\, 1991; 
Yusef-Zadeh \& Melia 1992) is seen  within the cone structure  at  5.5 GHz. 

The bent filament lies in a region with varying background emission, thus accurate determination of 
the spectral index  is not 
possible. Approximately accounting for the background flux, 
the surface brightness is $\sim$3--4 mJy beam$^{-1}$ at 5.5 and 1.5 GHz, 
implying a flat spectrum between  these frequencies.  The sensitivity at  34 and 44 GHz 
is not sufficient to 
detect the bent filament. The origin of the bent filament is unknown, though its morphology gives the best 
jet-like appearance that may be originated within 1$''$ of Sgr A*. 
 It is also possible that this filament is a member of the  
population of  nonthermal radio filaments found in the Galactic center
such as the  linear filaments found 
within the inner few arcminutes of Sgr A* where the Sgr A East SNR lies (Yusef-Zadeh \etal\, 2004; Nord \etal\, 2004). Future 
Future high resolution
measurements at low frequencies  including proper motion 
observations are required to constrain the nature of this bent structure.
 
\section{Discussion}

Here we discuss four  aspects of the new structures found near Sgr A*. First, we interpret the 
arc-like sources in terms of 
recent low-mass star formation activity within 2$''$ of Sgr A*. Second, we explain the origin of hot X-ray gas within the S 
cluster as arising from  colliding winds from mass-losing stars, either from  low-mass YSOs or 
young B   stars in the S cluster or both,  
and estimate the accretion rate onto Sgr A*. 
Third, we argue that the cometary radio structures and the Sgr A East tower 
are signatures of an   interaction between  a collimated 
outflow from Sgr A* with a PA$\sim50^\circ-60^\circ$ with  the stellar envelopes 
and ionized gas along the path  
of the jet. The consequence of 
this interaction infers  not only the mass loss rate from Sgr A* but also provides  a 
lower bound on  the accretion rate onto Sgr A*. Fourth, the
 ridge of  radio emission is interpreted as due 
to past flaring activity associated with Sgr A*.

\subsection{Proplyd Candidates  Near  Sgr A*?}

The nature of the radio sources within two arcseconds of Sgr A* is not clear. Two arc-like radio sources RS5 and 
RS6 resemble the bow-shock sources found at 34 GHz about $20''$ away from Sgr A* (Yusef-Zadeh \etal\, 2015c). As 
displayed in Figures 4b,c, RS5 and RS6 have near-IR counterparts. Although there is an early type star 
S1-22 that may be 
associated with RS6,  RS5 is not  identified as a near-IR star in various catalogs (e.g., Lu \etal\, 2009, 2013).  
There is no evidence for a known young and massive star associated with RS5.
RS5 has 3.8 and 2.18$\mu$m counterparts 
suggesting the the dust, traced by near-IR emission, and ionized gas, as traced at radio, 
are either intermixed, or else closely related through separate, 
as observed in 
photodissociating regions (PDRs). Because of the strong radiation field that produces 
high temperature of the ionized gas, as traced by radio emission, dust grains 
mixed in with ionized gas would have been evaporated. Thus, RS5 resembles the population of bow-shock sources that 
have recently been found 20$''$ from Sgr A* and are interpreted as candidate photoevaporative 
protoplanetary disks 
(proplyds) associated with newly formed low-mass stars (Yusef-Zadeh \etal\, 2015c). The short expansion time scale and 
the low density of ionized gas associated with the arc-like structures provide strong arguments in favor of 
protoplanetary disks (Li \& Loeb  2013; Yusef-Zadeh \etal\, 2015c).

To examine the protoplanetary disk scenario, we compared radio and near-IR flux from RS5 and another proplyd 
candidate, the so-called P8 source (Fig. 2c in Yusef-Zadeh \etal\, 2015). P8 is among 44 proplyd
 candidates found within 
25$''$ of Sgr A*. Figures 11a,b show a horizontal slice that crosses RS5 and P8, respectively, at 35 GHz, 
and the L$'$ and K$_s$ near-IR bands. The arrows point to the location of proplyd candidates. The emission from RS5 in the three panels 
lies at pixel 25.29 with peak surface brightness  of 0.31 mJy beam$^{-1}$ (beam = 88$\times45$ mas$^2$), 49 mJy pixel$^{-2}$ 
(pixel size = 27 mas) and 1.5 mJy pixel$^{-2}$ (pixel size = 27 mas) at 35 GHz, L$'$,  and K$_s$, respectively. 
Similarly, the intensity profile of P8 is shown in Figure 11b with a low signal-to-noise emission at K$_s$ band. The 
arc-like sources are likely photoionized externally by the UV radiation field from the 
$\sim100$ OB and WR stars that lie 
within 10$''$ of Sgr A*. We suggest that RS5 and P8 are proplyd candidates that are 
externally photoionized by strong 
sources of UV radiation. 

Assuming that radio emission traces the ionization front of proplyd candidates, we use the 
flux at L$'$ and K$_s$ to determine the temperature of dust emission from the outer surface of a 
hot disk.
We use the measured fluxes at K$_s$ and L$'$ bands to check  
that the emission is consistent with a protoplanetary disk irradiated by the intense radiation field in the inner Galaxy. 
The flux at frequency $\nu$ is given by 
\begin{equation} 
S_\nu = B_\nu(T_d)(1-e^{-\tau_\nu})\, \Omega \label{eqn:IR-flux} 
\end{equation} 
where $T_d$  is the dust temperature and 
$\tau_\lambda = 6\times10^{-27}\,N_H/\lambda\,\, \mathrm{H\,cm^{-3}}$
is the 
optical depth (Draine \& Lee 1984) and $\Omega$ is the solid angle of the source. Given $\Omega$, the observed 
fluxes in the 
two bands determine the dust temperature $T_d$ and associated hydrogen 
column $N_H$.

For RS5 we obtain extincted K$_s$ and L$'$ band fluxes of 1.5 mJy and 25\,mJy, respectively. 
Infrared emission from RS5 traces dust emission 
from the protoplanetary disk  candidate. 
The infrared emission  is unresolved at K$_s$ band, 
implying that the solid angle is less than (27 mas)$^2$, equivalent to radius $R\la110$\,AU at 8\,kpc. Adopting 3 
magnitudes of extinction in K$_s$ band and noting that $A_\lambda \propto 1/\lambda$, equation (1) 
yields the extinction corrected K$_s$ and L$'$ fluxes for 
a dust temperature of 740\,K 
in a layer at the disk surface with  $N_H \approx 1.5\times 10^{20}\mathrm{cm^{-2}} \, 
(R/100\,\mathrm{AU})^{-2}$. These values are reasonable 
providing  that the external radiation field is strong enough. 
To check this, we note that  the 
bolometric luminosity estimated from the 3.8\,$\mu$m flux is $4\pi d^2 \nu S_\nu 
\approx 200\,\mathrm{L}_\odot$. Adopting a total luminosity $L_* \approx 2\times 10^7\,\mathrm{L}_\odot$ in the inner 
0.1\,pc due to the population of hot stars surrounding Sgr A*, implies then 
 the disk intercepts $\sim L_* /(4\pi 
(0.1\,\mathrm{pc})^2) \times 2\pi R^2\sim 240\,(R/100\,\mathrm{AU})^2\,\mathrm{L}_\odot$, suggesting that $R\sim 
100\,$AU. Source P8 is less constrained with 
an extincted L$'$ band flux of 1.8\,mJy, and a  K$_s$ band 
upper limit of 0.5 mJy. 
This yields an 
upper limit on the dust temperature of $1250$\,K.
The high dust temperature associated with RS5 and P8 imply that these ionized features are associated with a layer of 
hot dust and that they are not blobs of ionized gas. The luminosity and column density of gas estimated from near-IR 
data here are very similar to those made toward proplyds found in Orion (Shuping \etal\, 2006). 
The EUV Lyman continuum ionization radiation from a smaller number of massive stars is estimated to 
be $\Phi\approx 8\times10^{49}$\,s$^{-1}$ (Genzel, Hollenbach \& Townes 1994) and for an assumed $\sim 0.5$ pc 
distance from the source of ionization, the incident ionizing photon flux is $\Phi/(4\pi\,(\mathrm{0.5\,pc})^2) 
\approx 2\times10^{12}$\,s$^{-1}$\,cm$^{-2}$. We also note that the peak flux densities of radio source RS5 is 
roughly five times stronger than those of distant proplyd candidates (P8 and P26 in Yusef-Zadeh \etal\, 2015), 
consistent with the suggestion that RS5 is photo-ionized by the stellar cluster near Sgr A*.

Another possibility is that these bowshock structure are  blobs of ionized gas and hot dust
that orbit  around Sgr A* and have a short photoevaporative lifetime.
As discussed 
by Yusef-Zadeh \etal\, (2015) 
the photoevaporation time scale is $\approx 300$\,yr, unless there is a reservoir of neutral material 
associated with a low-mass star. In addition,  the dusty blob 
must be bound by self gravity to avoid tidal disruption by  Sgr A*.  
A distance of 0.1 pc from Sgr A*, 
the  density of the blob must exceed $\nH
\ga 10^{11}\, \ut cm -3 $ with a $M \approx 0.5 \, r_{100} \, \msol$. 
These values are sufficient for the   collapse and formation of  stars. 
To support our argument,  we note that the G2 source
was considered to be a cloud of dust and gas  orbiting Sgr A* (Gillessen \etal\, 2012). However, recent 
measurements suggest that  G2 is unlikely to be an isolated  cloud  of gas 
and must have an embedded core, possibly a pre-main sequence star  (Scoville and Burkert 2013; Witzel \etal\, 2014).  
Future submm observations should test  this scenario 
by searching for 
emission from cool dust in the inner disks of the proplyd candidates. 

%One difference between the source RS5 and the large number of more distant proplyd candidates is that RS5 lies within 
%the ridge of emission from Sgr A*, thus an outflow from Sgr A* may influence  the morphology of RS5. Another 
%difference is that RS5 and RS6 are both embedded within the cluster of 100 massive young stars concentrated within 
%10$''$ of Sgr A*. 

\subsection{Stellar Mass Loss and the Hot Gas Associated with Sgr A*}

To first order, the diffuse X-ray emission centered on Sgr A* is fit by an optically-thin thermal plasma with $kT$ = 3.5 
keV, total X-ray luminosity in the 2-10 keV band $\sim 3\times10^{33}$ erg s$^{-1}$, and mass 
$\sim1\times10^{-3}\msun$ (Baganoff \etal\, 2003; Wang \etal\, 2013), implying  a mean number density $\nH\approx 
140\,\percc$.  This material is usually presumed to be accreting onto Sgr A*, and there is indeed 
evidence for an associated 
non-thermal component arising  from the gas falling in towards Sgr A* (Wang \etal\, 2013).  The 
Bondi accretion rate onto Sgr A* is $\sim 10^{-5} \msol$\,yr$^{-1}$, but this  is likely 
 an overestimate given that 
Sgr A* is not embedded in a uniform zero angular momentum medium.  Sub-mm polarization measurements indicate 
that the accretion rate close to the event horizon is $\sim 10^{-7}$ -- 10$^{-9}$ \msol\, yr$^{-1}$, depending on 
assumptions about the magnetic field (Marrone \etal\, 2007).

This gas is generally thought  to be supplied by the combined winds of the mass-losing young stars in the central 
parsec of the Galaxy, which are estimated to supply material to Sgr A* at a rate of a few times $10^{-6}\smpy$ (Coker 
\& Melia 1997; Rockefeller \etal\, 2004; Cuadra \etal\, 2006, 2008).  However, the discovery of populations of young 
stars within the S-cluster which consists of $\sim$16 B dwarfs and 3 O stars 
on highly-eccentric orbits within 1$''$, begs the 
question whether mass loss from these stars plays a dominant role in supplying gas to the vicinity of Sgr A*.  Loeb 
(2004) suggested that mass loss from the S-stars could explain this hot gas.  
In this picture winds from the 
stars are shocked to high temperature because of the high orbital speeds of the S-stars.  To attain a shock 
temperature of 3.5\,keV requires a shock speed $\sim 1700$\,\kms, which corresponds to the Keplerian speed at 6\,mpc 
(0.\arcs15) from Sgr A*.  Loeb (2004) suggested that the S-stars were OB or WR stars with powerful winds, but it is now 
known that they are mainly B stars. Theoretical calculations suggest that B stars have individual mass loss rates $\la 
10^{-8}$\,\msol\,yr$^{-1}$ (e.g.~Vink, de Koter \& Lamers 2000; Puls, Vink \& Navarro 2008), insufficient to replenish 
the hot gas.  However, the mass loss rates are known to be severely underpredicted, and recent modeling of the shells 
blown in molecular clouds by young A  and B stars suggests far higher rates, $\sim 10^{-7}$--$10^{-6}$\,\msol\,yr$^{-1}$ 
(Offner \& Arce 2015). Thus, the suggestion that 
the wind  created 
by the merging of individual stellar winds from the B stars in the S-star cluster expands and 
excludes the shocked winds from O and WR stars in the 
central parsec of the Galaxy (c.f. Loeb 2004)

The increasing evidence for YSOs within arcsecond of Sgr A*, such as the sources RS5 and P8 as discussed in this 
paper, 
suggests that YSOs may be intermingled with the disk and S-star populations. 
As pointed out earlier, the so-called G2 cloud could be another low-mass YSO candidate 
orbiting Sgr A* (Scoville and Burkert 2013). 

The young, massive stars within  $\sim$ 0.05pc - 0.5pc of Sgr A*, 
a fraction of which are found on the so-called clockwise rotating disk, 
are  formed between 2.5 and 6 Myrs ago (Lu et al 2009). 
Thus, it is reasonable to assume that a population of lower mass 
stars, with attendant protoplanetary disks, is also present.  The IMF is not as top-heavy as originally 
thought, but still somewhat flatter than the Salpeter or Kroupa IMFs, with $\phi(M) \propto M^{-1.75}$ (Lu et 
al.~2013).  
Of the 31 K$_s$ \le 16 magnitude stars that reside within the 
projected distance of 1$''$ from Sgr A*, 16 are B stars, 3 are O stars, and 12 are late-type stars, 
of which 16 B stars and 3 late type stars are likely to be 
true members after correcting for contamination by foreground and background stars (Genzel \etal\, 2010). 
Extrapolation down to solar-mass stars is fraught with uncertainty but
assuming that there are $\sim$16 young stars within the S cluster  with 
masses exceeding $3.5\msun$ implies about 120 stars with masses between 0.5 and 3.5 $\msun$.  
If these stars, which presumably have ages $\sim3\times10^6$ years still possess disks, then 
are exposed to the intense radiation from the OB stars in the central 0.5\,pc ($G_0\sim 10^5$) they will collectively 
lose mass by 
photoevaporation at a rate $\sim 10^{-6}\msol\,\mathrm{yr}^{-1}$.  In addition, material 
mass will be tidally stripped  from these 
disks. Here, we assumed that  B-type main sequence stars are formed the same time as 
the population of  OB  and WR stars beyond the inner 1$''$ of Sgr A*. 
 The truncation radius for stellar mass $M$ is $r_t \approx (2M_*/M_{BH})^{1/3}\,r \approx 
8\,\textrm{AU}\,(M/\msun)^{1/3}(r/5\,\mathrm{mpc}) $ where $r$ is distance from Sgr A*.
Figure 12 shows a schematic diagram of how low mass stars feed Sgr A* as well as create a high  pressure X-ray gas  
that prevent the material belong the inner 0.5 fall into  Sgr A*. 

%I think that there is a problem with this argument because you assume that the S-stars have formed in *the same star 
%formation event* than the O/B stars farther out, which formed 3-6 Myr ago. THis is not at all justified. The B-stars 
%near Sgr A* can be several tens of Myr old. They show clearly difference kinematicla properties than the younger stars 
%in the disk. Also, you would have to form the S-stars and or pre-MS stars at large distances from the SMBH and then 
%bring them in. They cannot have formed in situ. All of this must be discussed here if you want to claim that there is 
%an entire population of proplyds in the central arcsecond. A single one, maybe, but several ones would pen HUGE 
%theoretical problems, I think.

The X-ray emission extends slightly beyond the S-cluster boundary suggesting that the gas is not bound to Sgr A* and 
instead escapes as a supersonic wind (cf. Quaetaert 2004). At some point the escaping cluster wind encounters the material 
supplied by the combined winds of the massive young stars beyond 0.5\,pc.  In our proposed scenario, the outflowing 
material from the S-cluster prevents this material from reaching within 1-2$''$ of Sgr A*.  
Our interpretation of the 
diffuse X-ray emission associated with Sgr A*, due to an expanding hot wind fed by the S-cluster (whether B stars or 
low-mass YSOs or both), yields a very different estimate of the accretion rate onto Sgr A* than the standard picture.  
Instead, a small fraction of the stellar winds injected very close to Sgr A* is captured by the black hole; a rough 
estimate of the accretion rate is obtained by using the Bondi-Hoyle accretion rate appropriate for a medium of density 
$\rho$ moving by a point mass $M=4\times10^6$\,\msol\, at the typical stellar orbital 
speed $v=4000$\,km\,s$^{-1}$, i.e. $\dot{M} = 4\pi 
G^2M^2\rho/v^3 \approx 3\times10^{-7}$\,\msol\,yr$^{-1}$. This estimate is likely 
an upper limit because it neglects any net 
angular momentum. The proposed model naturally reduces the accretion rate 
to 
a level consistent with submm rotation measure constraints, 
and removes the need 
for inflow-outflow  solutions which magically turn most of the inflowing material around at very small radii and eject it to infinity.

\subsection{Jet Activity  of Sgr A*}

A  linear feature with an extent of $\sim$3 
pc at a position angle of $\sim60^\circ$ (Yusef-Zadeh \etal\, 2012) 
was recently suggested   to be tracing  a jet from Sgr A* interacting with the surrounding medium. 
Here we describe  two additional features, 
the Sgr A East tower and cometary sources, with PAs  $\sim60^\circ$. 
These features are suggestive  of a jet  interacting at sites near Sgr A*.

\subsubsection{The Sgr A East Tower}

We recently reported the tentative detection of a continuous linear structure 
symmetrically centered on Sgr A* with 
PAs$\sim60^\circ$ and 
240$^\circ$ 
(Yusef-Zadeh \etal\, 2012). The feature is terminated by linearly polarized 
structures P1 and P4 $\sim75''$ from Sgr A* at a PA$\sim60^\circ$ and $240^\circ$, respectively. 
This structure was interpreted as a mildly relativistic jet 
interacting with the ionized gas orbiting Sgr A*. We observed 
a larger region around Sgr A* 
and have identified  additional sites of 
of possible interaction between this jet candidate  
and  the Sgr A East SNR.
We presented the  striking tower structure 150$''$ NE of Sgr A*. 
This tower is an extension of  the Sgr A East shell distorted 
 toward the   NE. We note  multiple bow shock-like structures 
at the location where the tower terminates. 
In addition, structural details of P1 and P4 show linear structures 
that are aligned along the jet candidate. 
  Altogether, morphological structures presented at radio 
support a picture in which  
a energetic jet-driven outflow  is required to explain the distortion of the Sgr A East shell, 
as well as the alignment  of a number of sources at 
PAs$\sim50^\circ-60^\circ$ to the NE and
PAs$\sim230^\circ-240^\circ$ to the SW.

We adopt jet parameters $\gamma\sim 3$, $\dot{M}\sim1\times10^{-7}\smpy$, and opening angle 10 degrees.   
Then the jet pressure at the location of the tower, 150$''$ (or 6pc) from Sgr A*, is about 
$P_\mathrm{jet} = \dot{M}\, \gamma\, c\, / (4\pi\, d^2)\, \approx 
4\times10^{-8}\,$ dyn\,cm$^{-2}$.  
This is able to push through the thermal X-ray emitting gas filling 
the interior of  Sgr A 
East at speed $\sqrt{P_\mathrm{jet}/\rho}\approx 6\,00$\,\kms, implying a crossing time $\sim 
10^4$\,years assuming the  tower is 6pc away from Sgr A*. 

The jet is  also able to  drag a portion of the Sgr A East shell. 
The intensity of the synchrotron emission from the tower, i.e. 2.8\,mJy per 
$1\dasec39\times0\dasec6$ beam at 1.5\,GHz implies an equipartition field of 0.7\,mG and 
a total pressure of $2\times \mathrm{B^2}\,/8\pi\, \approx3.5\times10^{-8}$ erg\, cm$^{-3}$ 
(Here we have adopted a source depth 1\,pc, an $E^{-2}$ electron spectrum extending between 1\,MeV and 10\,GeV, and 
assumed that the energy density in protons is 100 times that for the electrons).
The jet pressure  is comparable to to  the magnetic pressure in the nonthermal shell 
of Sgr A East, thus the jet is able to push through.  
This picture implies that Sgr A East lies close to Sgr A*. The 
absorption of Sgr A West against Sgr A East (Yusef-Zadeh \& Morris 1987; Pedlar \etal\, 
1989) suggest that the NE jet must be moving away from us to be interacting with the shell of Sgr A East. 

An alternative possibility that the tower is generated by the passage of a neutron star that received a large 
velocity kick at birth.  The neutron star overtakes the remnant and produces a trail behind it as it interacts with 
the remnant. However, the trail behind the neutral star is expected to be narrow in this picture, unlike the observed 
structure of the extended base and the bow shock structures. Other SNRs also show jet-like or chimney-like structures at 
the edge of the remnant (e.g., Crab and CAS A), so it may be that the the Sgr A East tower is 
produced in the remnant by a generic mechanism  that   is not 
associated with Sgr A*.

%Before being ejected in a jet, this material must be accreted to the close vicinity of Sgr A*. So, would such a high 
%mass-loss rate in the jet not contradict directly te Bondi-Hoyle rate estimated in the previous section, which is a 
%factor of a few smaller?

%In this picture,  a nonthermal jet  from Sgr A* interacts  with stars and the medium in which 
%it is punching through. 

%If this jet is associated with the intrinsic elongation of Sgr A* on mas scale, 
%the jet must have changes in direction from a  PA$\sim90^\circ$ at AU   scale to 
% 50$^\circ$ at pc scale.   
%The features presented here do not show the actual jet structure but the interaction sites and 
%the working surface of the jet. 

\subsubsection{Cometary Sources}

Radio and infrared images, as described above, 
indicate alignment of a number of sources  within few arcseconds of Sgr A*  at a position angle around  50$^\circ$. 
These include the  cometary structures X7 and  X3 located SW, within 0.8$''$  and 3.4$''$ of Sgr A*, respectively
 (Muzic \etal\, 2007, 2010). 
Both X3 and X7 show proper motions in the direction away from the direction toward Sgr A* 
suggesting that the bow-shock morphology of these sources is produced by an outflow from the direction of Sgr A* 
and not by their   motion (Muzic \etal\, 2007, 2010).  
We now detect two additional cometary sources F1 and F2, within 4.8$''$  NE of Sgr A*. 
These sources lie within 5$''$ of Sgr A* aligned at a PA$\sim50^\circ$. 
Muzic \etal\, (2007, 2010) explain the origin of  X3 and X7 in terms of an outflow either 
from the cluster wind 
interacting with the atmosphere  of stars, or an outflow from Sgr A* driving a shock that produces the 
cometary morphology. 
Given that almost all the cometary sources, X3, X7,  source F1 (radio cometary) and F2 (MIR cometary), 
 have the same position angle to that 
of  polarized sources P1 and P4 and the Sgr A East tower, 
a collimated outflow from Sgr A* could be better alternative to explain the 
origin of cometary structures.

The cometary sources are marginally resolved, implying stand-off distances equivalent to $\sim 50$\,mas at 8\,kpc, ie. 
$\sim6\times10^{15}$\,cm.  The ionized gas mass derived from the 35\,GHz flux density $\sim 0.2$\,mJy beam$^{-1}$ is 
$\sim 4\times10^{-5}\msun$, and adopting a wind speed of 750\, \kms\, we obtain mass loss rates $\sim 10^{-5}\smpy$ 
and ram pressures at the stand-of distance $\sim 10^{-4}$\,dyn\,cm$^{-2}$.  This is comparable to the jet ram pressure 
at $\sim0.1$\,pc from Sgr A* and compatible with the projected separations of X3, F1 and X7 sources, i.e., 0.13, 0.19, 
0.03 pc.

%, but the projected separation of X7 from Sgr A* is only 
%$0\dasec8$.

\subsection{Flaring Activity of Sgr A*}

One possible  explanation for the  ridge of radio emission 
is related the flaring activity of Sgr A*. 
Past monitoring campaigns  have found evidence for time 
delay of $\sim30$ minutes between the peak emission at 43 and 22 GHz 
(Yusef-Zadeh \etal\,  2006, 2009; 
Brinkerink \etal\, 2015).   This time delay is 
consistent with a picture in which an overpressured synchrotron emitting plasma blob  at these 
wavelengths is initially optically thick. 
The  blob then expands subrelativistically,  peaks and declines at each frequency once it 
becomes optically thin.  
In this picture, the blob-like  and arc-like  structures noted   in the east-west ridge of emission 
detected at radio wavelengths are interpreted to be  expanding blobs
produced by flaring activity escaping from Sgr A*. 
If the thermal density of expanding blobs is 
 sufficient to overcome the external pressure, the  outflow rate of thermal blobs of plasma is 
estimated to be $\la 2\times10^{-8}$\,\msol\,yr$^{-1}$ (Yusef-Zadeh \etal\, 2006). 
The expansion blob model has 
successfully been applied to flaring activity of microquasars where 
outflows  have  been detected  (e.g., Fender and Belloni 2004). 

%If the expanding blobs have high bulk motion, the plasma blob can exceed the escape speed of Sgr A* where the 
%gravitational potential of Sgr A* is weak and become unbound. Assuming a plasma blob is formed at 1000 AU from Sgr A*, 
%the escape speed is 2.6$\times10^3$\, \kms.

Another possibility is that  the ridge of radio emission results from 
the interaction of  thermal winds from 
the cluster of young massive stars   with the gravitational potential of Sgr A*. 
Wardle and Yusef-Zadeh (1992) considered a picture in which the IRS 16 stars  are
the source of ionized winds interacting with Sgr A*.  
In this picture, 
thermal winds are  focused by Sgr A* and form
blobs of hot gas in the Sgr A* ridge.
Given that  recent observations indicated that 
mass-losing stars are members of a cluster that  lie in a disk orbiting Sgr A*, 
it is not clear how the ridge is produced asymmetrically on one side of 
Sgr A*.  

VLBA measurements on mas scale  have identified  elongated structure associated 
with Sgr A* at 43 GHz with a position angle of 95$^\circ$ (Bower \etal\, 
2014). These authors explain this elongation in terms of both jet and 
accretion disk models. The radio plume and the ridge of emission closest 
to Sgr A* in Figure 1a lie along a  position angle similar to the 
intrinsic elongation of Sgr A*. This  suggests that the 
elongation of Sgr A* might be  associated with blobs of radio emission  
detected within the ridge but on a scale  roughly 2000 times 
smaller than the ridge of emission from Sgr A*. The inference 
is that the radio sources in the plume-like and ridge structures are indirectly tracing 
the interaction of an outflow due to flaring activity of Sgr A* and stars located in the S cluster.  
The orientation of the outflow from flaring activity is not collimated by the disk so the opening angle could be 
large, as evidenced in the plume structure.  On the other hand, the jet from Sgr A* is presumably 
collimated by the disk 
and has a different PA than that of blobs of gas ejected from the 
corona of the accretion disk  of Sgr A* due to its flare activity.

%If the ridge of emission is indeed associated with the elongation of Sgr A*, as measured by VLBA, the position angle of the 
%outflow from Sgr A* must be changing from east-west to south-west directions, as plasma blobs move away from Sgr A*.

\subsection{Conclusions}

\subsubsection{Morphology}

In  summary, several new morphological features are 
revealed within the inner 30$''$ of Sgr A*. On a scale of few arcseconds, an east-west 
plume-like ridge of emission appears to arise from Sgr A* toward SW. This diffuse feature shows eight radio sources 
within 2$''$ of Sgr A*, two of which are extended with arc-like morphology. One of the arc-like sources RS5 faces in 
the general direction of Sgr A* whereas the other RS6 faces away from the direction of Sgr A*. Although the ridge of 
emission is seen on one side of Sgr A*, we also detect new  cometary and head-tail structures (sources F1, F2 \& F3) 
within 13$''$ of Sgr A*  to the NE at the 
position angle of $\sim50^\circ-60^\circ$, 
pointing in the direction of Sgr A*. Two previously identified cometary sources X3 and 
X7 located 3$''$ and 0.7$''$ to the SW of Sgr A*, respectively, at a position angle of $\sim50^\circ$. These five 
cometary sources X3, X7 and F1. F2 and F3 lie within $\sim5^\circ$ of a line passing through Sgr A*. 
In addition, the ionized gas 
along the Northern arm shows a discontinuity as the gas approaches Sgr A*. The discontinuous structure is $\sim2.5''$ SE 
of Sgr A*. We also note that some of the ionized gas from the Northern arm shows a semi-linear feature curving around Sgr 
A* and giving the appearance of a cloud leaving a trail of ionized gas along its path. On a scale of $20''$, we note a 
hollow cone-like structure, within which a striking bent filament is detected. On a scale of two arcminutes from 
Sgr A*, we detect a new tower-like feature which appears to be associated with the Sgr A East shell. 
This remarkable structure becomes narrower as it extends to the NW with a PA$\sim50^\circ$ and terminated with 
two bow shock like structures.  

\subsubsection{Interpretation}

The identification of new structural and kinematic features within a few arcseconds of Sgr A* motivates a 
new scenario for the origin of the hot gas responsible for the diffuse X-ray emission associated with Sgr A*. We 
argued the presence of low-mass YSO candidates within the inner 2$''$ of Sgr A* based on the 
morphology, size and their dust and ionized properties. We  also  argued 
that stellar winds from low mass YSOs  within the S cluster  merge to 
form a cluster wind that interacts with the surrounding orbiting ionized gas and excludes the combined winds from the 
massive stars in the central parsec of the Galaxy. 
The  accretion rate of the S-cluster wind onto Sgr A* is 
$\la 
3\times10^{-7}$\,\msol\,yr$^{-1}$, 
helping to explain the low luminosity of Sgr A*. 
This is an alternative to models 
in which the low luminosity is due to ejection of the bulk of the material accreted from larger radii. 
In our picture,  the cluster wind accounts  the  observed mass of  X-ray gas within 1$''$ of 
Sgr A* has a residence time of  $\sim10^3$ years. 
This X-ray gas plays two major roles. 
One  is that the  injected material during the cooling time 
of X-ray gas $\sim10^5$ years expands away and leaves the inner 1$''$ of Sgr A*. 
The injected ionized winds 
 from the low-mass and/or young stars prevents the material  beyond  1$''$ of Sgr A* to reach Sgr A*. 
The second role of the X-ray gas surrounding Sgr A* 
is that the accretion rate to the injected material by stars is 
 at least two orders of magnitude  lower than the Bondi accretion rate  
of $\sim10^{-5}$ \msol\, yr$^{-1}$. This is because the injected mass from stars 
has a much higher velocity due to their orbital motion  
 and is highly stirred near Sgr A*. Thus, a very small fraction of the X-ray gas 
accretes onto Sgr A*. 
  
On a larger scale, we interpreted a number of cometary features
and the distorted shape of the Sgr A East SNR in terms of the interaction of a collimated jet-driven outflow  with
the surrounding medium and impinging on stellar envelopes  as well as a portion of the nonthermal Sgr A East shell. 
Lastly, we interpret that the east-west ridge of radio emission results from flaring activity of 
Sgr A*. Flares are explained in terms of expanding, over-pressured,   plasma blobs that 
escape  the gravitational 
potential of Sgr A*. Unlike the jet-driven outflows,     the outflowing material should not be highly 
collimated. Finally, we note that  many of these  suggestions 
can be tested  by high resolution proper motion and submm  measurements. 

%Over-pressured bubble of hot gas as it expands into its surroundings.  

Acknowledgments:
This work is partially supported by the grant AST-0807400 from the NSF.
The research leading to these results has also received 
funding from the European Research Council under the European Union's Seventh 
Framework Programme (FP/2007-2013)/ERC Grant Agreement No. [614922].

\begin{deluxetable}{llcccc}      
\tablecaption{VLA Observations of Sgr A*}
%\rotate
\tabletypesize{\scriptsize}            
\tablecolumns{10}
\tablewidth{0pt}
\setlength{\tabcolsep}{0.03in}
\tablehead{       
\colhead{Date} & \colhead{Frequency} & \colhead{Bandwidth} & \colhead{Number of IFs} & 
\colhead{Number of Channels} & \colhead{{$ \theta_{a} \times \theta_{b} $ (PA)}} \\
 & GHz & GHz & & & \sl{arcsec} $ \times $ \sl{arcsec} (\sl{deg}) }
\startdata
2014 Feb. 21 & 44.6 & 8 & 16 & 64 & 0.074 $ \times $ 0.034 (-4.0)\\
2014 March 9 & 34.5 & 8 & 16 & 64 & 0.089 $ \times $ 0.046 (-1.6)\\
2014 March 10 & 14.1 & 8 & 16 & 64 & 0.187 $ \times $ 0.073 (0.6)\\
2014 April 17 & 8.9 & 2  & 16 & 64 & 0.187 $ \times $ 0.073 (0.6)\\
2014 April 02 & 5.5 & 2  & 16 & 64 & 0.59 $ \times $ 0.27 (-0.2)\\
2014 April 02 & 1.52 & 1  & 16 & 64 & 1.87 $ \times $ 0.91 (0.8)\\
\enddata
\end{deluxetable}
\vfill\eject

\begin{deluxetable}{lccccccccc}
\tablecaption{Predicted Positions of Young  Stars in the S Cluster (Gillessen \etal\, 2009) at the 2014.19 Epoch}
\tabletypesize{\scriptsize}
\tabletypesize{\scriptsize}
\tablecolumns{7}
\tablewidth{0pt}
\setlength{\tabcolsep}{0.04in}
\tablehead{
\colhead{Source Name} & RA (IR)  & Dec (IR) & \colhead{($\sigma_X$)} & \colhead{($\sigma_Y$)}\\ 
%&   \colhead{X (NIR)} & \colhead{Y (NIR)} & \colhead{X (Radio)} & \colhead{Y (Radio)}\\
                                &  \sl{$17^{h} 45'$} & \sl{$-29^\circ 00'$} & \sl{arcsec} & \sl{arcsec} 
%&\sl{pixel number} & \sl{pixel number} & \sl{pixel number} & \sl{pixel number}
}
\startdata
%P#  Name        RA (IR)       Dec (IR)    dx(IR) dy(IR) 
%E19 & IRS16NW &  40$^s$.0442 & $26'$.843  & 0.078 &  1.226 \\
S1 &  40.0286 & 28.1151 & 0.0144 & 0.0187 \\  
S2 & 40.0411 & 28.0174 & 0.0008 & 0.0013 \\
S4 &  40.0570 & 28.0474 & 0.0216 & 0.0086 \\
S5 & 40.0348 & 28.0605 & 0.0075 & 0.0077 \\
S6  & 40.0337 & 28.0758 & 0.0212 & 0.0024 \\
S8 &  40.0793 & 28.5629 &  0.0052 & 0.0048 \\
S9 &  40.0358 & 28.0184 & 0.0058 & 0.0090 \\
S12 &  40.0490 & 27.9601 & 0.0186 & 0.0144 \\
S13 &  40.0342 & 28.3788 & 0.0221 & 0.0156 \\
S14 & 40.0529 & 27.8995 & 0.0225 & 0.0193  \\
S17 &  40.0425 & 27.7964 & 0.0015 & 0.0060 \\
S18  & 40.0414 & 28.0197 & 0.0125 & 0.0149 \\
S19 &  40.0421 & 28.1146 & 0.0203 & 0.0077 \\
S21 &  40.0417 & 28.0590 & 0.0086 & 0.0050 \\
S24  &  40.0393 & 28.1468& 0.0031& 0.0166 \\
S27  &  40.0397 & 27.9761 &0.0033 &0.0160 \\
S29  & 40.0304 & 27.9426 &0.0866 &0.1822 \\
S31   & 40.0387 & 27.9182 &0.0416 &0.1664 \\
S33  & 39.9993 & 28.1424 &0.1097 &0.0620 \\
S38   & 40.0404 &  28.0739 &0.0218 &0.0226 \\
S66  & 39.9561 & 27.8625 &0.1121 &0.1450 \\
S67  & 39.9990 & 28.3748 &0.0648 &0.0858 \\
S71  & 40.0204 & 28.2907 &0.5941 &1.1285 \\
S83  & 40.1387 & 28.6212 &0.9300 &0.1324 \\
S87  & 39.9140 & 28.3996 &0.2083 &0.3649 \\
S96  & 39.9402 & 27.6973 &0.1739 &0.2426 \\
S97  & 40.0403 & 29.0882 &2.0261 &0.4675 \\
\enddata
\end{deluxetable}

\vfill\eject

%\begin{deluxetable}{lccccccccc}
%\tablecaption{Predicted Positions of Young  Stars in the S Cluster (Lu et al. 2009) at the 2014.19 Epoch}
%\tabletypesize{\scriptsize}
%\tabletypesize{\scriptsize}
%\tablecolumns{7}
%\tablewidth{0pt}
%\setlength{\tabcolsep}{0.04in}
%\tablehead{
%\colhead{Source Name} & RA (IR)  & Dec (IR) & \colhead{($\sigma_X$)} & \colhead{($\sigma_Y$)}\\ 
%                                &  \sl{$17^{h} 45'$} & \sl{$-29^\circ 00'$} & \sl{arcsec} & \sl{arcsec}\\

\begin{deluxetable}{lccccccccc}
\tablecaption{Predicted Positions of S Stars (Lu et al. 2009) at the 2014.19 Epoch}
\tabletypesize{\scriptsize}
\tabletypesize{\scriptsize}
\tablecolumns{7}
\tablewidth{0pt}
\setlength{\tabcolsep}{0.04in}
\tablehead{
\colhead{Source Name} & RA (IR)  & Dec (IR) & \colhead{($\sigma_X$)} & \colhead{($\sigma_Y$)}\\ 
                                &  \sl{$17^{h} 45'$} & \sl{$-29^\circ 00'$} & \sl{arcsec} & \sl{arcsec} 
}
\startdata
S0-14 &  39.9808 & 28.3463 & 0.0040 & 0.0017 \\  
S0-15 &  39.9635 & 27.9216 & 0.0039 & 0.0016 \\
S1-3  &  40.0587 & 27.1722 & 0.0012 & 0.0021 \\
S1-2  &  40.0473 & 29.0847 & 0.0017 & 0.0063 \\
S1-8  &  39.9958 & 28.9944 & 0.0031 & 0.0040 \\
IRS16NW &  40.0465 & 26.8338 & 0.0009 & 0.0033 \\
IRS16C  &  40.1151 & 27.4796 & 0.0023 & 0.0012 \\
S1-12   &  39.9839 & 29.0952 & 0.0032 & 0.0038 \\
S1-14   &  39.9395 & 28.456  & 0.0041 & 0.0013 \\
IRS16SW &  40.1246 & 29.0023 & 0.0025 & 0.0023  \\
S1-21   &  39.9143 & 27.9777 & 0.0060 & 0.0013  \\
S1-22   &  39.9214 &28.5818  &0.0049  &0.0018   \\
S1-24   & 40.0941  & 29.7428  &0.0019   &0.0039  \\
S2-4    & 40.1553  & 29.5114  &0.0036  &0.0036  \\
IRS16CC &40.1886  &27.4463   &0.0036  &0.0013   \\
S2-6    &40.1663  &29.3915  &0.0027  &0.0023  \\
S2-7  &40.1074  &26.2056  &0.0030  &0.0051  \\
IRS29N  &39.9219  &26.7034   &0.0038  &0.0034  \\
IRS16SWE  &40.1835  &29.1707  &0.0029  &0.0019  \\
IRS33N  &40.0362  &30.3219   &0.0016  &0.0065  \\
S2-17  &40.1421  &29.9418   &0.0033  &0.0047  \\
 S2-16  &39.9546  &25.9938  &0.0024  &0.0046  \\
S2-19  & 40.0654 &25.7507   & 0.0015  &0.0047  \\
S2-66  &39.9283  &25.8824  &0.0086   & 0.0116   \\
S2-74  & 40.0452  &25.2775  &0.0020  &0.0063  \\
IRS16NE  &40.2596  &27.1622  &0.0034  &0.0015  \\
S3-5  &40.2636  & 29.2064 &0.0049  &0.0022   \\
IRS33E  &40.0932  &31.1956  &0.0054   &0.0107   \\
S3-19  &39.9227  & 30.8233  &0.0073  &0.0105  \\
S3-25  &40.1437  &25.1221  &0.0055  &0.0089  \\
S3-30  &40.1623  &30.9647  &0.0065   &0.0106  \\
S3-10  &40.2908  &29.1575  &0.0060  &0.0026  \\
\enddata
\end{deluxetable}

%P#  Name        RA (IR)       Dec (IR)    dx(IR) dy(IR) 
%E19 & IRS16NW &  40$^s$.0442 & $26'$.843  & 0.078 &  1.226 \\
%&   \colhead{X (NIR)} & \colhead{Y (NIR)} & \colhead{X (Radio)} & \colhead{Y (Radio)}\\
%&\sl{pixel number} & \sl{pixel number} & \sl{pixel number} & \sl{pixel number}

\vfill\eject

\begin{deluxetable}{lccccccccc}
\tablecaption{Predicted Positions of S Stars (Yelda \etal\,  2014) at the 2014.19 Epoch}
\tabletypesize{\scriptsize}
\tabletypesize{\scriptsize}
\tablecolumns{7}
\tablewidth{0pt}
\setlength{\tabcolsep}{0.04in}
\tablehead{
\colhead{Source Name} & RA (IR)  & Dec (IR) & \colhead{dX} & \colhead{dY}\\
&  \sl{$17^{h} 45'$} & \sl{$-29^\circ 00'$} & \sl{arcsec} & \sl{arcsec}
%&\sl{pixel number} & \sl{pixel number} & \sl{pixel number} & \sl{pixel number}
}
\startdata
S1-3        &      40.0586     &     27.173     &     0.2669  &     0.8961    \\
S0-15       &    39.9637      &  27.928         &     -0.9786 &     0.1414  \\
IRS16C      &    40.1152     &  27.479          &  1.0093     &    0.5898  \\
S1-12       &   39.9844      &  29.099    & -0.7068  &   -1.0301  \\
S1-14       &   39.9394      &  28.465     &  -1.2979   &  -0.3955  \\
IRS16SW     &  40.1252     &  28.999    &  1.1404   &  -0.9304  \\
S0-14       &  39.9811      &  28.352    &  -0.7504 &    -0.2828  \\
S1-1        &  40.1194      &  28.022    &  1.0634  &    0.0475  \\
IRS16NW     &  40.0461    &  28.838     &  0.1021  &    1.2310  \\
S1-33       &  39.9432     &  28.042    & -1.2481  &    0.0273  \\
S1-18       &  39.9757     &  26.554  & -0.8217  &     1.5152  \\
S1-22       &  39.9210 & 28.595 &  -1.5386  &   -0.5256  \\
S2-4        &  40.1565 & 29.508 &  1.5503   &   -1.4388  \\
S2-7        &  40.1069 & 26.204 &  0.9001   &   1.8649  \\
S2-6        &  40.1673 & 29.386 &  1.6927   &  -1.3167  \\
IRS16SW-E   & 40.184  & 29.164 &  1.9187   &  -1.0944   \\
S2-22       &   40.2131   &  28.241 &  2.2934 &    -0.1724  \\
S2-58       &   40.2015 & 29.161 &   2.1414  &    -1.0924  \\
S1-2        &   40.0482 & 29.081 &  0.1296  &   -1.0118  \\
S1-8        &   39.9966 & 29.001 & -0.5473  &   -0.9320  \\
S1-21       &   39.9145 & 27.990 & -1.6247  &    0.0789  \\
S1-19       &   40.0738 & 29.712 &  0.4655   &  -1.6432  \\
S1-24       &   40.0951 & 29.740 &  0.7451   &  -1.6709  \\
IRS16CC    &  40.1888 & 27.436 &  1.9748  &     0.6325  \\
IRS29N      & 39.9213 & 26.7314 & -1.53534 &     1.3548  \\
IRS33N      & 40.0372 &  30.3426 &  -0.01349  &   -2.2574  \\
S2-50       & 40.1686 & 29.5458  & 1.70947 & -1.4895  \\
S2-17       & 40.1436 & 29.9439  & 1.38141 &    -1.8701 \\
S2-16       & 39.9538 & 26.0307  & -1.10837 &  2.0622  \\
S2-21       & 39.9178 & 29.7348  & -1.58133 &    -1.6793  \\
S2-19       & 40.0646 & 25.7450  & 0.34541  &    2.3194  \\
S2-74       & 40.0442 & 25.2478  & 0.07746  &    2.7909  \\
S2-76       & 40.0220 & 25.2451  & -0.21348 &     2.8184  \\\
IRS16NE     & 40.2599 & 27.1443  & 2.90649  &    0.9260  \\
S3-2        & 40.2748  & 27.5406   & 3.10341 &     0.5632  \\
S3-3        & 40.2750  & 28.6486   & 3.10449 &    -0.6172  \\
S3-5        &  40.2648 & 29.191    & 2.9708  &   -1.1224  \\
S3-96       &  39.7994  & 28.666  & -3.1334   &  -0.5972  \\
S3-19       &  39.9227  & 30.863  & -1.5164   &  -2.7942  \\
IRS33E      &  40.0942  & 31.203  & 0.7334    & -3.1336  \\
S3-25       &  40.1433  & 25.105   &1.3774     & 2.9641  \\
S3-26       &  39.8438  & 30.125  & -2.5510  & -2.0555  \\
S3-30       &  40.1646  & 30.979 &   1.6565    & -2.9103  \\
IRS13E1     &  39.8099  & 29.729 &  -2.9957   &  -1.6596  \\
S3-190      &  39.7939  & 26.660 & -3.2056  &    1.4093  \\
S3-10       &  40.2928  & 29.148 &   3.3391 &    -1.0790  \\
IRS13E4     &  39.7893  & 29.462 &  -3.2661 &    -1.3927  \\
IRS13E2     &  39.7918  & 29.786 & -3.2340  &   -1.7167  \\
S3-314      &  40.3317  & 28.133 &  3.8484  &   -0.0639  \\
S3-331      &  39.9466  & 24.390 & -1.2027  &    3.6787  \\
S3-374      &  39.8279  & 30.929 & -2.7602  &   -2.8595  \\
S4-36       &  39.7548  & 26.300 & -3.7193  &    1.7690  \\
S4-71       &  40.0970  & 32.171 &  0.7697  &   -4.1025  \\
IRS34W      &  39.7271  & 26.530 & -4.0827  &    1.5393  \\
S4-169      &  40.3739  & 27.767 &  4.4024  &    0.3022  \\
IRS3E       &  39.8623  & 24.245 & -2.3086  &    3.8236  \\
IRS7SE      &  40.2682  & 24.599 &  3.0162  &    3.4705  \\
S4-258      &  39.7011  & 29.683 & -4.4234  &   -1.6136  \\
S4-262      &  40.3640  & 30.039  & 4.2724 &    -1.9700  \\
IRS34NW     &  39.7483 & 25.252   & -3.8045 &     2.8174  \\
S4-287      &  40.0493  & 32.826  & 0.1440  &   -4.7570  \\
S4-364      &  40.2106  & 23.603  & 2.2608  &    4.4663  \\
S5-34       &  39.7065  & 30.811  &-4.3528  &   -2.7421  \\
IRS1W       &  40.4383  & 27.392  &  5.2471 &     0.6768  \\
S5-235      &  40.2497  & 23.540  & 2.7737  &    4.5289  \\
S5-237      &  40.4569  & 27.026  & 5.4915  &    1.0431  \\
S5-236      &  39.6174  & 29.340  & -5.5214 &    -1.2707  \\
S5-183      &  40.3870  & 31.513  & 4.5749  &   -3.4439  \\
S5-187      &  39.9073  & 33.626  & -1.7185 &    -5.5571  \\
S5-231      &  40.4814  & 27.937  &  5.8131 &     0.1319  \\
S5-191      &  40.2804  & 32.963  & 3.1755  &   -4.8942  \\
S6-89       &  40.4545  & 25.086  & 5.4602  &    2.9830  \\
IRS9W       &  40.2609  & 33.638  & 2.9201  &   -5.5687  \\
S6-90       &  39.7367  & 23.163  & -3.9571 &     4.9056  \\
S6-96       &  39.5767  & 29.948  & -6.0548 &    -1.8790  \\
S6-81       &  40.5221  & 27.770  & 6.3469  &    0.2995  \\
S6-95       &  39.8557  & 22.060  & -2.3955 &      6.0087  \\
S6-63       &  40.1824  & 34.364  & 1.8900  &   -6.2947  \\
S6-93       &  40.3795  & 23.101  & 4.4765  &    4.9678  \\
S6-100      &  40.1551  & 21.527  & 1.5319  &    6.5418  \\
S6-82       &  40.5512  & 28.501 &   6.7288 &    -0.4320  \\
S7-30       &  40.5303  & 30.770 &  6.4538  &   -2.7012  \\
S7-161      &  39.4753  & 28.020 & -7.3861  &    0.0491  \\
S7-16       &  40.1630  & 35.285 &  1.6362  &   -7.2159  \\
S7-19       &  39.7511  & 21.539 & -3.7671  &    6.5295  \\
S7-180      &  39.4757  & 29.704 & -7.3807  &   -1.6345  \\
S7-10       &  39.9517  & 20.445 & -1.1363  &    7.6236  \\
S7-36       &  40.5245  & 32.471 &  6.3780  &   -4.4018  \\
S7-216      &  39.4498  & 26.607 &  -7.7205 &     1.4621  \\
S7-20       &  39.7581  & 21.097 & -3.6762  &    6.9717  \\
S7-228      &  39.4492  & 26.339 & -7.7281  &    1.7296  \\
S7-236      &  39.4958  & 24.488 & -7.1167  &    3.5812  \\
S8-15       &  39.9146  & 20.042 & -1.6228  &    8.0272  \\
S8-7        &  39.7592  & 35.481  & -3.6613 &    -7.4118  \\
S8-181      &  39.4564  & 31.661 & -7.6335  &   -3.5919  \\
S8-4        &  40.0363  & 19.487  & -0.0264 &     8.5824  \\
S8-196      &  39.4219  & 30.966  & -8.0861 &    -2.8968  \\
S9-143      &  39.4005  & 31.424  & -8.3670 &    -3.3548  \\
S9-20       &  40.3676  & 36.091  & 4.3203  &   -8.0216  \\
S9-23       &  39.9392  & 18.939  & -1.3003 &     9.1305  \\
S9-13       &  39.8091  & 19.226  & -3.0070 &     8.8430  \\
S9-1        &  40.7579  & 27.803  & 9.4396  &    0.2660  \\
S9-114      &  39.5431  & 34.936  & -6.4968 &    -6.8671  \\
S9-283      &  39.3065  & 30.602  & -9.6001 &    -2.5329  \\
S9-9        &  40.4686  & 36.258  & 5.6447  &   -8.1888  \\
S10-50      &  40.7688  & 31.252  & 9.5831  &   -3.1832  \\
S10-136     &  39.3801  & 33.320  & -8.6351 &    -5.2510  \\
S10-5       &  39.9176  & 18.037  & -1.5838 &    10.0321  \\
S10-4       &  40.0433  & 17.806  & 0.0653  &   10.2635  \\
S10-32      &  40.8174  & 29.741  & 10.2201 &    -1.6723  \\
S10-34      &  40.7155  & 33.673  & 8.8832  &   -5.6042  \\
S10-7       &  40.7783  & 23.667  & 9.7069  &    4.4018  \\
S10-48      &  39.9987  & 17.326  & -0.5190 &     10.7429  \\
S11-21      &  40.2329  & 17.133  &  2.5531 &    10.9360  \\
S11-5       &  40.1426  & 16.359  & 1.3684  &   11.7096  \\
S13-3       &  40.9456  & 22.124  & 11.9027 &     5.9454  \\
\enddata
\end{deluxetable}

\begin{deluxetable}{llcccclcc}
\tablecaption{Positions of Radio Sources}
%\rotate
\tabletypesize{\scriptsize}
\tablecolumns{9}
\tablewidth{0pt}
\setlength{\tabcolsep}{0.03in}
\tablehead{
\colhead{ID} & \colhead{Alt Name} & \colhead{RA (J2000)} & 
\colhead{Dec (J2000)} & \colhead{Dist. from Sgr A*} & \colhead{Pos. Accuracy} & \colhead{$ \theta_{a} \times \theta_{b} $ (PA)} & \colhead{Peak Intensity} 
& \colhead{Integrated Flux} \\
& & ($17^{\rm h}45^{\rm m}$) & ($-29^{\circ}00^{\prime}$) & (\sl{arcsec}) & (\sl{mas}) & \sl{mas} $ \times $ \sl{mas} 
(\sl{deg}) & \l{(mJy beam$^{-1}$)} & \sl{(mJy)}
}
\startdata
RS1 &  & 40.0070 & 28.0275 & 0.41 & 2.72 & 21 $ \times $ 0 (7)
 & 0.178 $ \pm $ 0.018 & 0.133 $ \pm $ 0.026  \\
RS2 &  & 39.9986 & 27.9628 & 0.53 & 5.94 & 180 $ \times $ 68 (1)
 & 0.238 $ \pm $ 0.017 & 0.985 $ \pm $ 0.084  \\
RS3 &  & 39.9955 & 28.1008 & 0.56 & 2.90 & ---
 & 0.197 $ \pm $ 0.018 & 0.102 $ \pm $ 0.021  \\
RS4 &  & 39.9569 & 28.4896 & 1.15 & 8.27 & 230 $ \times $ 106 (180)
 & 0.206 $ \pm $ 0.016 & 1.485 $ \pm $ 0.133  \\
RS5 &  & 39.9248 & 28.2151 & 1.50 & 7.08 & 326 $ \times $ 76 (173)
 & 0.329 $ \pm $ 0.016 & 2.492 $ \pm $ 0.139  \\
RS6 &  & 39.9139 & 28.5265 & 1.70 & 2.46 & 140 $ \times $ 81 (11)
 & 0.471 $ \pm $ 0.017 & 1.848 $ \pm $ 0.080  \\
RS7 &  & 39.8933 & 28.1996 & 1.91 & 2.97 & 59 $ \times $ 53 (88) & 0.254 $ \pm $ 0.017 & 0.492 $ \pm $ 0.048  \\
RS8 &  & 39.8881 & 28.3757 &  1.99  & 8.58   & 60 $ \times $ 20 (14) & 0.075 $ \pm $ 0.014 & 0.101 $\pm  $ 0.030  \\                    &  
\enddata
\end{deluxetable}

\begin{figure}[tbh]
\centering
\includegraphics[scale=0.6, angle=0]{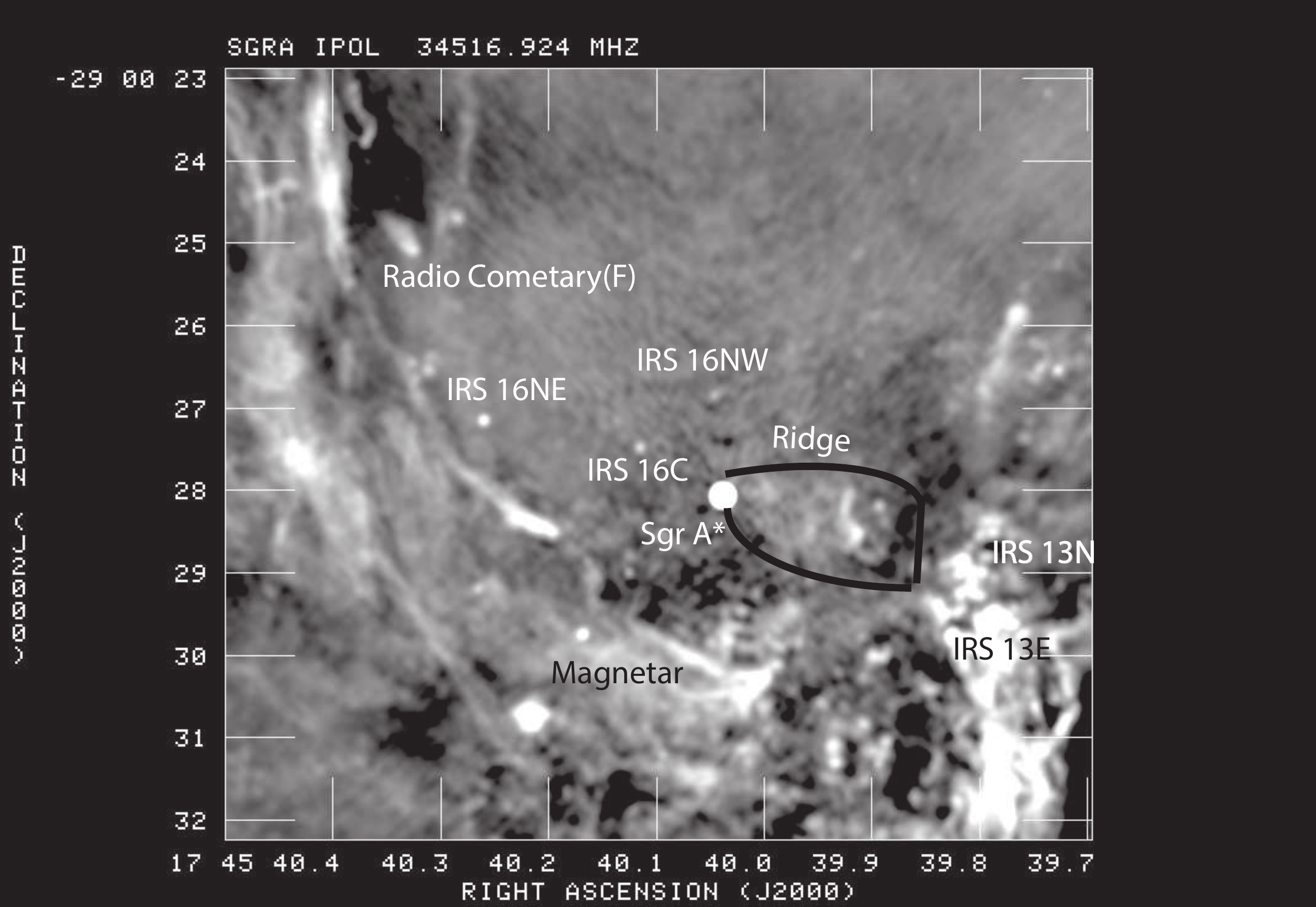}
\includegraphics[scale=0.6, angle=0]{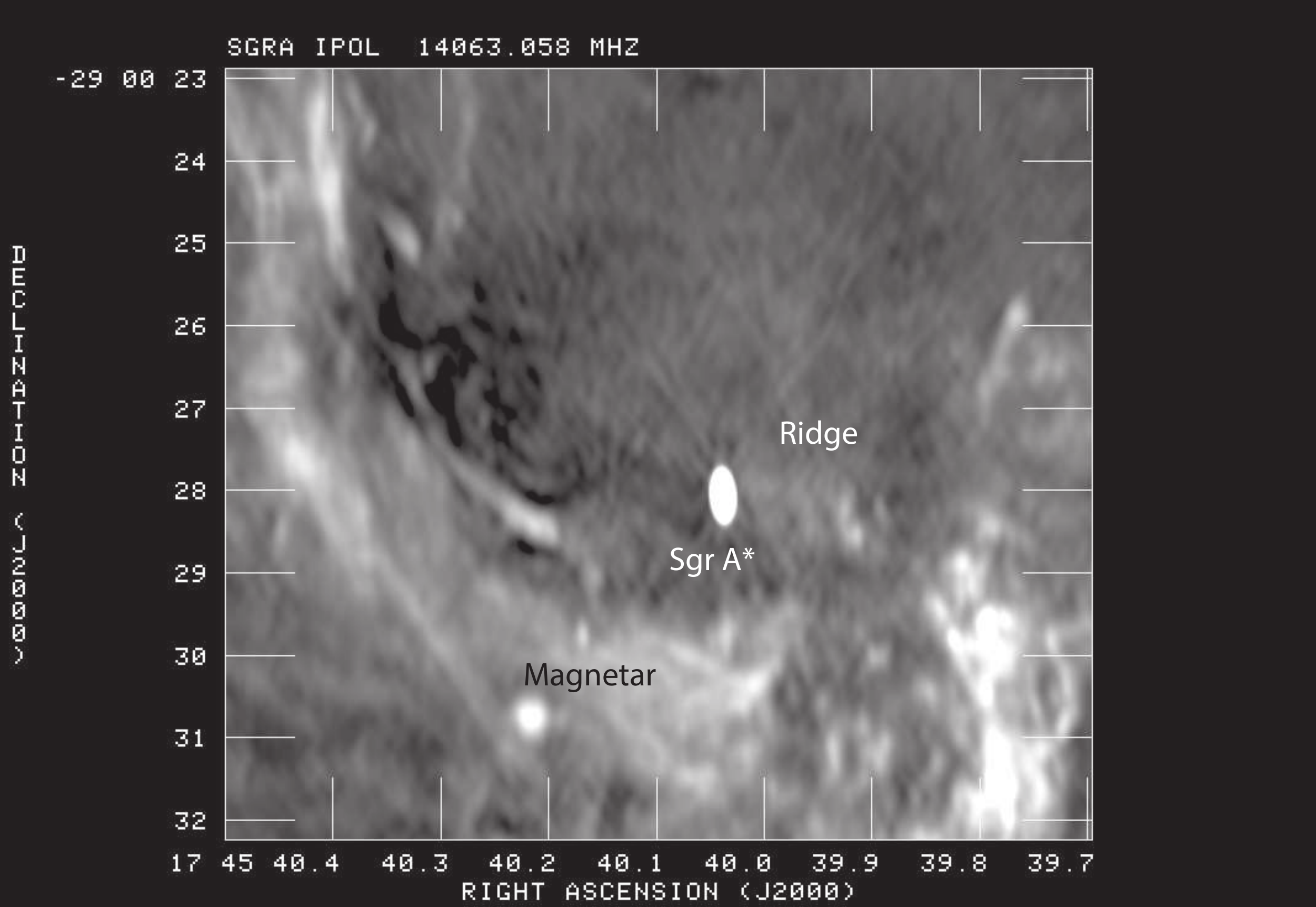}
\caption{\small\small{
{\it (a)}        
A grayscale 34.5 GHz  image convolved to a resolution of 
0.1$''\times0.1''$ with grayscale range -0.3 and 1 mJy.  
{\it (b)}
The same as (a) except at the 14.1 GHz with a resolution of 
234$\times108$ mas. 
{\it (c)}    
Similar to (a) except that the image display ranges between 
-0.31 and  0.1 mJy.  
{\it (d)}    
Similar to (a) except in reverse color with a resolution   88$\times46$ mas. 
%{\it (e)}
%The same as (b) except at 5.8 GHz with a resolution of
%0.59$''\times0.27''$.
}}
\end{figure}

\begin{figure}[tbh]                 
\centering
\includegraphics[scale=0.6, angle=0]{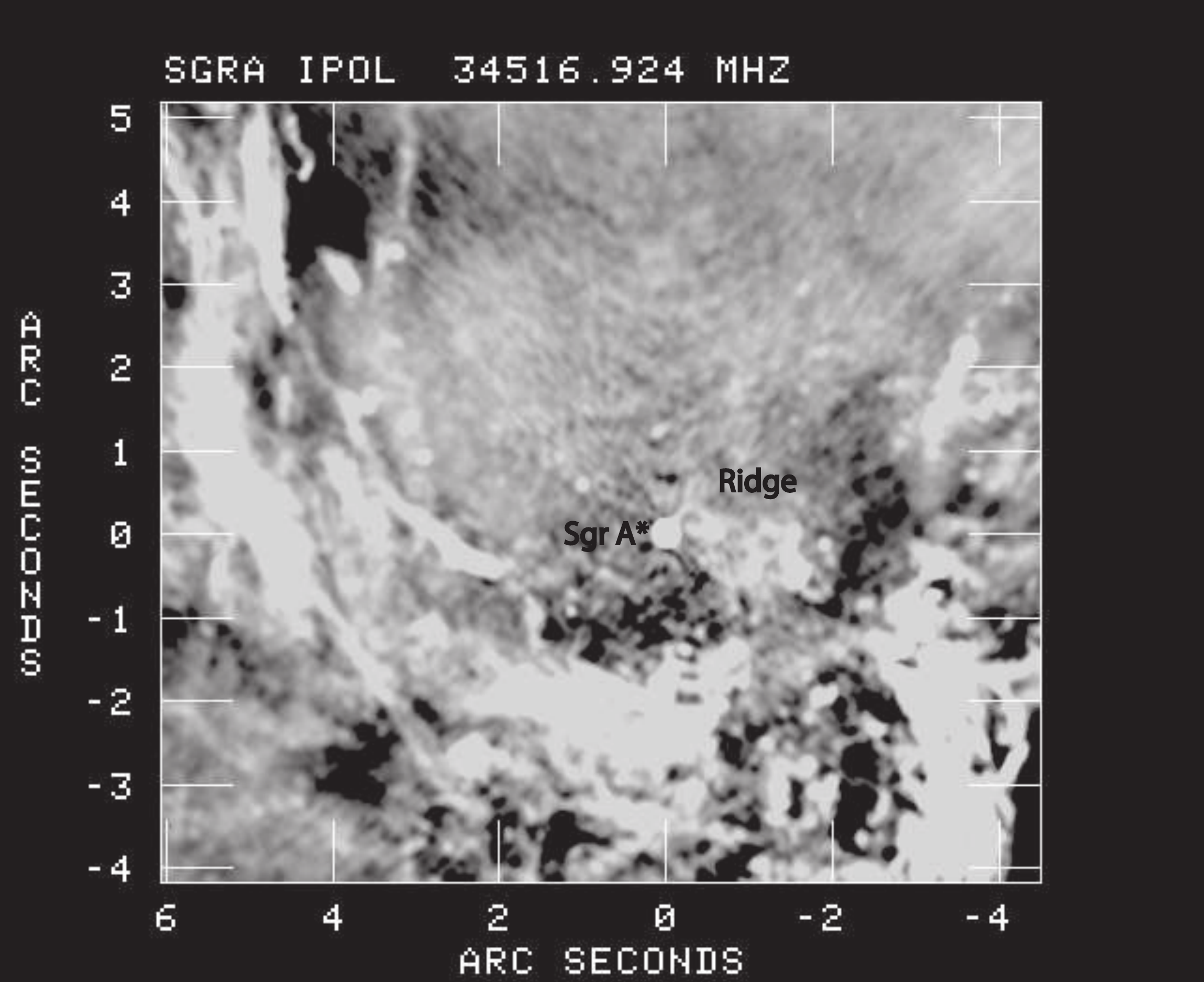}
\centering
\includegraphics[scale=0.6, angle=0]{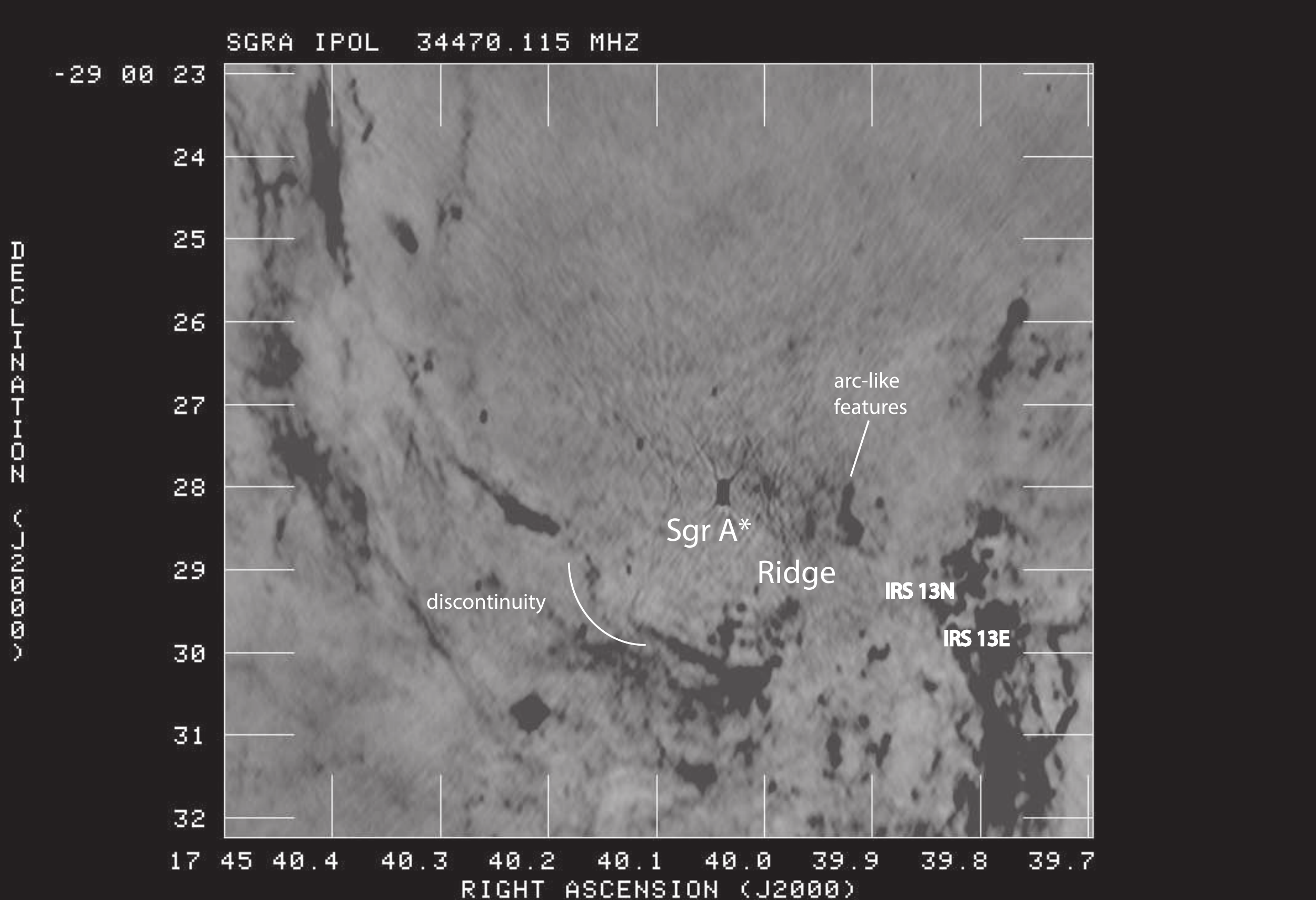}
%\centering
%\includegraphics[scale=0.4, angle=0]{f1e_CC_3291+3543_46934787.pdf}
\end{figure}

\begin{figure}[p]
\center
\includegraphics[scale=0.8, angle=0]{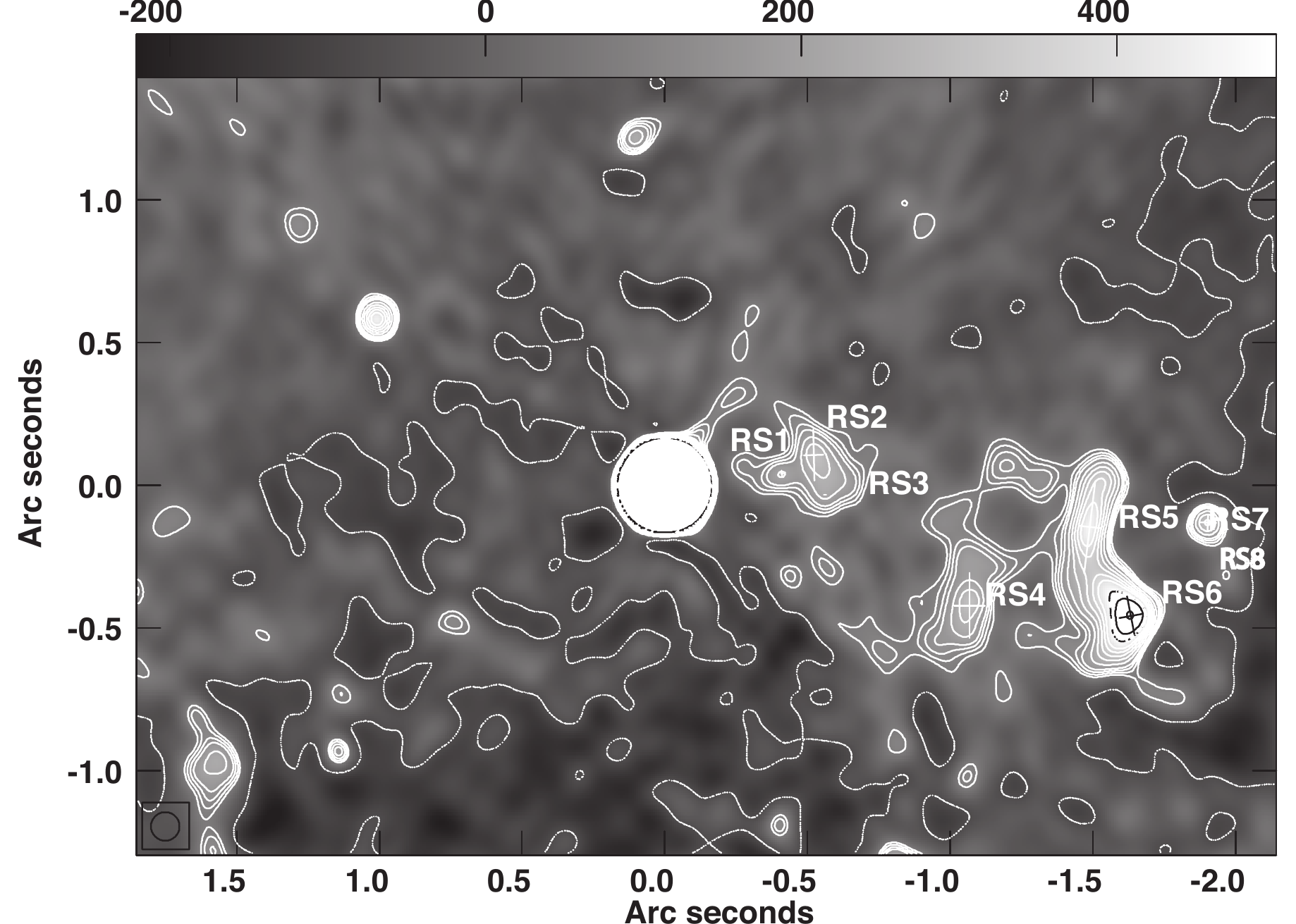}
\includegraphics[scale=0.8, angle=0]{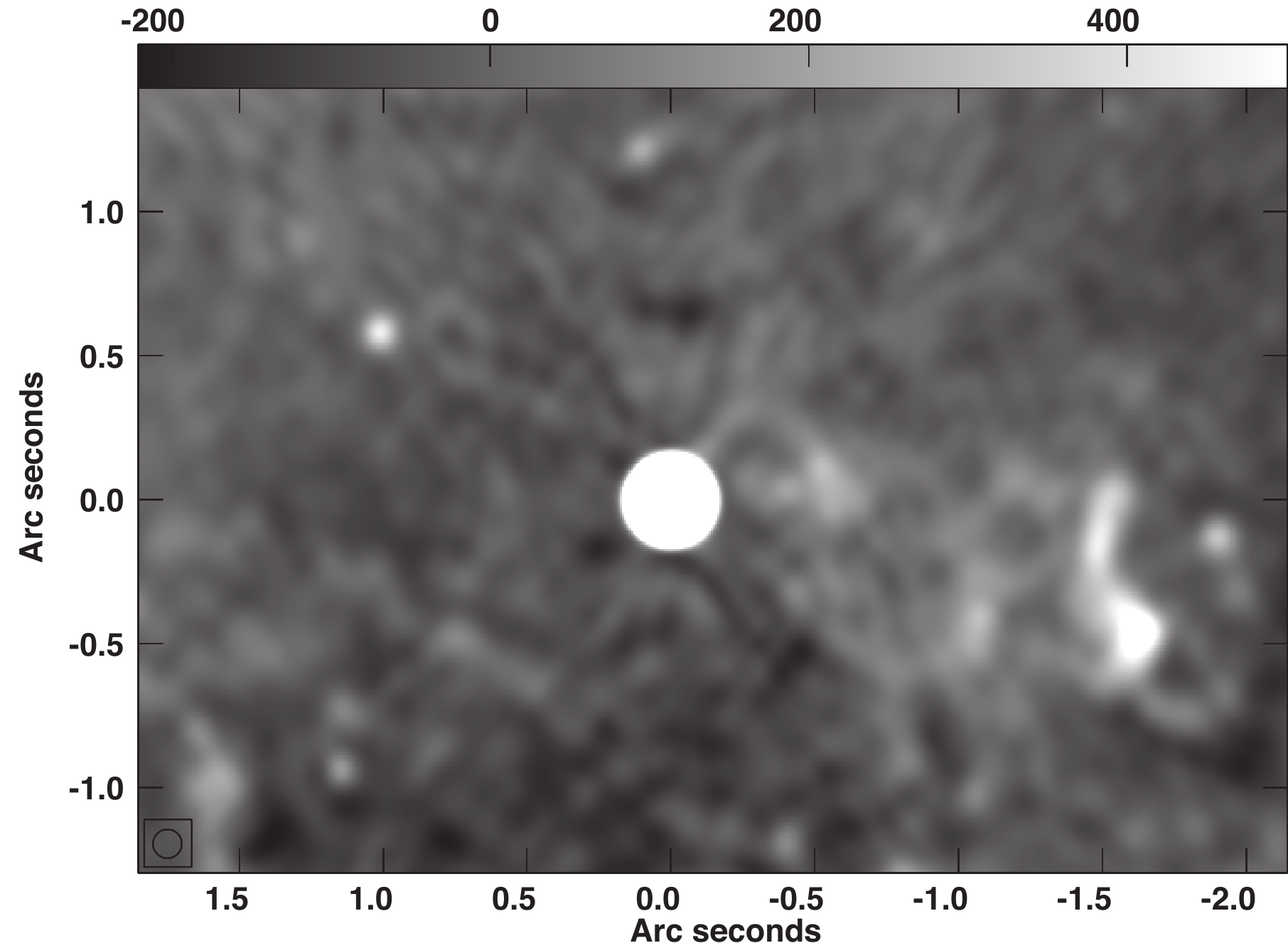}
\caption{\small\small{
{\it (a)}  
A 34.5 GHz grayscale image similar to  Figure 1a
except that the image is convolved to a resolution of 0.1$''\times0.1''$. 
The coordinate labels are given relative to the position of 
Sgr A*.  
Eight   radio sources are detected to the W of Sgr A* in the east-west ridge within  
2$''$ of Sgr A* (see Table 5). 
The grayscale  ranges between -220 and 500 $\mu$Jy beam$^{-1}$. 
{\it (b)}
The same as (a) except no contours are drawn.  
{\it (c)}  
Similar to (a) expect at 5.5 GHz with a resolution of 
0.59$''\times0.27''$ (PA=$-0.2^\circ$). 
The grayscale  ranges between -2.5 and 5 mJy beam$^{-1}$. 
}}
\end{figure}

\vfill\eject

\begin{figure}[tbh]                 
\centering
\includegraphics[scale=0.8, angle=0]{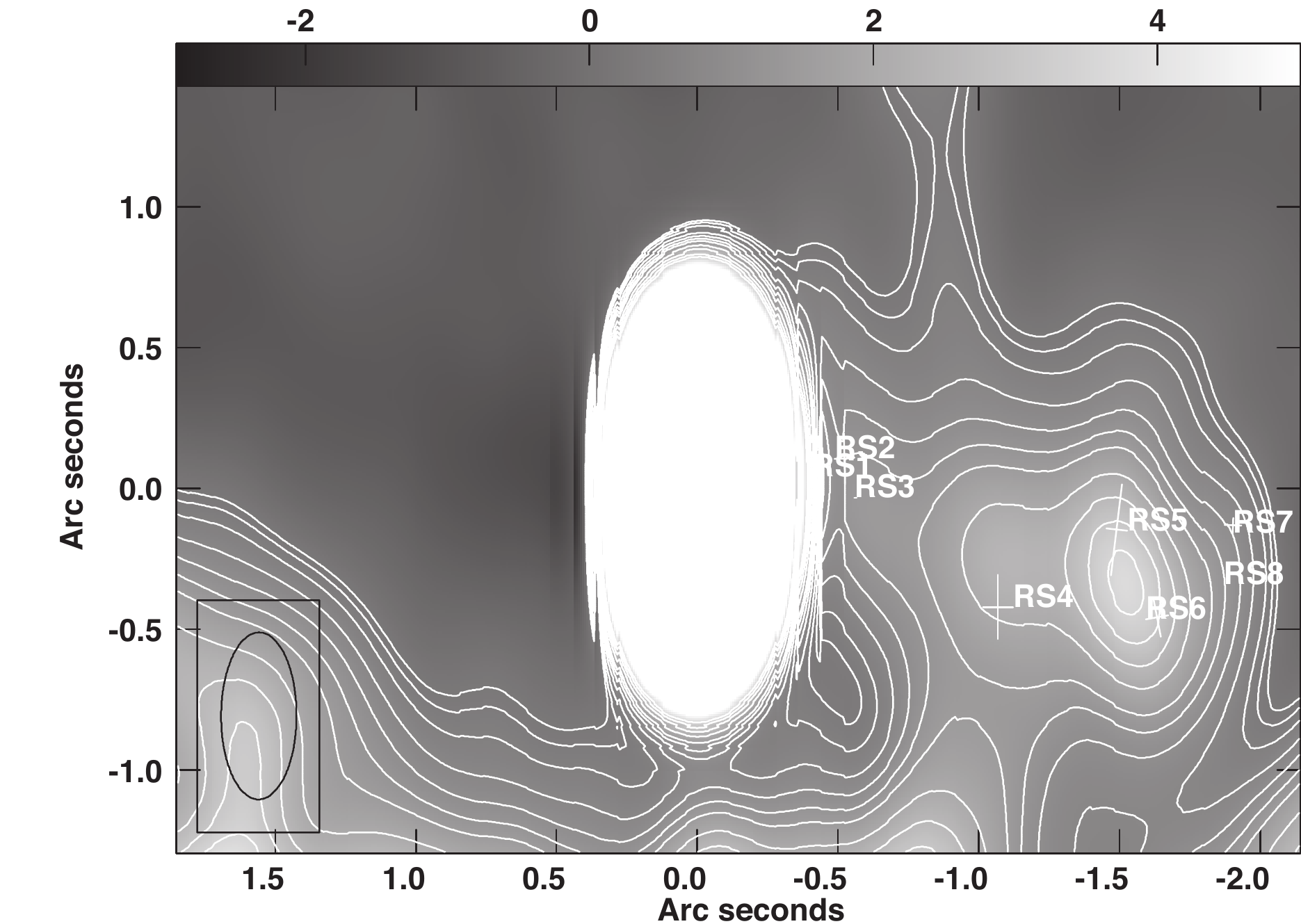}
\end{figure}

\vfill\eject

\begin{figure}[tbh]
\centering
\includegraphics[scale=0.8, angle=0]{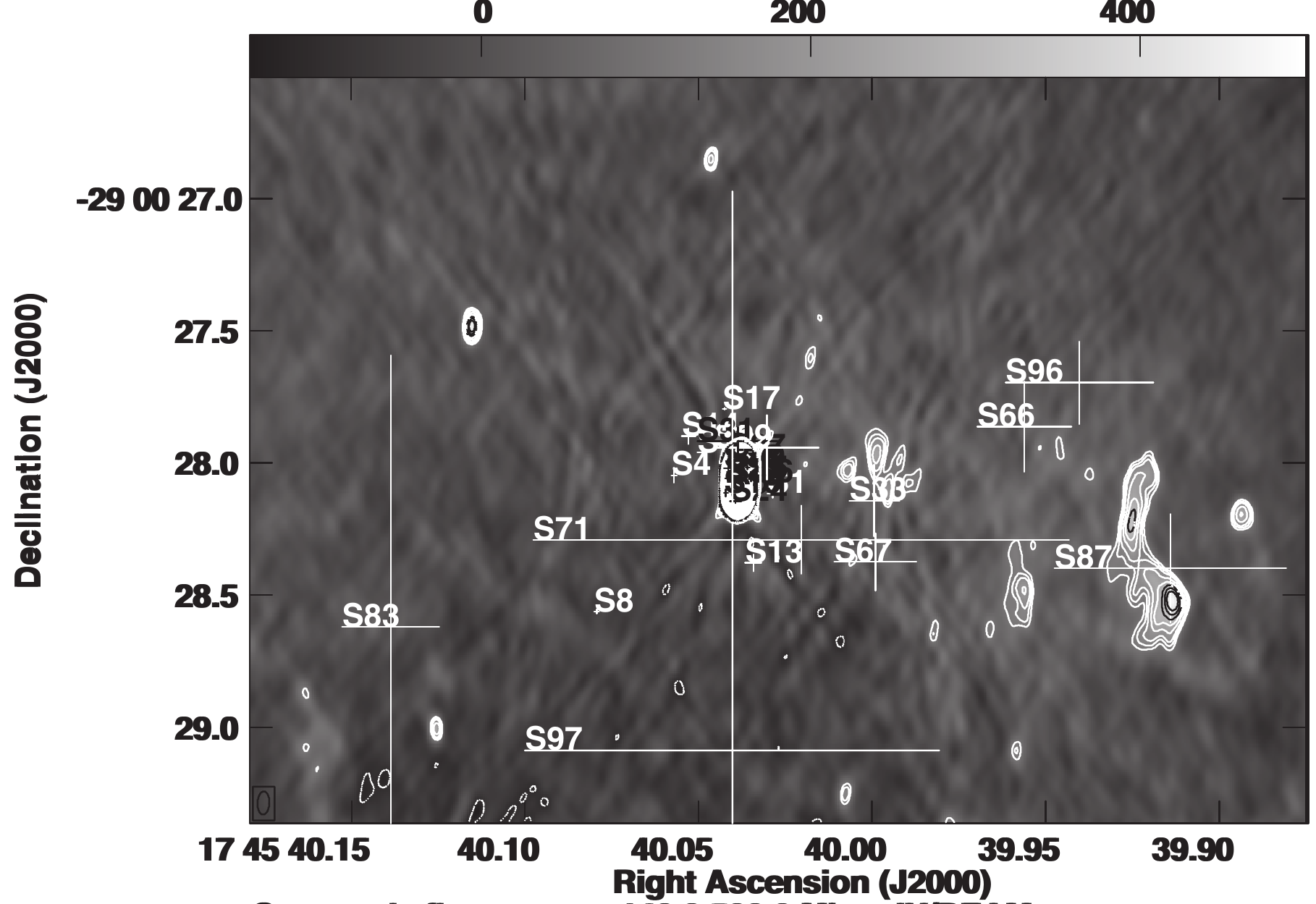}
\includegraphics[scale=0.8, angle=0]{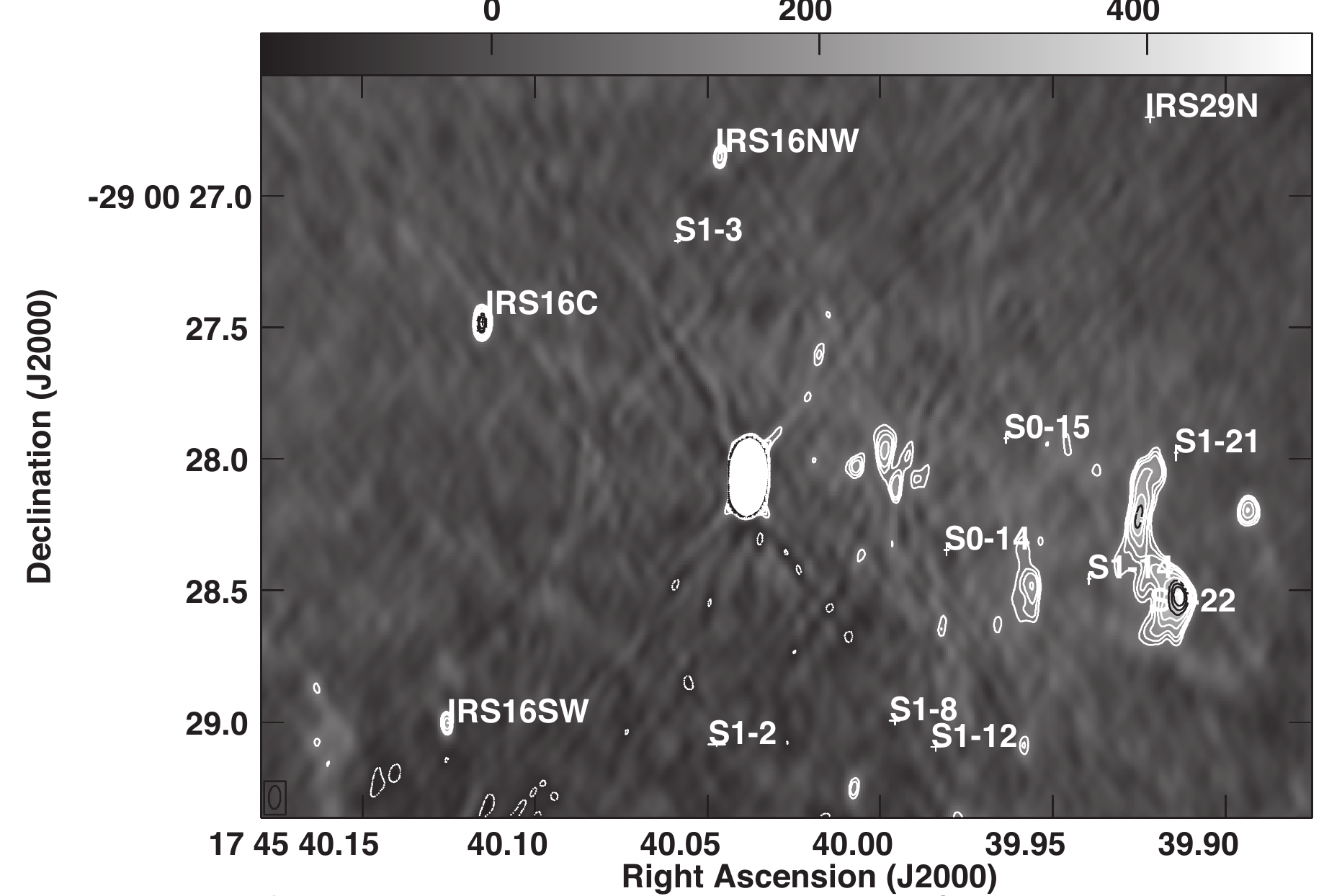}\\
\caption{\small\small{
{\it (a)}        
Grayscale contours of  34.5 GHz emission 
at (-100, 100, 125, 150 200, 250, ..., 
400) $\times \mu$Jy beam$^{-1}$. 
The error bars correspond to the predicted 
position of  S-stars (Gillessen \etal\,  2009) 
at the epoch 2014.19. 
The resolution of the image is 879$\times448$ mas (PA=-2.5$^\circ$). 
{\it (b)}        
Similar to (a) except  that the crosses correspond to the 
position of stars given by  Lu et al. (2009). 
The drawn error bars are twice  the estimated values.
The grayscale  ranges between -140 and 500 $\mu$Jy beam$^{-1}$ in both (a) and (b). 
{\it (c)}  
The same region as  shown in (a)
except that the image is convolved to a resolution of 0.1$''\times0.1''$. 
The coordinate labels are given relative to the position of 
Sgr A*.  The crosses  correspond to radio sources detected in 
the S-cluster RS1--8 (Table 4). 
The grayscale  ranges between -200 and 500 $\mu$Jy beam$^{-1}$. 
{\it (d)}
Similar to (c) except an L$'$ image   
}}  
\end{figure}

\begin{figure}[tbh]                 
\centering
\includegraphics[scale=0.8, angle=0]{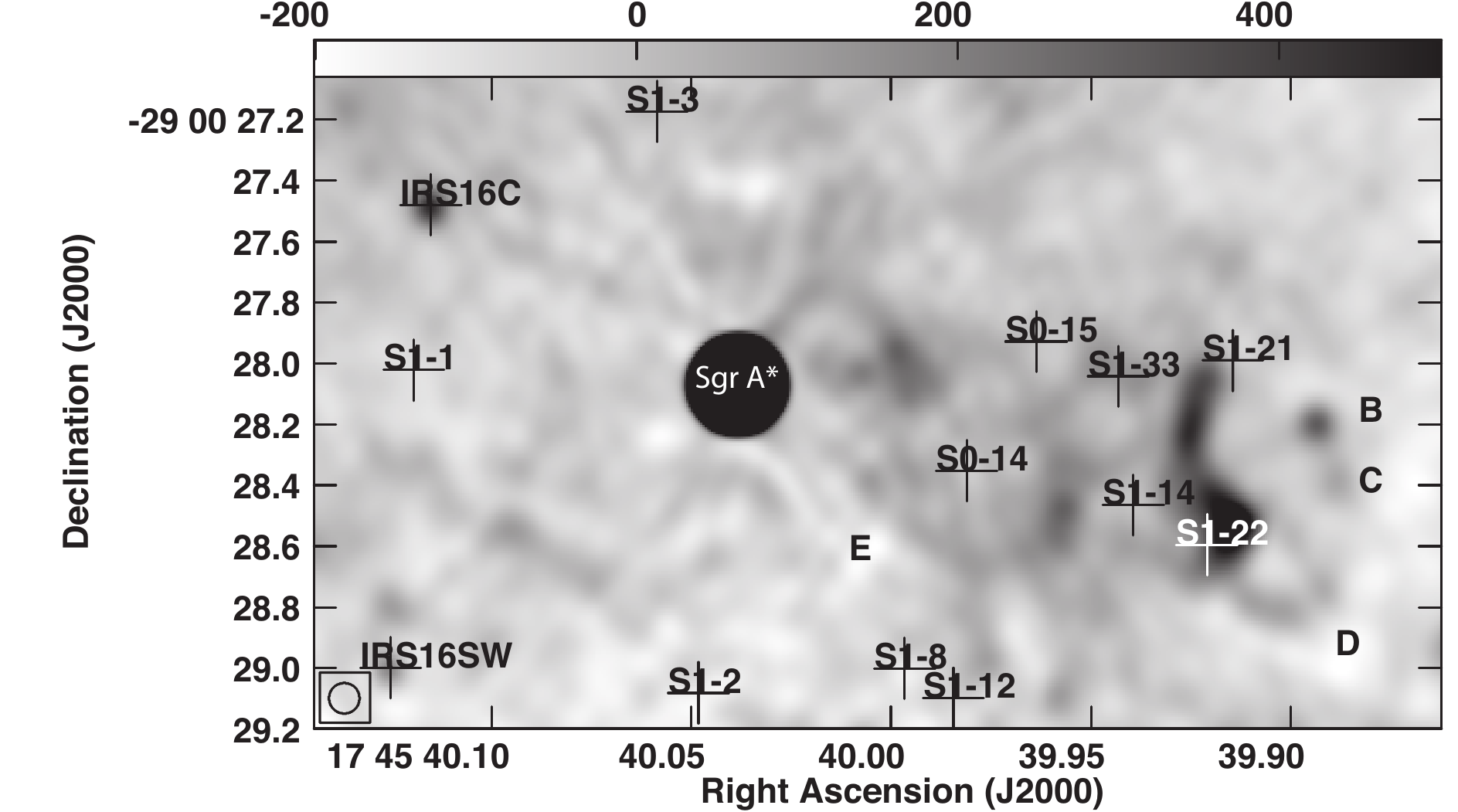}
\includegraphics[scale=0.8, angle=0]{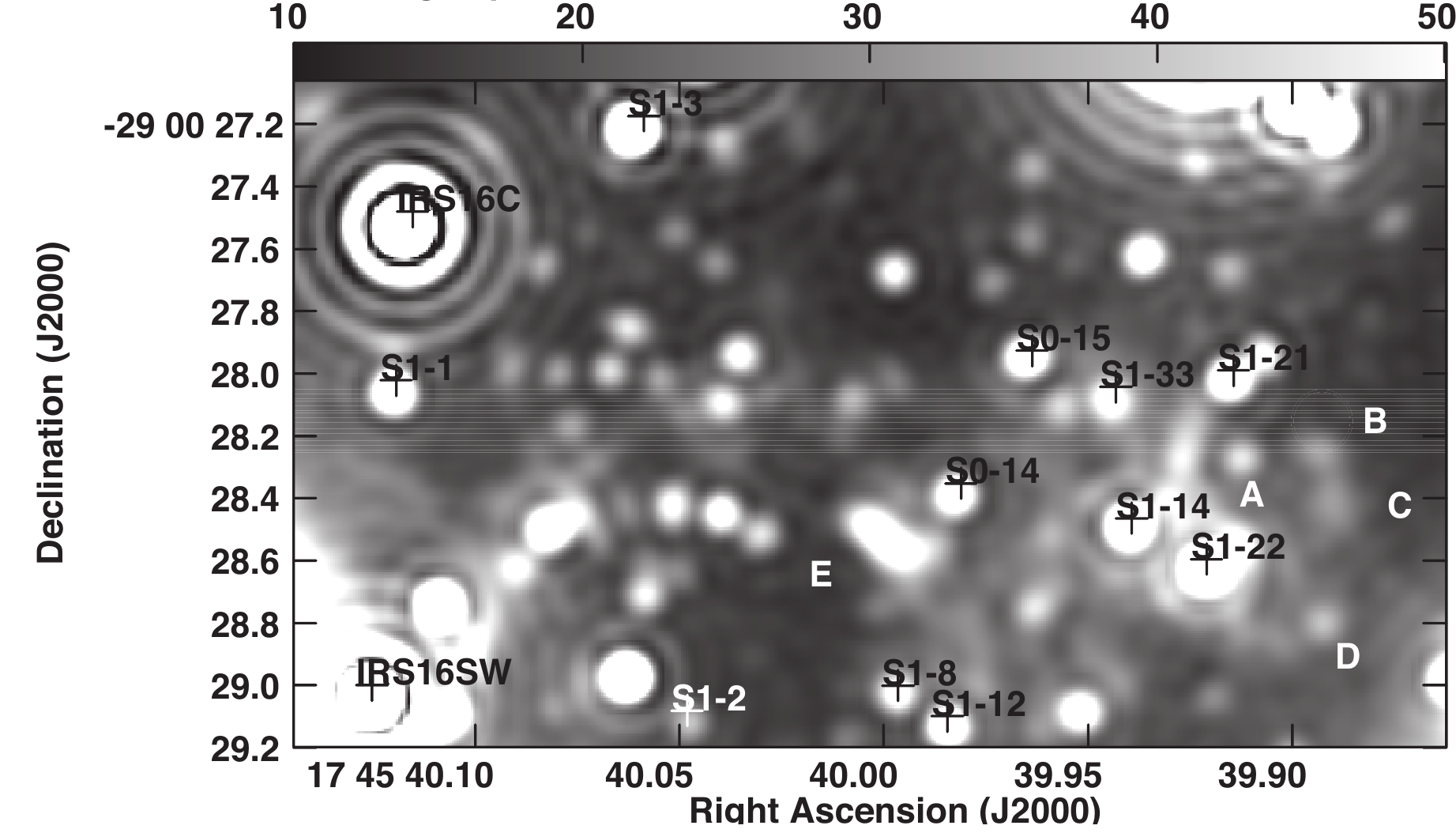}
\end{figure}

\begin{figure}[tbh]
\centering
\includegraphics[scale=0.7, angle=0]{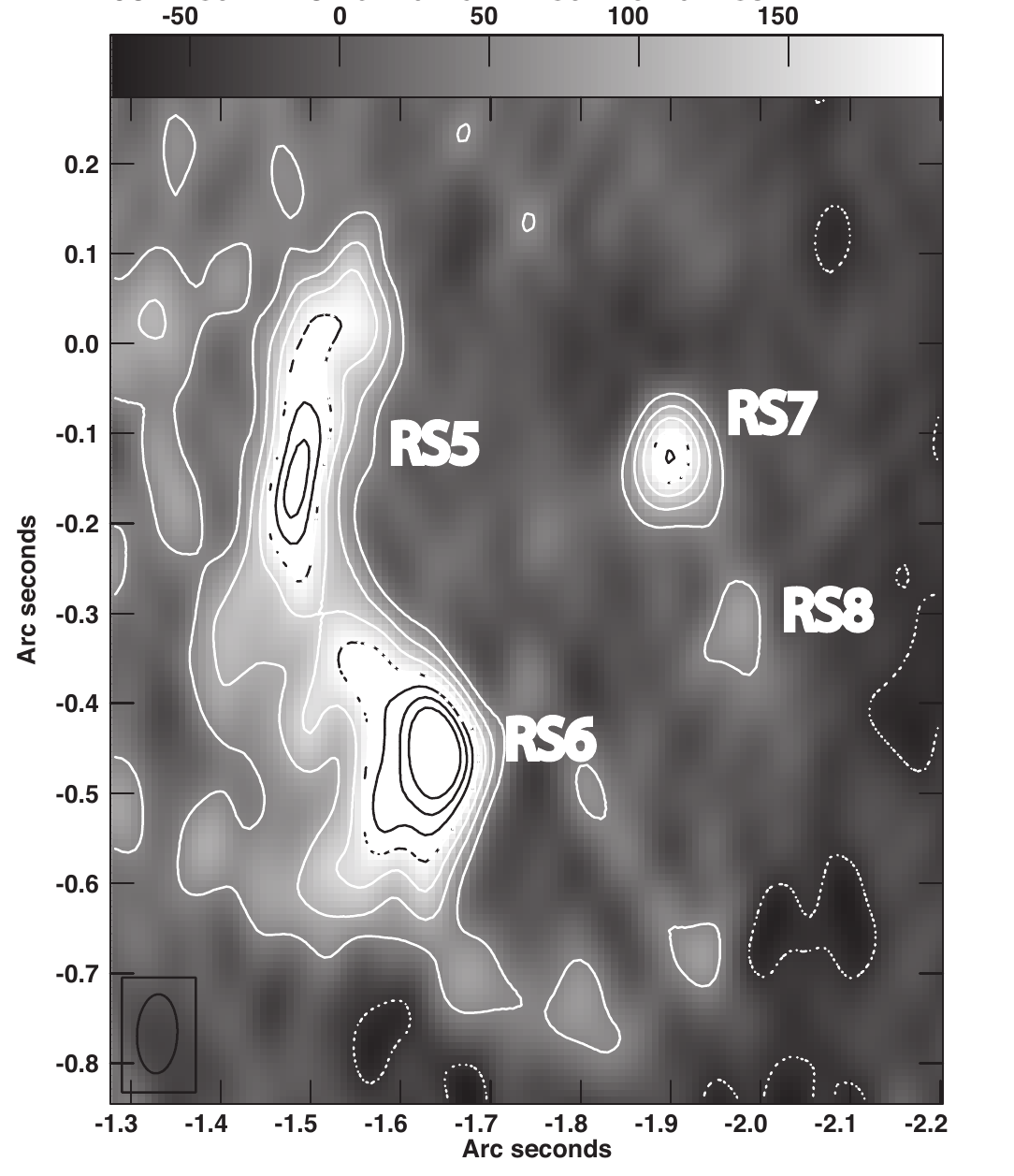}\\
\includegraphics[scale=0.4, angle=0]{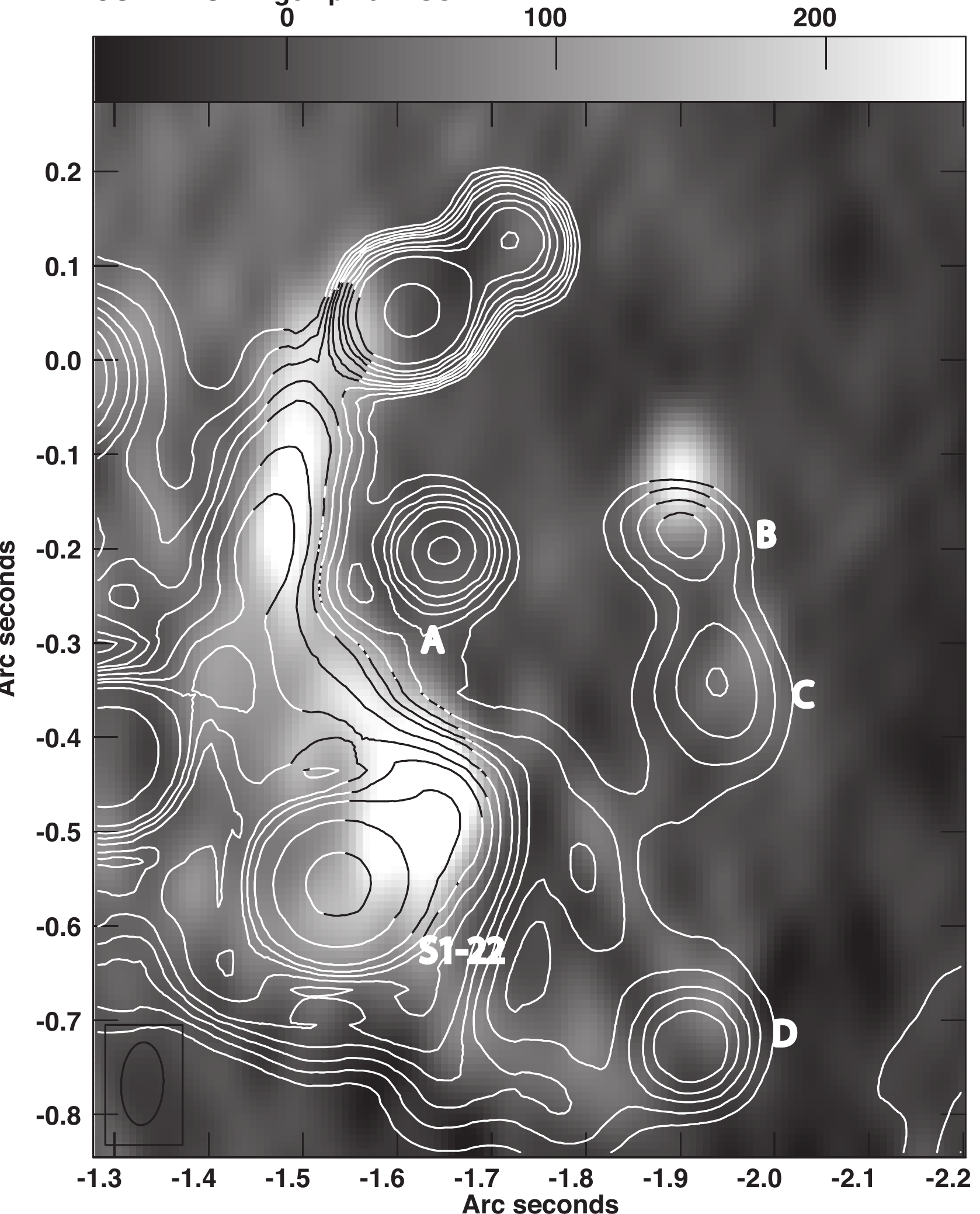}
\includegraphics[scale=0.4, angle=0]{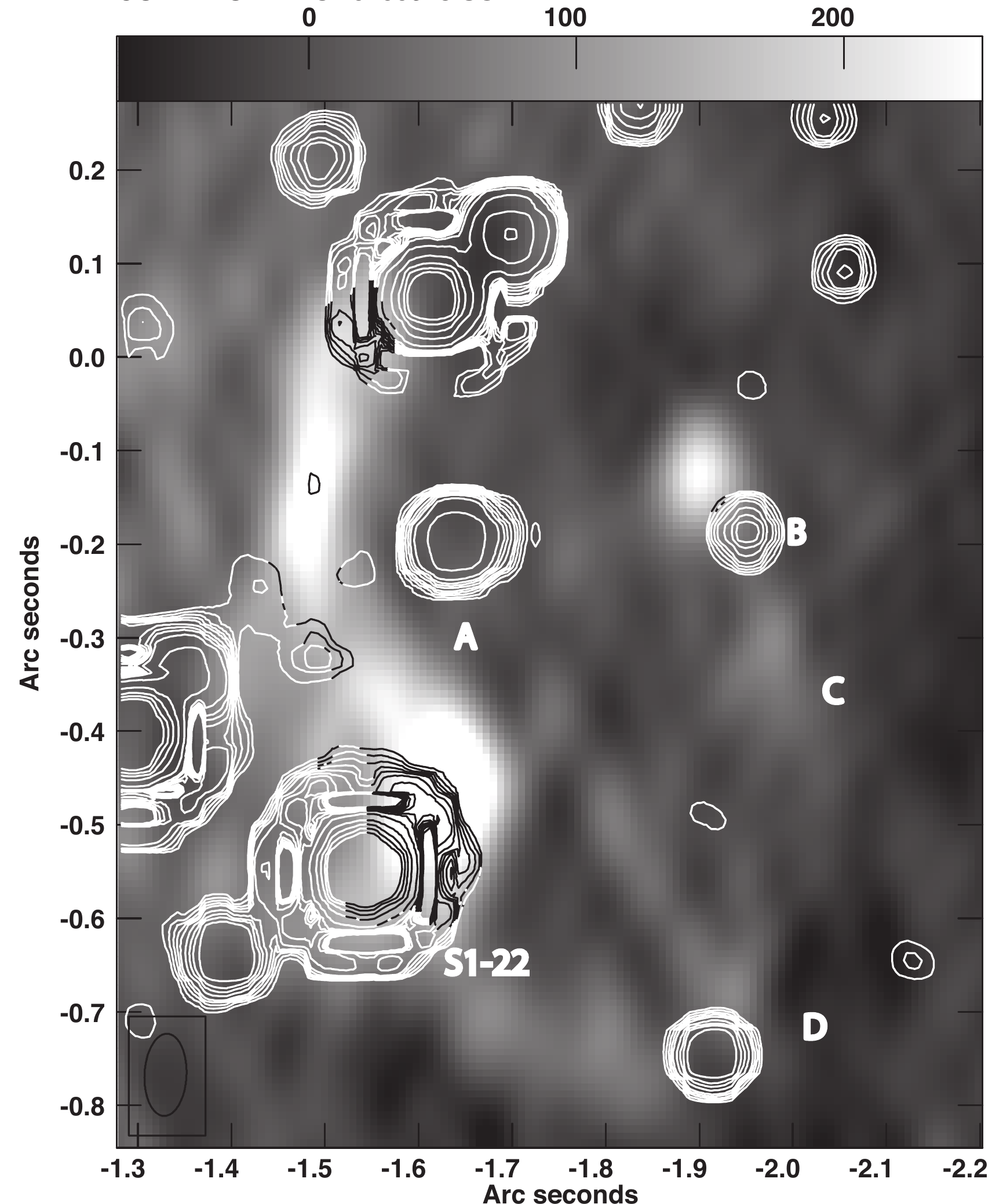}
\caption{\small\small{
{\it (a)}        
Grayscale contours of  34.5 GHz emission 
at -50, 50, 100, 150, ....350\, $\mu$Jy beam$^{-1}$
with a resolution of  879$\times448$ mas (PA=-2.5$^\circ$).
Radio sources are labeled. 
{\it (b)}
The same area as in (a) except that 
contours of L$'$   emission at
2.5, 2.75, 3, 3.25, 3.5, 4, 4.5, 5, 6, 10, 20, 30, 
40, 50) $\times$ 10 mJy
 are superimposed on a  grayscale  34.5 GHz image. 
{\it (c)}
The same as (b) except that 
contours of K$_s$  emission with levels  
2.5, 2.75, 3, 3.25, 3.5, 4, 4.5, 5, 6, 10, 20, 30, 
40, 50) $\times$ 6.5 mJy 
 are superimposed on a grayscale  34.5 GHz image. 
The bar at the top shows the grayscale range in $\mu$Jy. 
}}
\end{figure}

\begin{figure}[tbh]
\centering
\includegraphics[scale=0.7, angle=0]{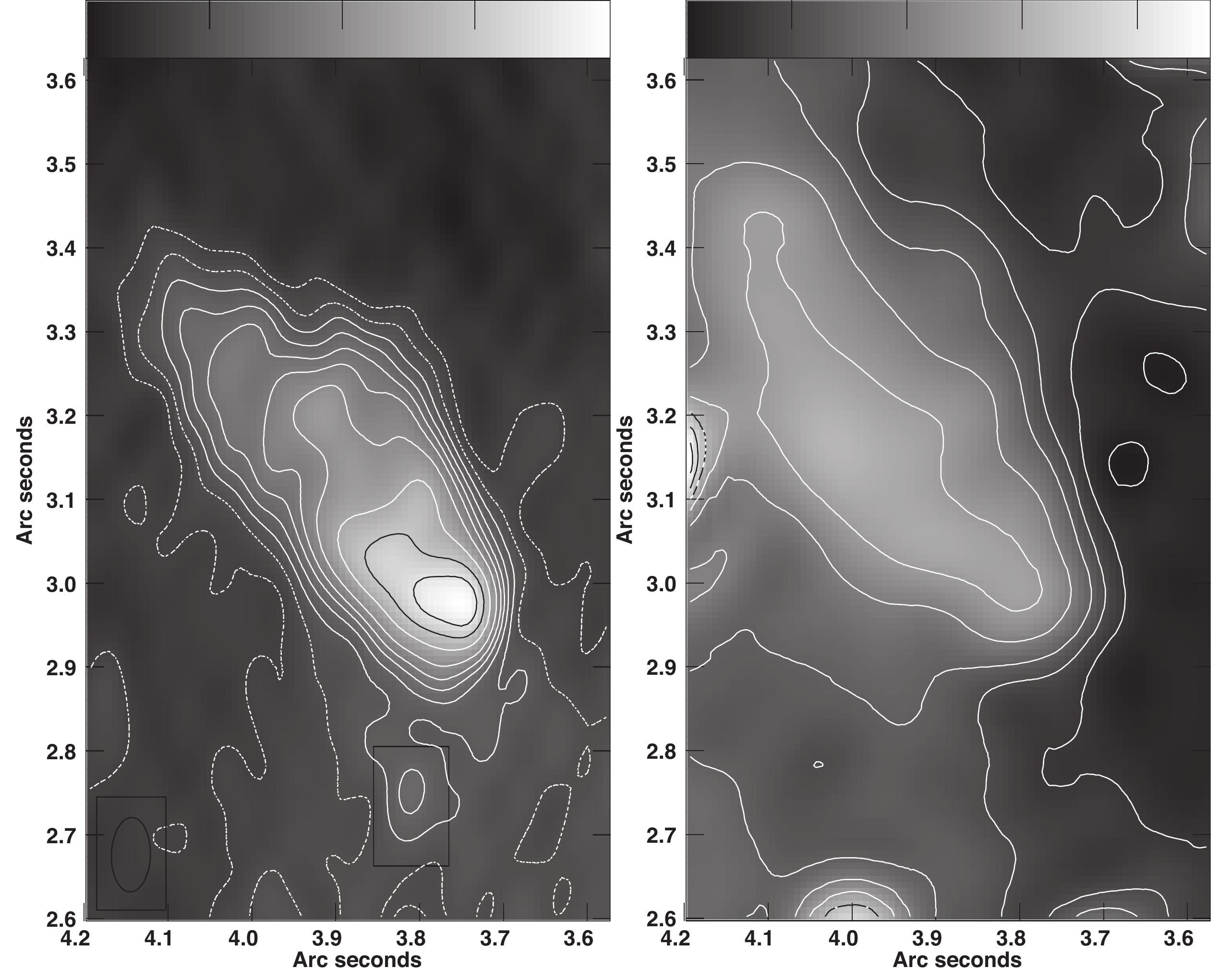}
\caption{
{\it (a)}        
Grayscale contours of 34.5 GHz emission from 
a  cometary radio source (source F1) that points toward Sgr A* 
at (-3, -1, 1, 3, 5, 7, 10, 15, 20, 25, 30) $\times$ 10 $\mu$Jy beam${-1}$. 
with a resolution of  879$\times448$ mas (PA=-2.5$^\circ$).
The range in grayscale is between -100 and 300 $\mu$mJy beam$^{-1}$. 
{\it (b)}  
Similar to (a) except at 3.8$\mu$m with contour levels 
(16, 18, 20, 22, 25, 28, 34, 37) mJy per 27 mas$^2$). 
The range in grayscale is between 15 and 38 mJy. 
{\it (c)} 
A M$'$ band grayscale image at 4.68$\mu$m in reverse color shows 
dusty stars and clouds near Sgr A*.  The inset  to the right shows a blow-up 
of a rectangular  region marked by dashed lines.  Contour levels are set at 
0.02, 0.04, 0.06, 0.08  Jy per arcsecond$^2$.    The labeled MIR cometary is also 
called source F2.      
{\it (d)} 
A grayscale contours of emission from the cometary source F3 at 34.5 GHz at 
at -6, 6, 10, 15, 18, 22, 30, 40, 50 and  60 $\mu$Jy beam$^{-1}$.  The grayscale range 
is between -40 and 200 $\mu$Jy. 
}
\end{figure}

\begin{figure}[tbh]                 
\centering
\includegraphics[scale=0.7, angle=0]{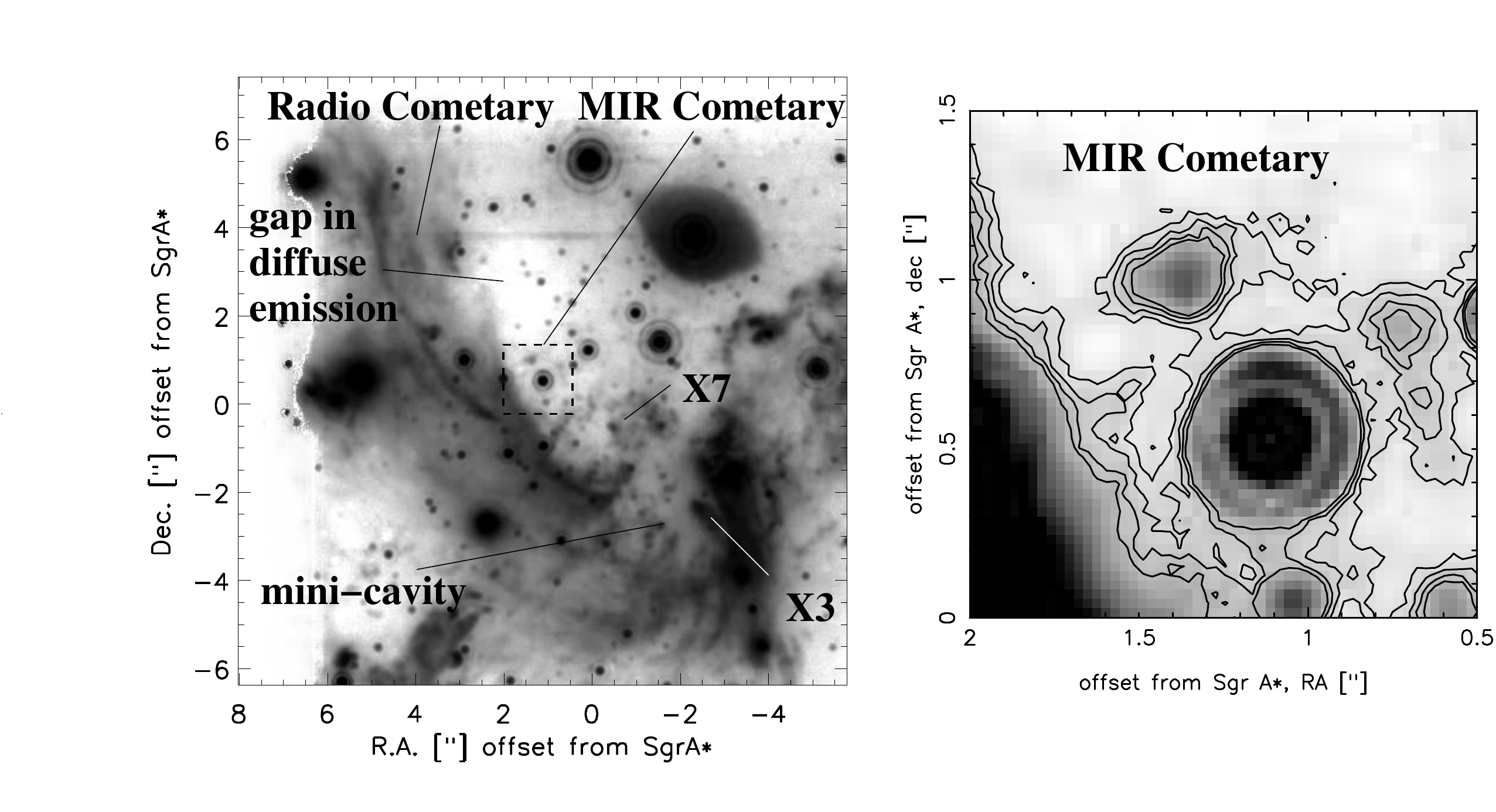}
\includegraphics[scale=0.7, angle=0]{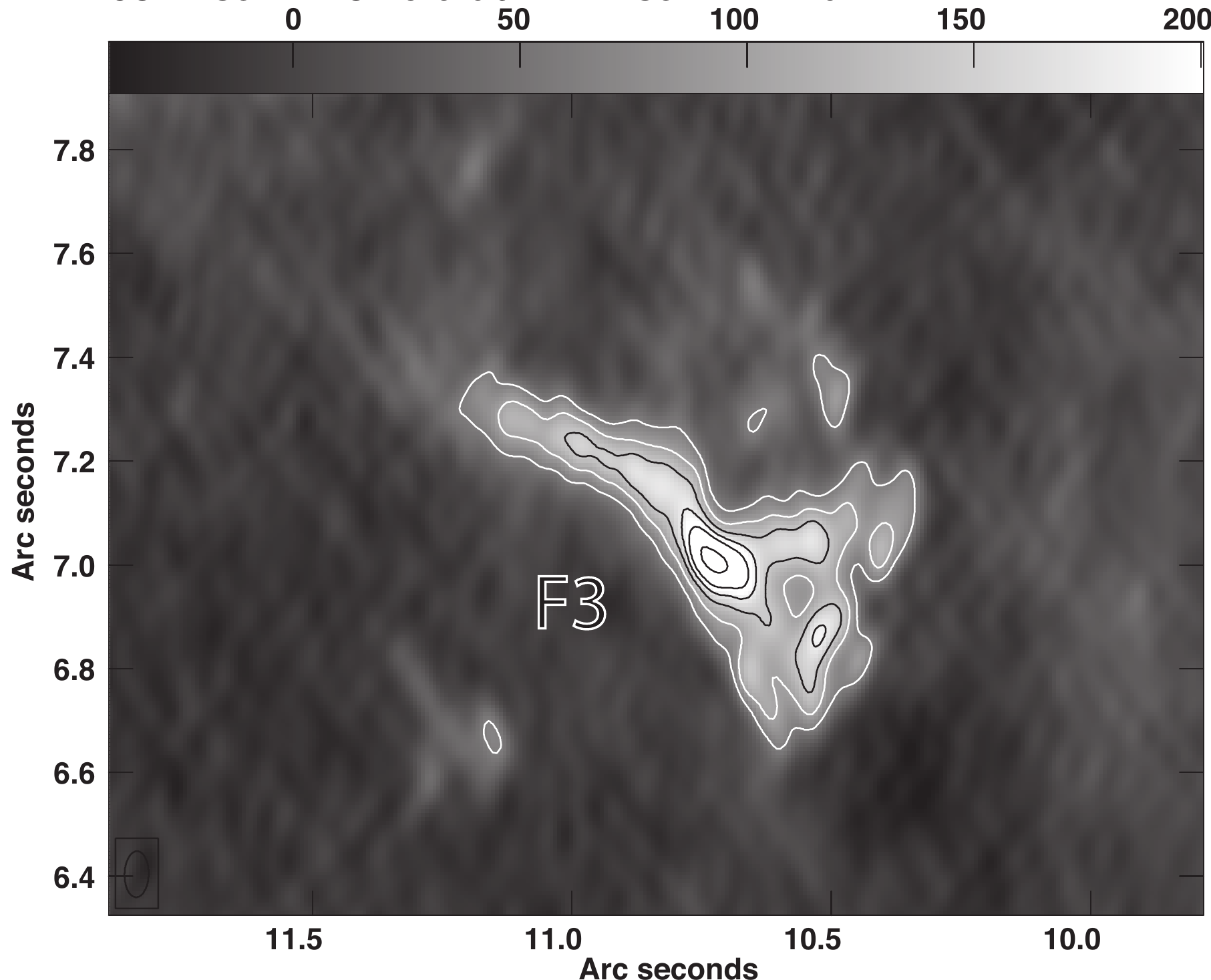}
\end{figure}

\begin{figure}[tbh]
\centering
\includegraphics[scale=0.6, angle=0]{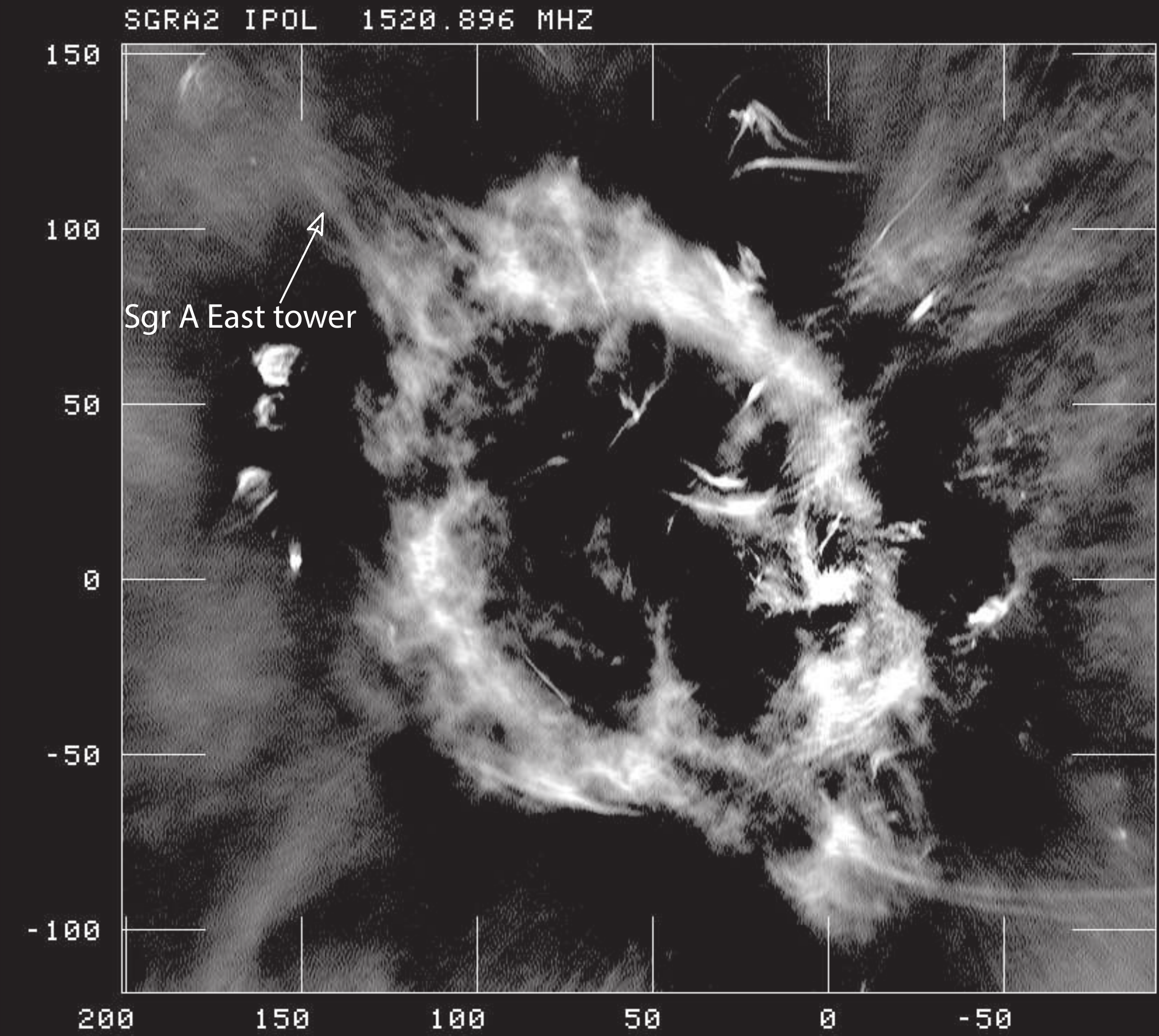}
\includegraphics[scale=0.6, angle=0]{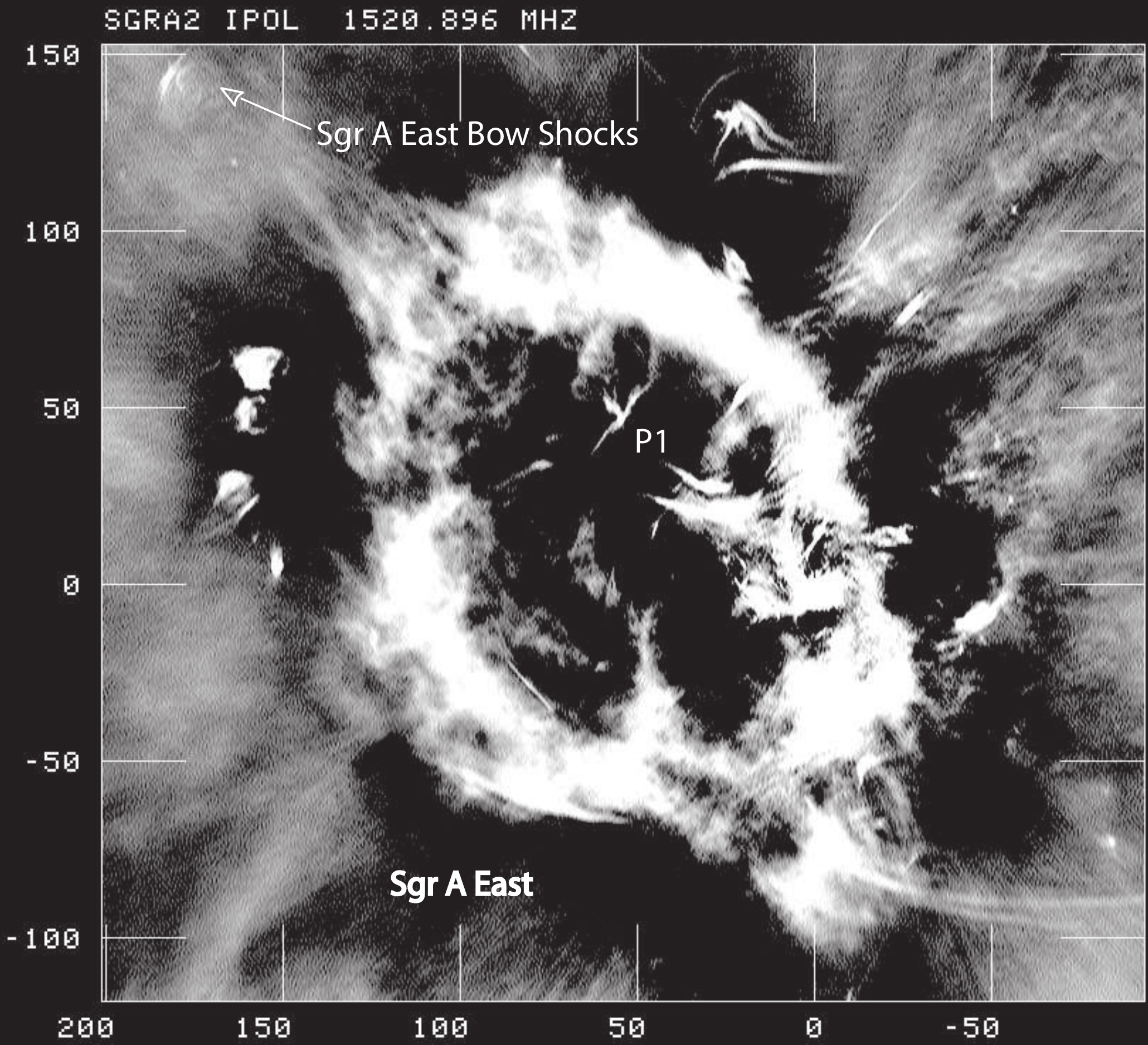}
\caption{
{\it (a)}        
A grayscale scale image of Sgr A East and West at 1.5 GHz. 
with a resolution of  1.4$''\times0.6''$  (PA=-1$^\circ$).
The brightness range of the  grayscale is between 
-1.5$\times10^{-4}$  and 0.002 Jy beam$^{-1}$. 
{\it (b)}  Similar to (a) except that the  grayscale ranges is 
 between -1.5$\times10^{-4}$  and 5$\times10^{-4}$  Jy beam$^{-1}$.  The 1$\sigma$ rms noise
is 53 $\mu$Jy beam$^{-1}$. 
{\it (c)}        
A schematic diagram of features noted at 1.5 GHz. 
}
\end{figure}

\begin{figure}[tbh]
\centering
\includegraphics[scale=0.8, angle=0]{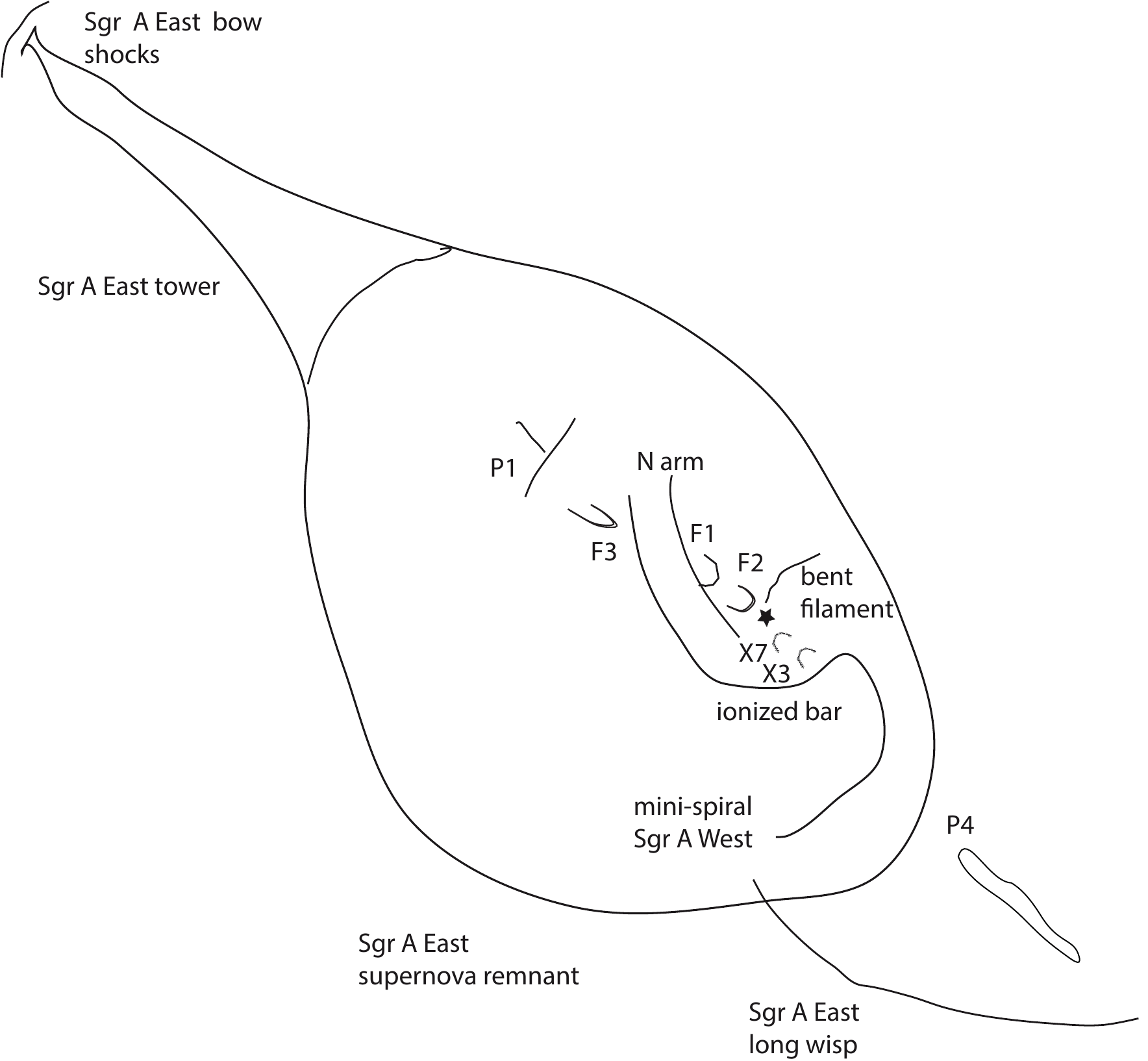}
\end{figure}

\begin{figure}[tbh]
\centering
\includegraphics[scale=0.35, angle=0]{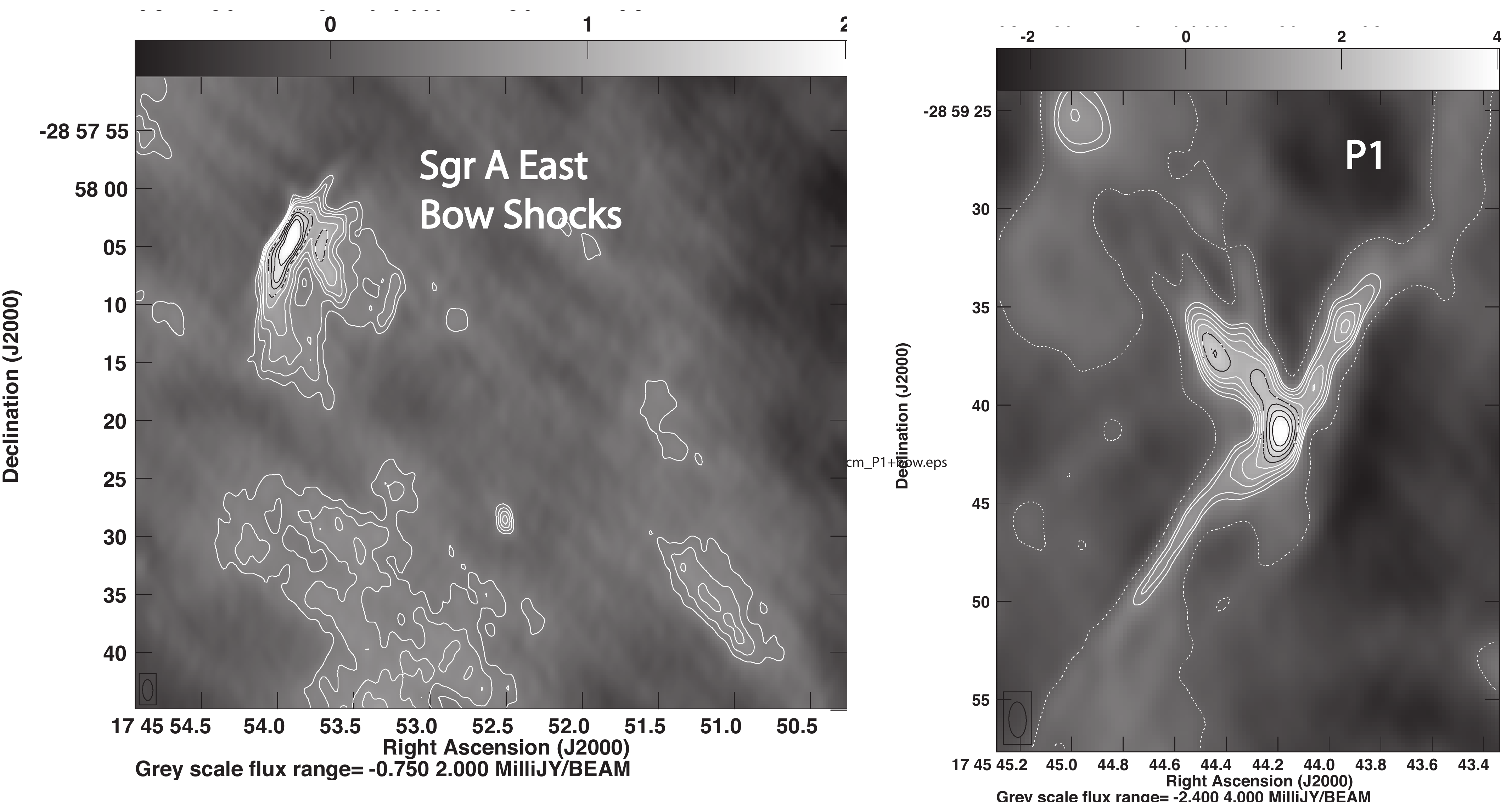}
\includegraphics[scale=0.35, angle=0]{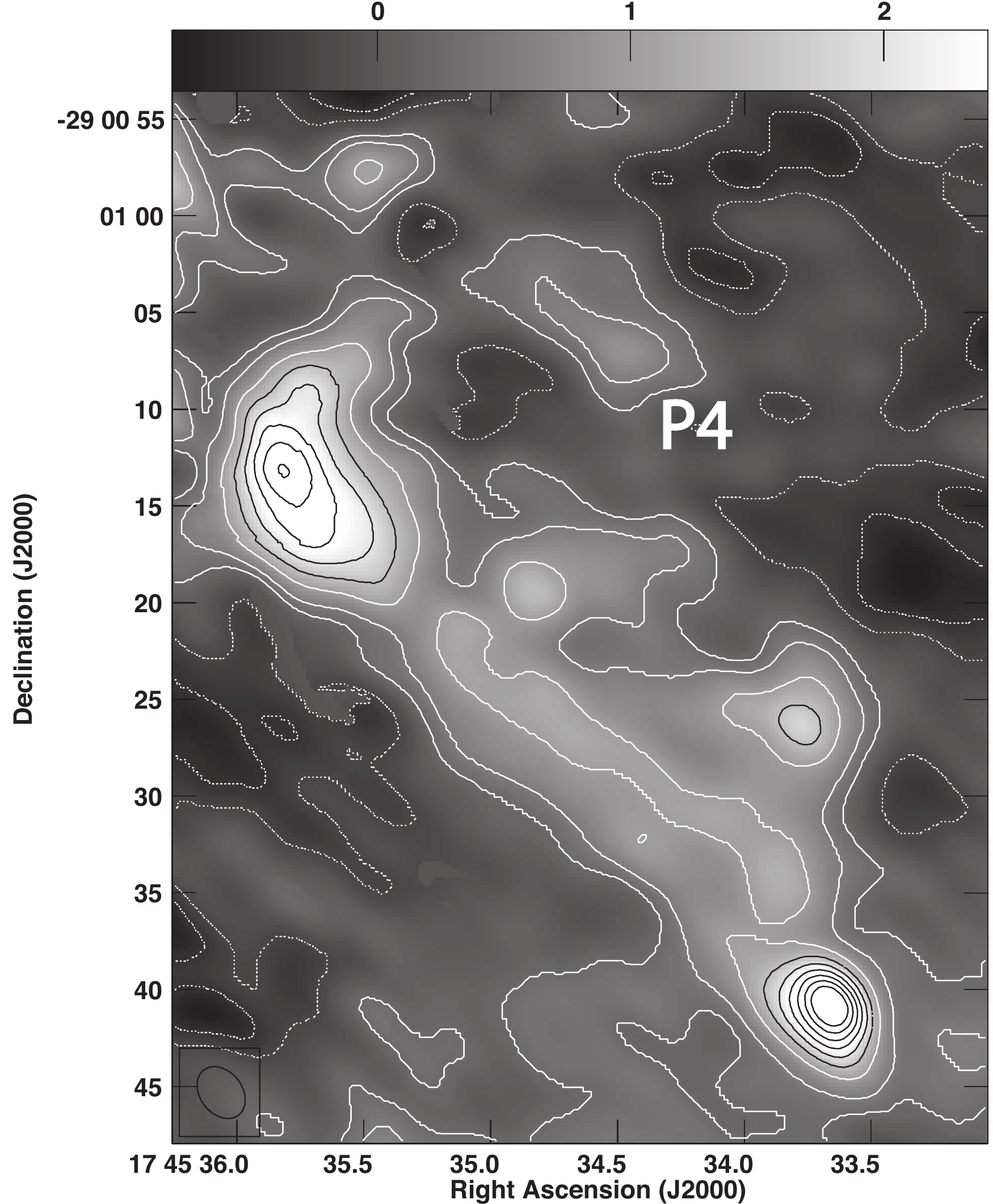}
\caption{
{\it (a - top left)}        
Grayscale contours of the bow shock region  where the Sgr A East 
tower terminates at 1.52 GHz
with a resolution of  1.81$''\times0.89''$ (PA=1.65$^\circ$).
The levels are set at 
 -0.9, 0.4, 0.5, 0.6, 0.7, 0.9, 1.1, 1.3, 1.5, \&  1.7 
 mJy beam$^{-1}$. 
{\it (b - top right)}        
Similar to (a) except that the grayscale contours of  P1
 -0.5, 0.5, 0.7, 1, 1.3, 1.6, 2.0, 2.5, 3.0 \& 3.5
 mJy beam$^{-1}$. 
{\it (c - bottom)}  
The P4 region at 1.52 GHz with 
a resolution of  2.9$''\times2.2''$ (PA=-38$^\circ$).
Grayscale range is between -0.8 and 2.4 mJy beam$^{-1}$.
Contour levels are set at (-1, -0.5, 1, 2, ...,8)$\times0.5$ mJy beam$^{-1}$. 
}
\end{figure}

\begin{figure}[tbh]
\centering
\centering
\includegraphics[scale=0.4, angle=0]{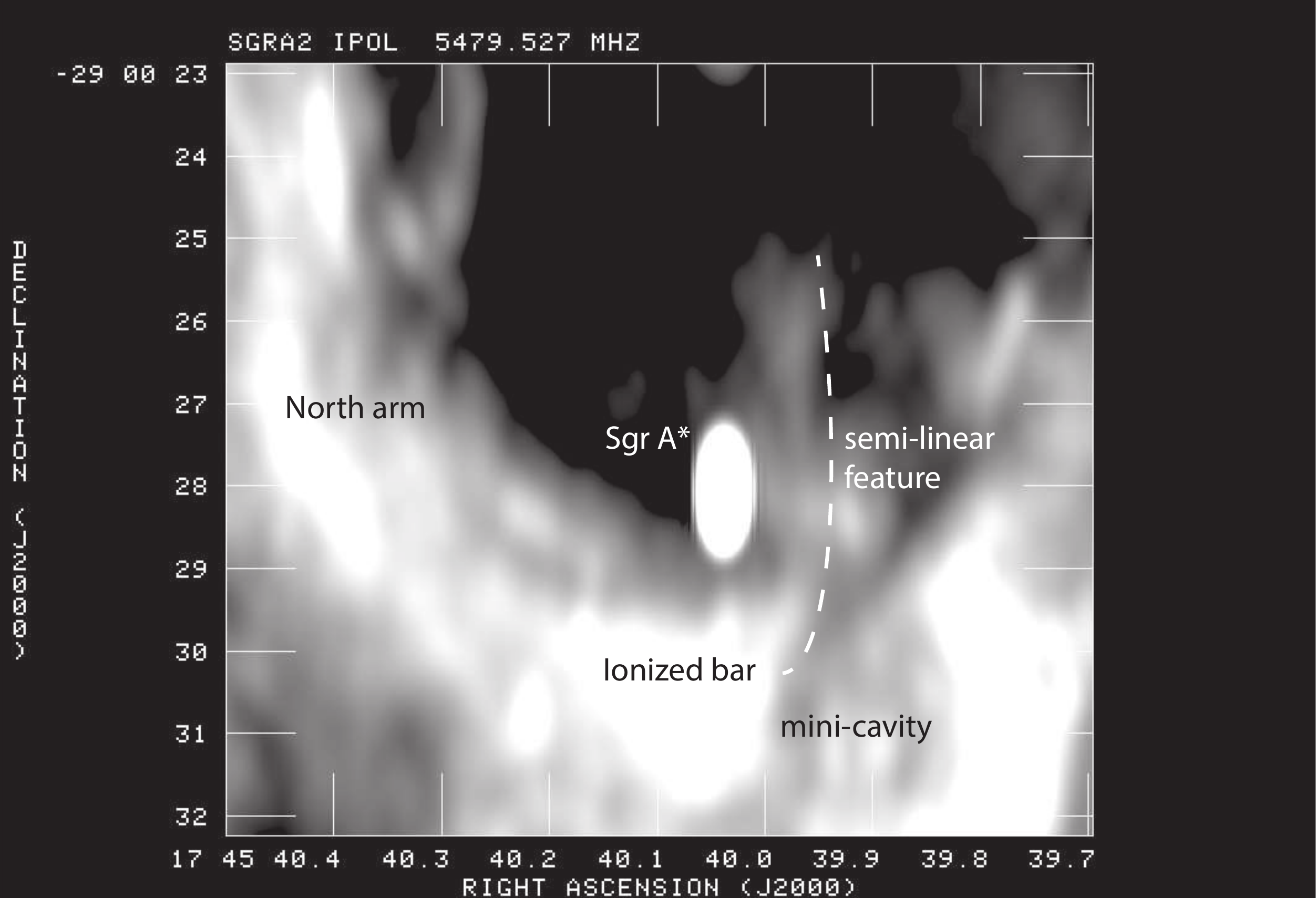}
\includegraphics[scale=0.6, angle=0]{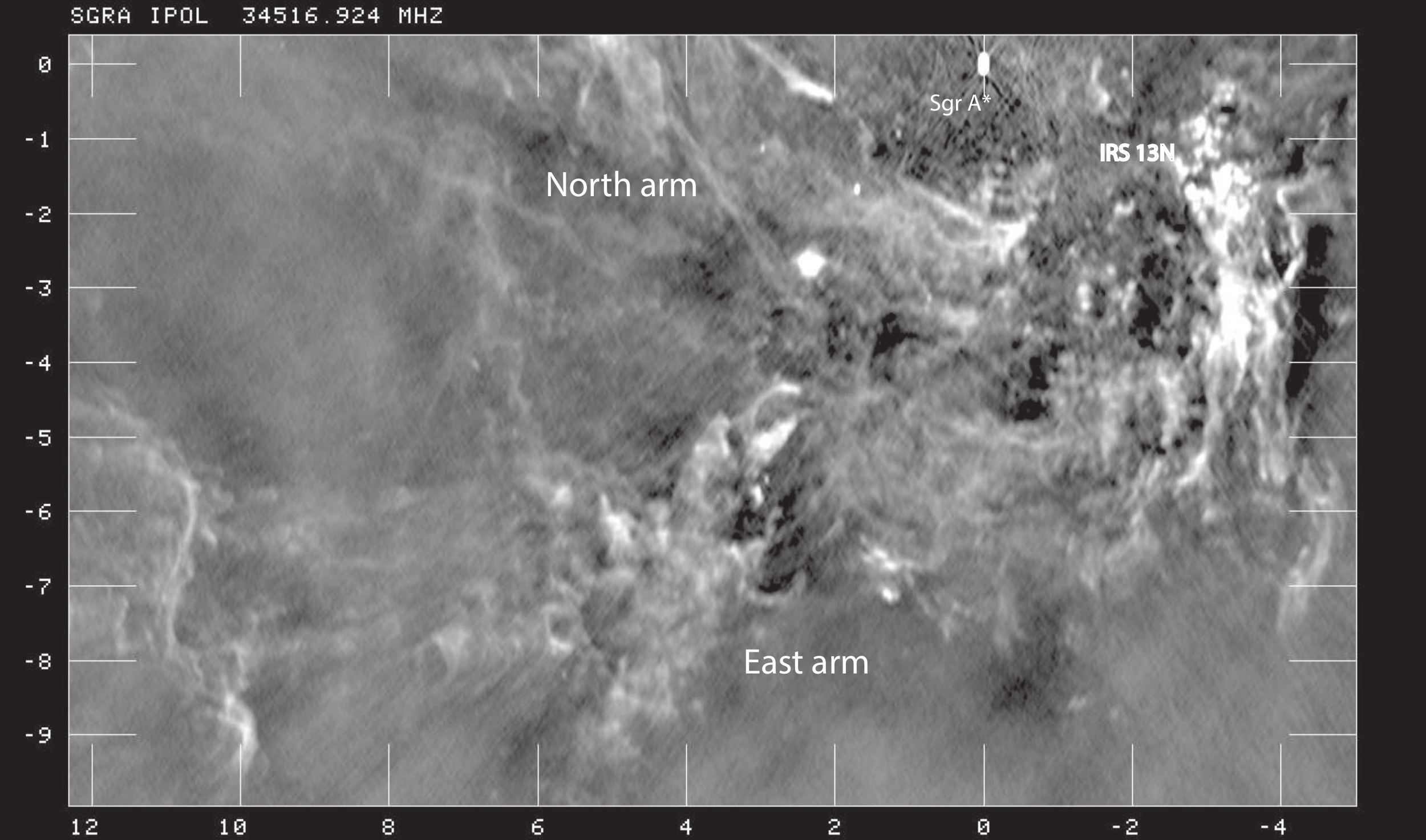}
\caption{{
{\it (a)}        
The same as (b) except at 5.8 GHz with a resolution of
0.59$''\times0.27''$.}
{\it (b)}        
A grayscale 34 GHz continuum image of ionized gas showing  the East  ad North arms  of Sgr A West. 
The spatial resolution of this image is 88.6 $\times 46.5$ mas (PA=-1.56$^\circ$). 
Axes are in the direction of RA and Dec and the coordinates are given in arcseconds with respect to 
Sgr A*.  
}
\end{figure}

\begin{figure}[tbh]
\centering
\includegraphics[scale=0.3, angle=0]{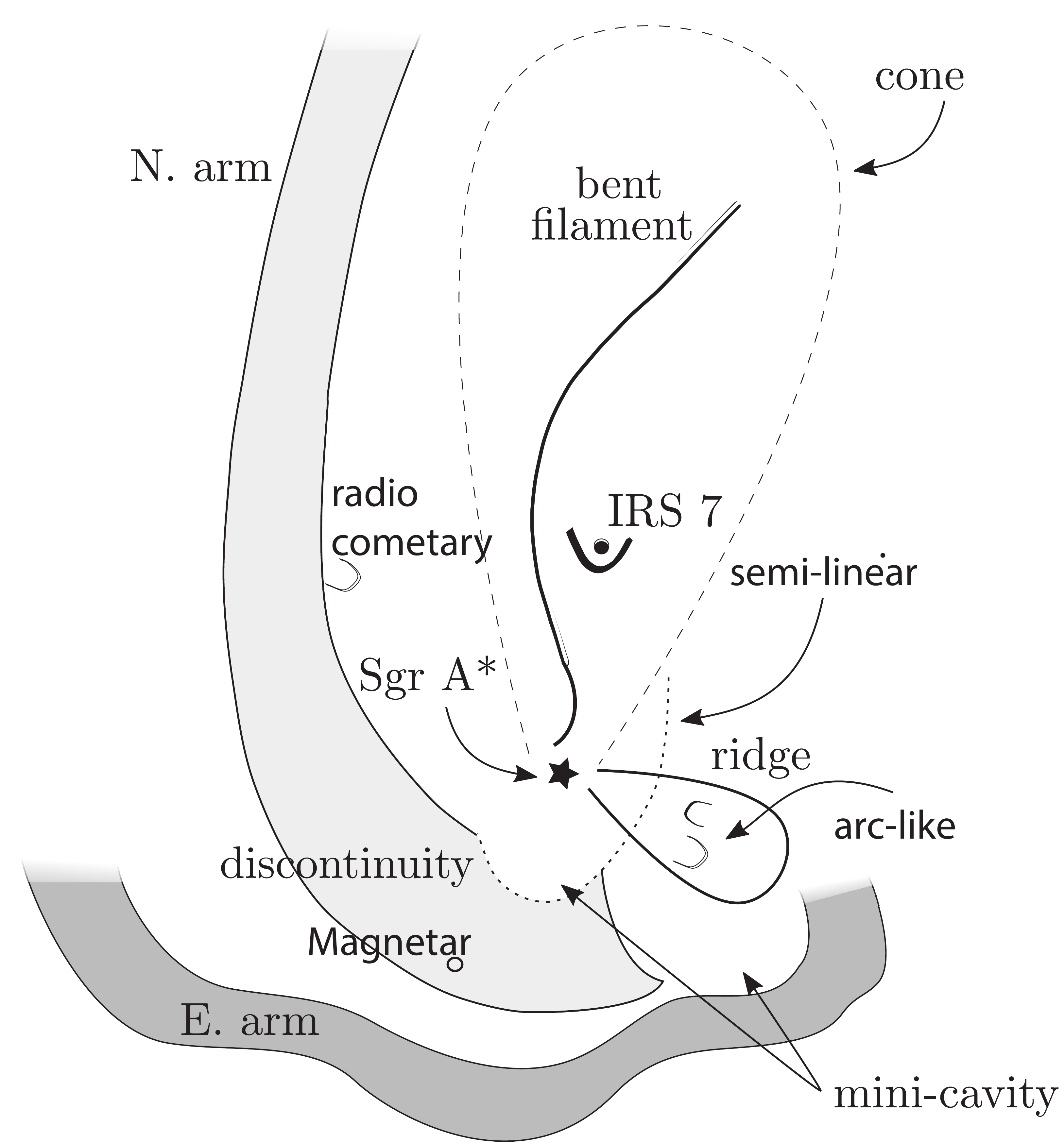}
\caption{\small\small{
%{\it (b)}   
A  diagram of prominent features in Sgr A West within  50$''$ of Sgr A*. 
}}
\end{figure}

%\begin{figure}[p]
%\ContinuedFloat
%\center
%\caption{\small\small{xx}}
%\end{figure}

%\begin{figure}
%\ContinuedFloat
%\center
%\caption{\small\small{xx}}
%\end{figure}

\clearpage

\begin{figure}[tbh]
\centering
\includegraphics[scale=0.5, angle=0]{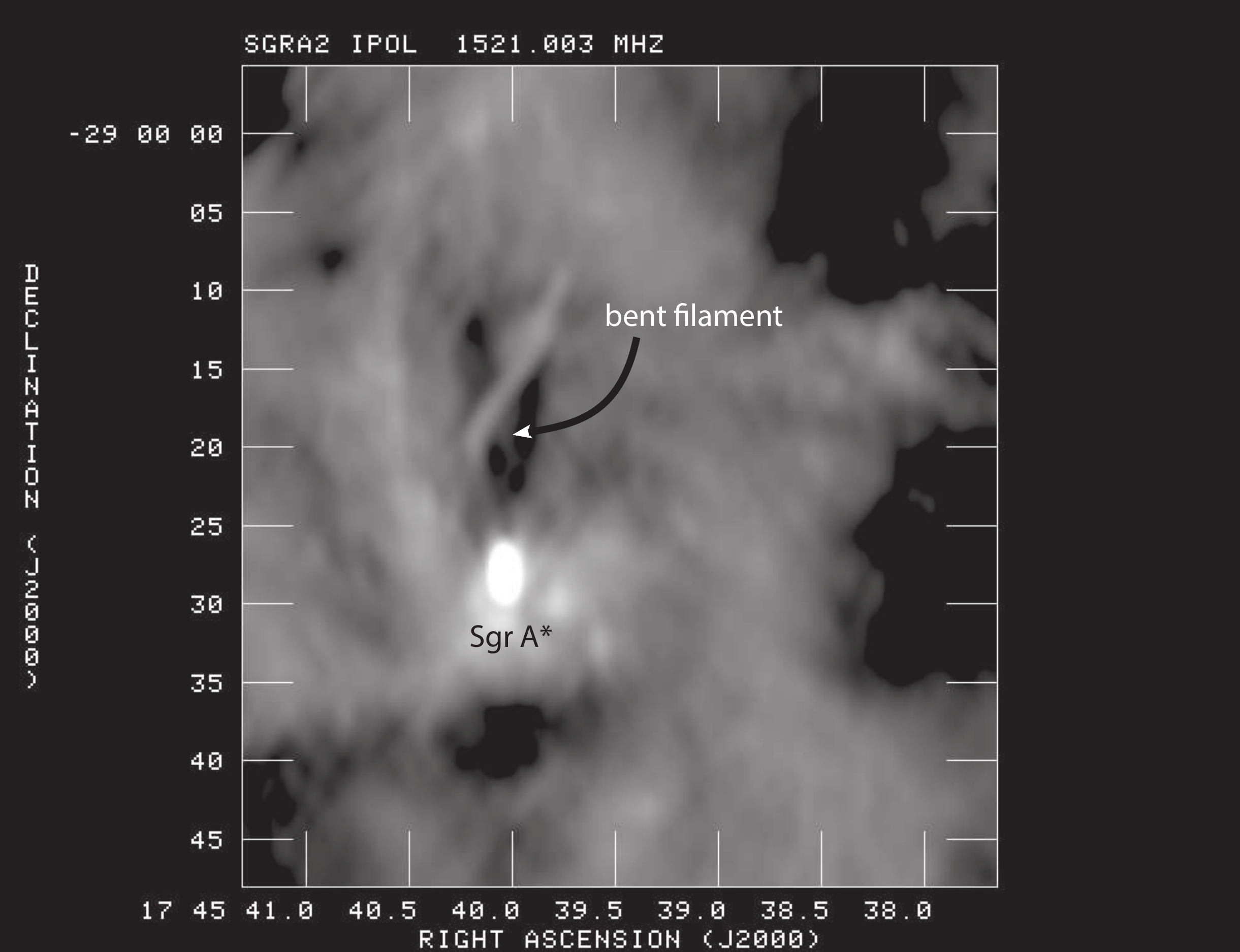}
\includegraphics[scale=0.5, angle=0]{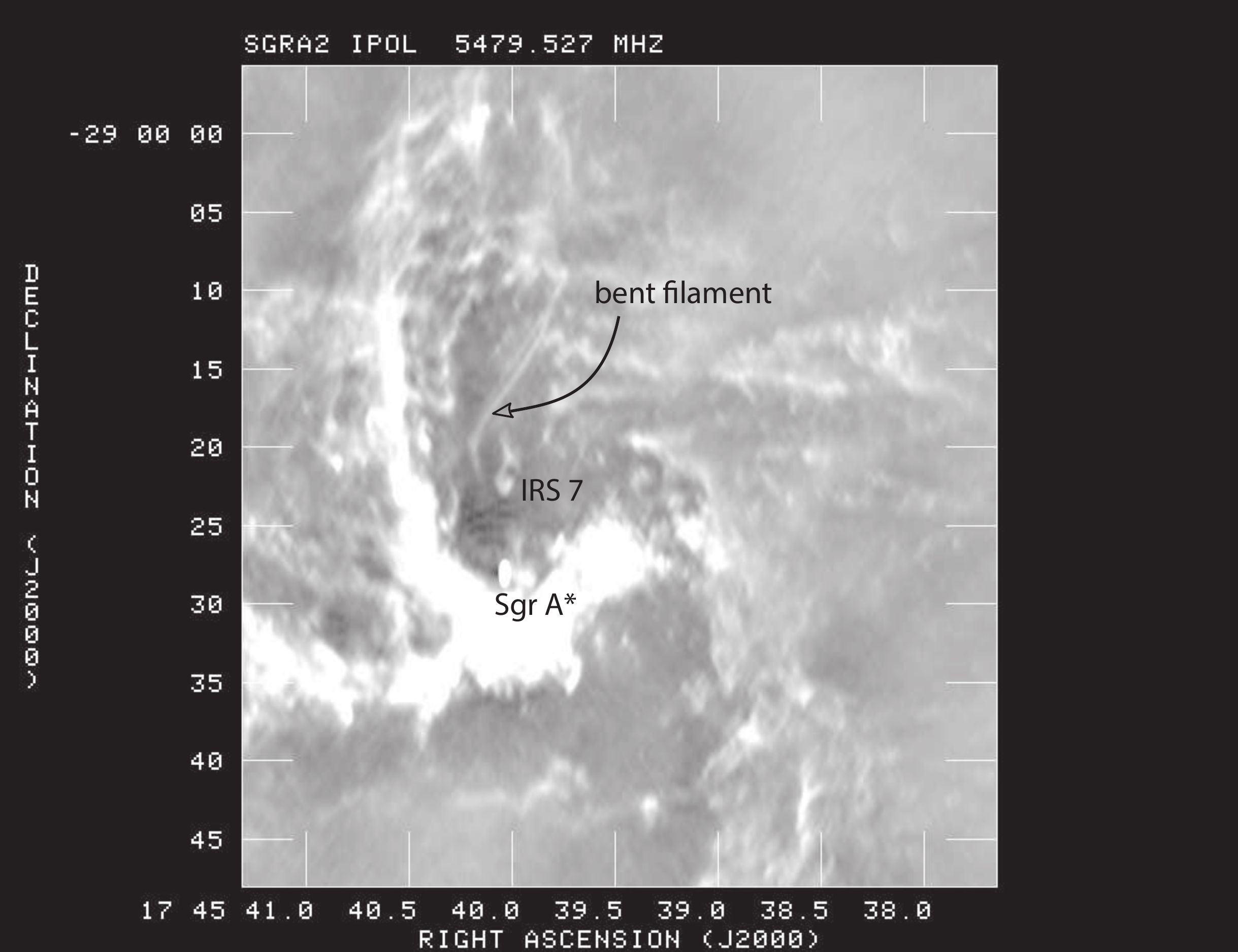}
\caption{{
{\it (a)}        
 A 1.5 GHz continuum  image with a resolution of 
1.87$''\times0.9''$ (PA=0.8$^\circ$) shows Sgr A West and a striking filamentary structure
adjacent to the North arm. 
{\it (b)}        
Similar to (a) except at 5.5 GHz with a spatial resolution of 
0.59$''\times0.27''$. 
{\it (c)} 
A 1.5 GHz image of the bent filament and Sgr A* with a resolution of 
$1.8''\times0.9''$ (PA$1.6^\circ$). 
Sgr A* is marked with a star. 
{\it (d)}
Similar to c) except with a resolution of
$1.4''\times0.6''$ (PA$1.6^\circ$). 
{\it (e)} 
Contours of 8.9 GHz emission from the bent  filament with a resolution of 
$0.34''\times0.19''$ are set at (-0.05, 0.05, 0.1,  0.2,.., 0.7, 1, 1.5, 2) $\times$ 
0.5 mJy beam$^{-1}$. 
}}
\end{figure}

\begin{figure}[tbh]
\centering
\includegraphics[scale=0.3, angle=0]{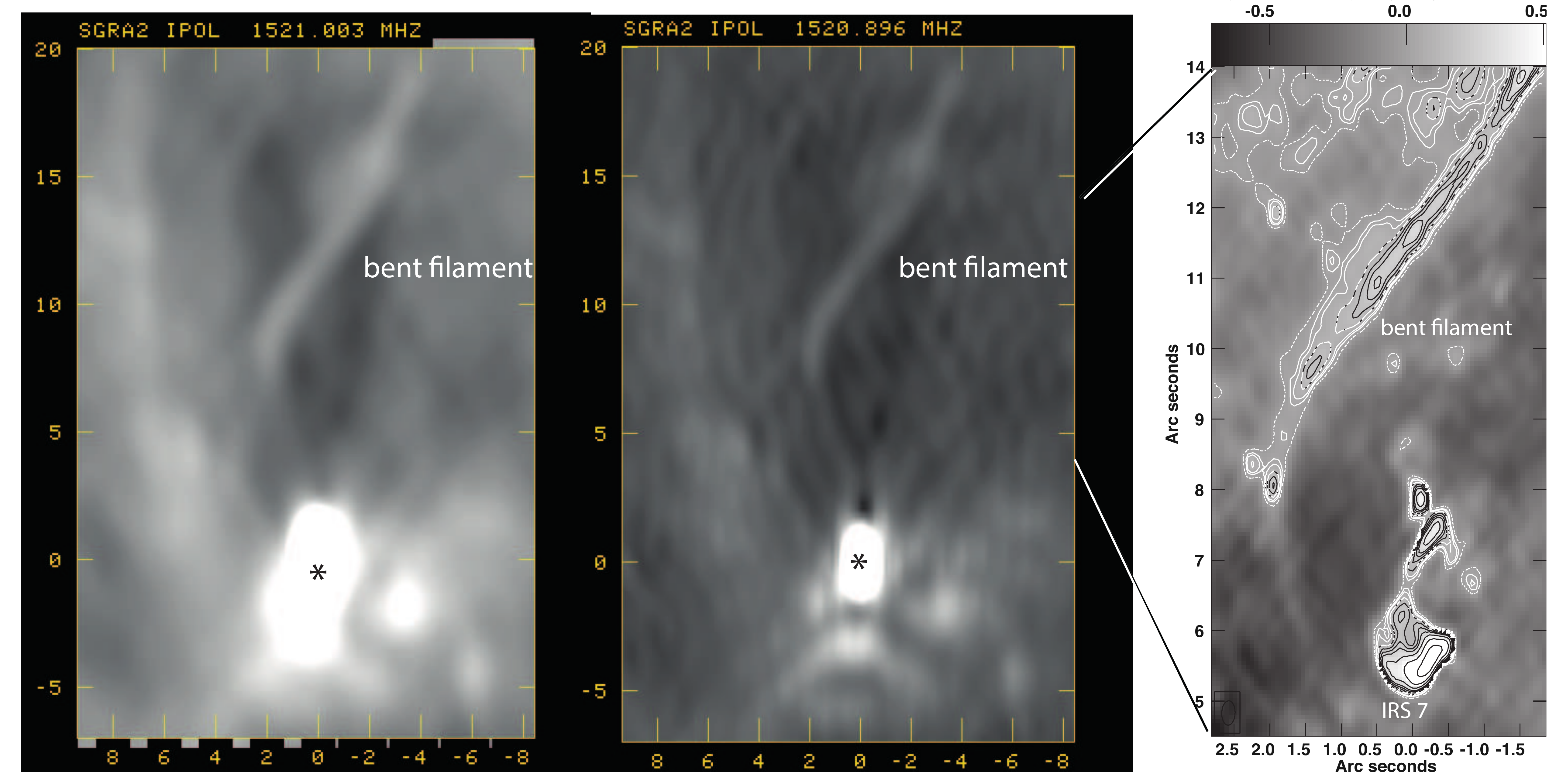}
\end{figure}

%\includegraphics[scale=0.6, angle=0]{f3c_filament_xband.pdf}
%\vfill\eject

\begin{figure}[tbh]
%\begin{figure}[p]
\centering
\includegraphics[scale=0.4,angle=0]{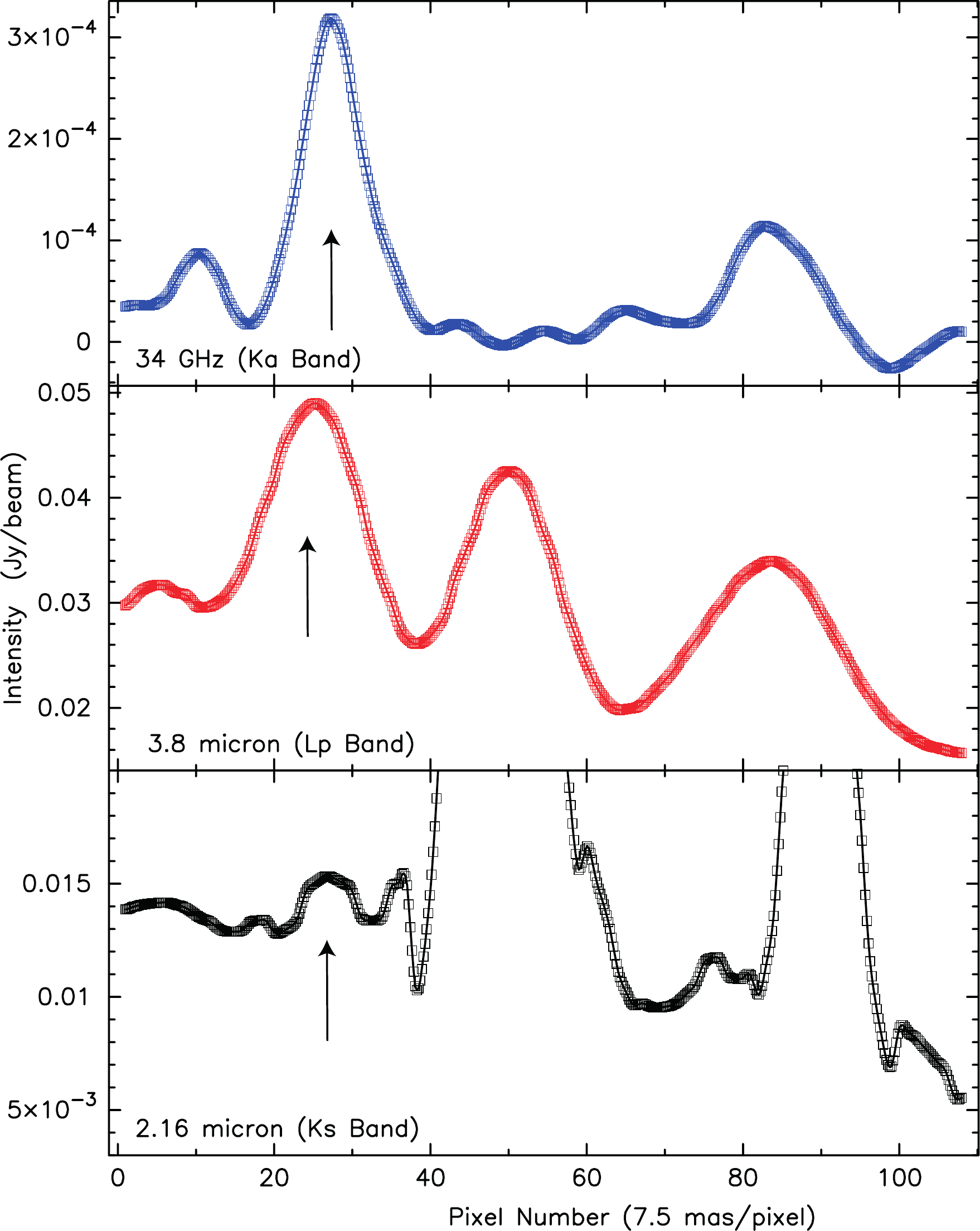}
\includegraphics[scale=0.4,angle=0]{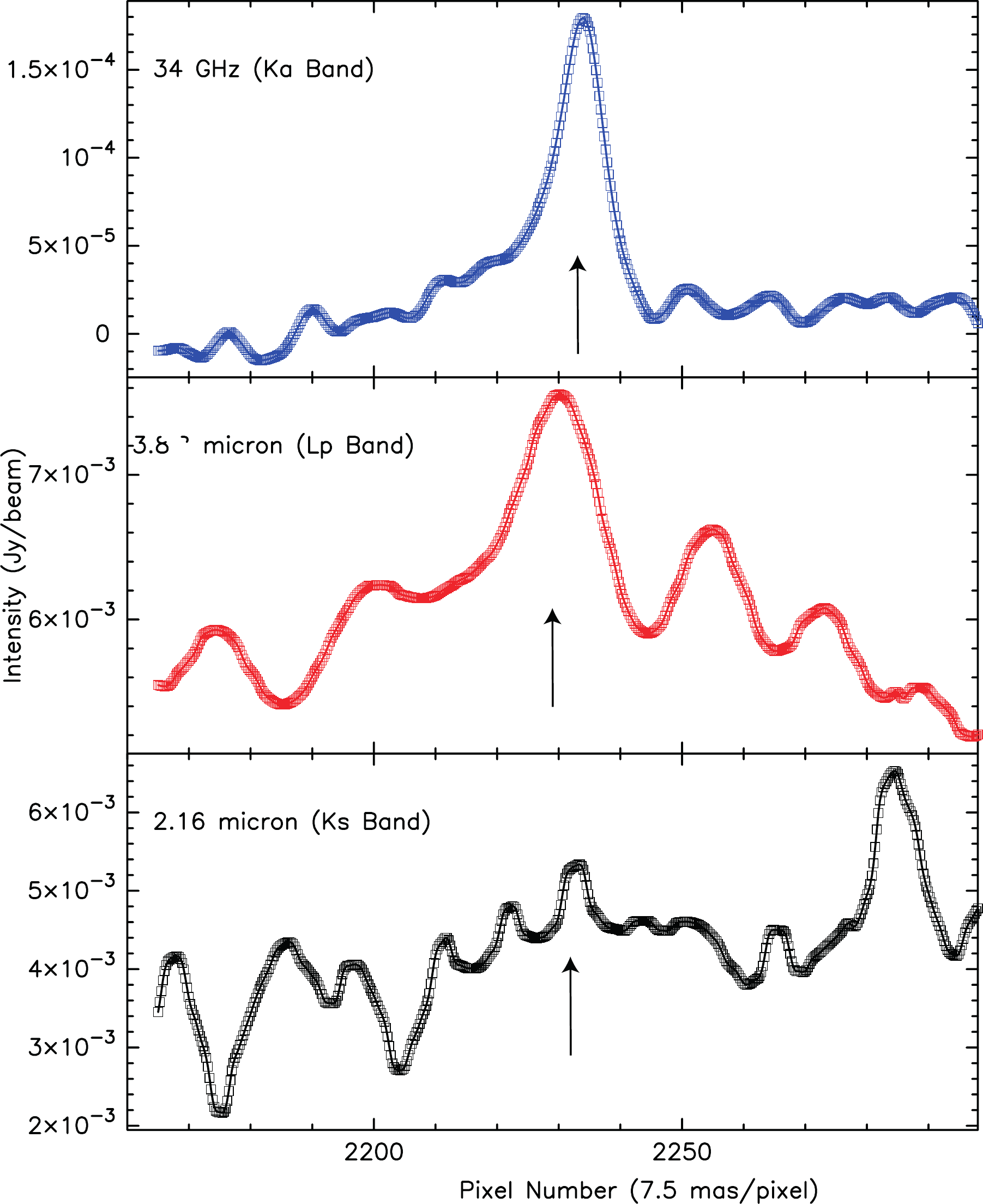}
\caption{
Three panels show the flux density of RS5 (left) and P8 (right) 
as a function of the 
pixel number based on three different registered images
at 34 GHz, 3.8 $\mu$m and 2.16 $\mu$m.  
The  slice for RS5 was made at a constant declination  $-29^\circ\,  00'\,  28''.25$ between 
$17^h\,  45^m\,  39^s.879$  and $17^h\, 45^m\, 39^s.940$ for an extent of ~$\sim0.81''$.
} 
\end{figure}

\vfill\eject

%\begin{figure}[p]

\begin{figure}[tbh]
%\begin{figure}[p]
\centering
\includegraphics[scale=0.9,angle=0]{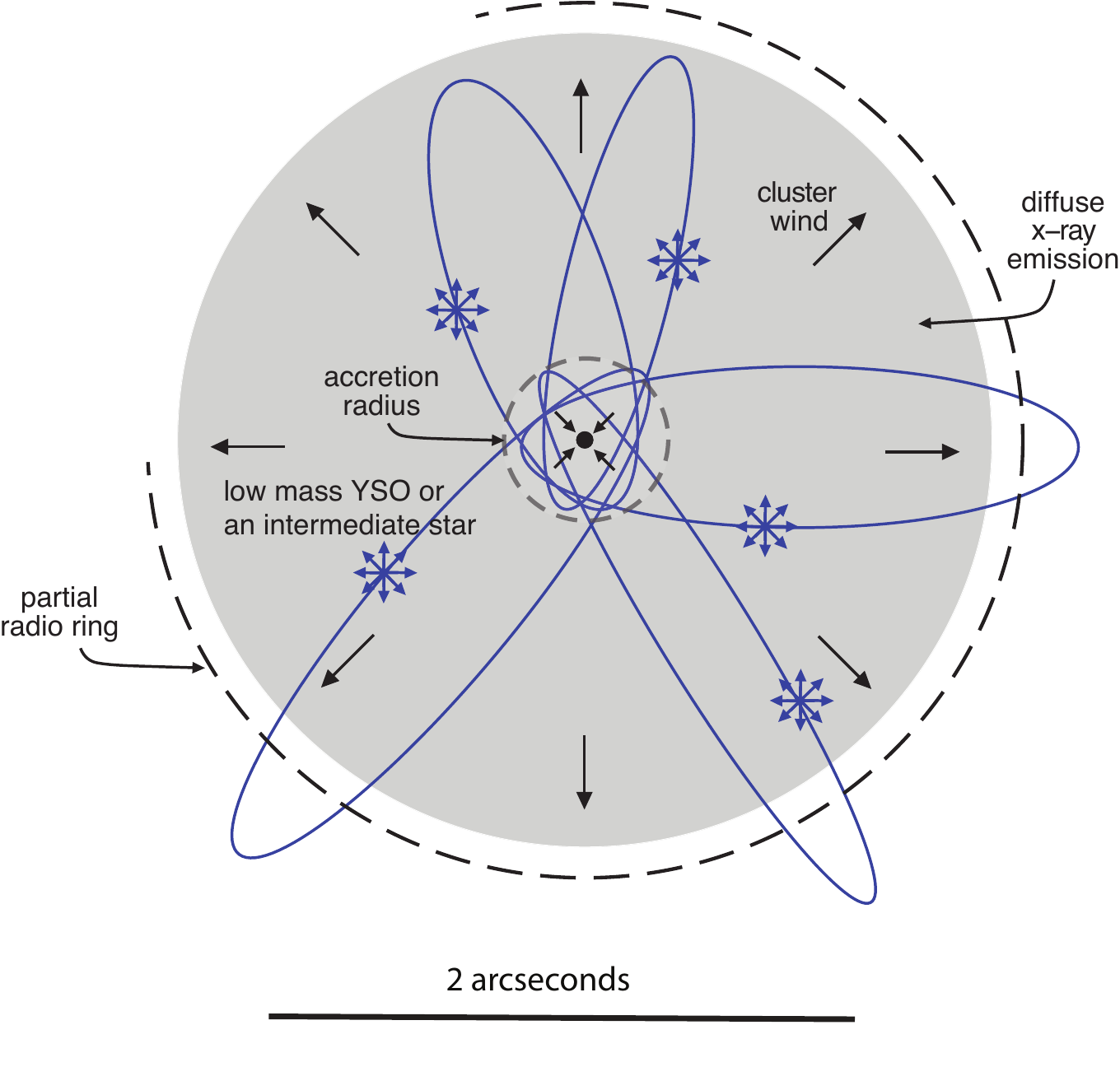}
\caption{ 
A schematic diagram of a  model of the emission 
within  the inner 1$''$ of Sgr A* where the S cluster is located.
The diffuse X-ray  cluster wind   is produced  from merged winds 
of  mass-losing B  stars or material  photoevaporating 
from the disks of young low mass stars that orbit Sgr A*. 
The X-ray cluster  wind plays two roles. First, its pressure  
is sufficient to prevent external material beyond 1$''$ of Sgr A* to approach  Sgr A*. 
Second, the inner  material feeds Sgr A* (c.f. Loeb 2004; Quataert 2004). 
} 
\end{figure}

\end{document}